\documentclass{aa}
\usepackage{graphicx}
\usepackage{txfonts}
\usepackage{natbib}
\bibpunct{(}{)}{;}{a}{}{,} % to follow the A&A style

\newcommand{\M}{\mathrm}

\begin{document}

\title{Mt. Wendelstein imaging of comet 41P/Tuttle-Giacobini-Kresak during the 2017 perihelion arc}
 \subtitle{}
 \titlerunning{Imaging of Comet 41P during the 2017 Perihelion}
 \authorrunning{Boehnhardt et al.}
 
 \author{Hermann Boehnhardt \inst{1}
  \and  Arno Riffeser \inst{2}
  \and  Christoph Ries \inst{2}
  \and  Michael Schmidt \inst{2}
  \and  Ulrich Hopp \inst{2}
}

\institute{Max-Planck-Institut f\"ur Sonnensystemforschung, Justus-von-Liebig-Weg
  3, D-37077 G\"ottingen, Germany
  \thanks{Correspondence: Hermann Boehnhardt, MPI
    f\"ur Sonnensystemforschung, Justus-von-Liebig-Weg 3, D-37077 G\"ottingen,
    Germany; Email: boehnhardt@mps.mpg.de} \and Universit\"ats-Sternwarte,
  Ludwig-Maximilian-Universit\"at M\"unchen,  Scheinerstr. 1, D-81679 M\"unchen,  Germany }
  
\date{Received February 05, 2020 / Accepted April 02, 2020}

\abstract
% Context
{In April 2017 comet 41P/Tuttle-Giacobini-Kresak (41P) passed perihelion close 
to Earth.}
% Aims
{We present observations of the comet and aim to use these to characterize its dust environment
  in the coma and of physical properties of the nucleus.}
% Methods
{41P was imaged in the visible at the Mt. Wendelstein observatory over 45
  nights between 2017 March 11 and 2017 July 17. Photometry of the coma provides
  information on the dust activity of the comet. Nucleus dimensions are
  determined by comparing radial profiles of observed images with those from
  models of the brightness distribution in the near-nucleus coma. Coma and tail
  structures in the images are enhanced using adaptive Laplace filtering and
  image normalization techniques. The geometry of the coma structures found
  are compared to results from dust models of the comet in order to obtain
  information on the nucleus rotation axis and the location of active regions on
  the nucleus.}
% Results
{Analyzing the evolution of the dust activity of the comet versus solar distance
  gives different results depending on the assumed phase angle corrections for
  the dust. Contamination from light emissions of the coma gas modified the
  colors in the coma and the slopes of the radial profiles. The light
  distribution in the very inner coma indicates some dust fragmentation close to
  the nucleus. The appearance of a prominent coma fan during the time period from March to May 2017
  constrains the rotation axis of the nucleus (52 deg obliquity, 50 deg rotation
  angle of the perihelion meridian). This fan is produced by an extended region
  beyond about 50 deg latitude on the nucleus. Two narrow regions at low
  northern to equatorial latitudes were responsible for the time-variable
  occurrence of linear dust jets in the coma. An outburst of about 0.6 mag on
  2017 May 11 generated various dust structures in the coma. The outburst originated from a
  region extending over a latitude range from +30 deg to -5 deg and released
  about 3\,000 tons of dust into the coma. The  brightness of the nucleus indicates an
  equivalent radius of about 600 m and a minimum axis ratio larger than two. 
  }
% Conclusions
{}


\maketitle

\section{Introduction}\label{sec_1}

Comet 41P/Tuttle-Giacobini-Kresak (text abbreviation = 41P) was originally
discovered by H.~P. Tuttle in 1858 and was observed twice more: in 1907 by M. Giacobini and in 1951 by L. Kresak
\citep{1986cspc.book.....B,1964pcc..book.....V} with no
observations in between. 41P is a low inclination (9.2
deg) Jupiter family comet (JFC) with an orbital period of 5.4 years and a current
perihelion distance (1.045 AU) close to the Earth orbit. The orbit evolution of
41P over the past approximately 200 years \citep{1986cspc.book.....B} shows several
encounters with Jupiter from moderate (0.37 AU in 1975) to large (1.60 AU in
1963) distances. Triggered by the initial analysis of
\citet{2015P&SS..118...14F} and based on their own analysis of the orbit
dynamics of 41P over several tens of thousands of years, \citet{2018A&A...615A.154P}
suggest that this comet may belong to the group of near-Earth JFCs with possible
origin in the main asteroid belt.

Despite 41P occasionally passing the Earth at relatively close proximity, only limited
information on its physical properties were collected before the 2017 perihelion
passage of the comet. The absolute coma magnitude (15.3 mag from Jet Propulsion Laboratory 
ephemeris\footnote{ \texttt{https://ssd.jpl.nasa.gov/horizons.cgi}}) indicates
41P to be a moderately active comet. Based on nucleus brightness estimations
published in the literature, \citet{2000A&AS..146...73T} deduced an absolute
magnitude of the nucleus of 18.5 mag. This corresponds to a nucleus radius of
0.7 km for an assumed geometric albedo of 0.04. In late May and early June 1973
two major outbursts were reported with a brightness amplitude of about 9 mag and a coma
expansion speed of more than 500 m/s (\cite{1991kap..book.....K},
\cite{1991kap..book.....H}). These are among the most violent events observed in
comets so far \citep{1976ARA&A..14..143W}. 41P is therefore discussed in the context
of split comets \citep{1991kap..book.....S}, although only indirect evidence
(outburst, nongravitational forces) and no direct subnucleus detection is
reported \citep{1991kap..book.....H}.

The 2017 perihelion passage of 41P allowed a very favorable observing window for
the comet, as it passed close to Earth (closest distance of 0.14 AU on
2017 April 01) and was clearly observable from Earth for 6 months in
2017. From this apparition of the comet, \citet{2018Natur.553..186B} provide
evidence that the apparent rotation period of the nucleus may have increased
from about 20 h to more than 46 h between March and May 2017. The increase of the
rotation period is analyzed in greater detail by \citet{2019AJ....157..108S}
based on their  observing campaign of the
comet of about 4.5 months. \citet{2018A+A...619A.156M} provide results on the gas (OH, CN, C$_2$,
C$_3$, NH) and dust activity ($\M{af}\rho$ values) of
41P. \citet{2018A&A...615A.154P} also present first modeling results on the dust
coma. Both of these latter papers are based on observations of 41P made with the Trappist 0.6m
telescopes between 2017 February 16 and 2017 July 20. \citet{2019MNRAS.485.4013L} found
rapid variations of the dust color in 41P. \citet{2019SoSyR..53...91R} presented
polarization measurements of 41P from a single epoch in April 2017. Other partly
preliminary results were reported on volatile species \citep[C$_2$H$_2$, HCN,
C$_2$H$_6$ and H$_2$CO; ][]{2017DPS....4930509M,2018AAS...23114413M}, the OH
production rate and gas expansion velocity \citep{2017DPS....4942004L}, and the
radar size of the nucleus \citep{2017DPS....4941424H}.
 
Here, we describe the results from the 2017 imaging campaign toward comet 41P using
the 2m Fraunhofer telescope at the Mt. Wendelstein Observatory, Bavaria/Germany
(section \ref{sec_2}). We discuss our findings with respect to the coma environment and dust
activity of the comet (section \ref{sec_3.1}) as well as to the size of the nucleus 
(section \ref{sec_3.2}). Furthermore, we use coma structures to
assess the dust tail in the coma (section \ref{sec_3.6}) as well as the rotation
axis and the locations of active regions on the nucleus (section
\ref{sec_3.7}). In section \ref{sec_3.8} we analyze a singular outburst event
seen in the coma of 41P on 2017 May 11. Section \ref{sec_4} presents the
conclusions.

\section{Observations, data reduction, and analysis methods}\label{sec_2}

\subsection{Observations}\label{sec_2.1}

Observations of comet 41P were performed with the 2m Fraunhofer Telescope at
the Mt. Wendelstein Observatory of the Ludwig-Maximilian University Munich
over 45 nights between 2017 March 11 and 2017 July 17 (Table~\ref{tab1}). During
that time interval the comet moved from 1.14 AU pre- to 1.61 AU
post-perihelion. 41P was closest to Earth on 2017 April 01 (0.14 AU). The
observations cover a phase angle range of the comet from 14 to 69 deg.

\begin{table}[!]
  \centering
  \tiny
  \begin{tabular}{c c c c}       
    \hline\hline 
    & $\lambda_{\M{eff}}$ [nm] & $\lambda_{\M{blue}}$ [nm] & $\lambda_{\M{red}}$ [nm] \\
    \hline
    \multicolumn{4}{c}{g band} \\      
    \hline
    WWFI      &   483.4  &  418  &  550 \\
    PS1       &   483.2  &  414  &  552 \\
    SDSS      &   462.7  &  402  &  533 \\
    \hline
    \multicolumn{4}{c}{r band} \\      
    \hline
    WWFI      &   620.6  &  557  &  695 \\
    3KK       &   630.3  &  562  &  697 \\
    PS1       &   618.8  &  548  &  694 \\
    SDSS      &   613.9  &  559  &  675 \\
    \hline
    \multicolumn{4}{c}{i band} \\      
    \hline
    WWFI      &   761.2  &  698  &  836 \\
    3KK       &   759.5  &  682  &  841 \\
    PS1       &   752.5  &  688  &  824 \\
    SDSS      &   746.7  &  690  &  817 \\
    \hline
  \end{tabular}
  \caption{Filters used at the Mt. Wendelstein 2m Fraunhofer telescope instruments WWFI 
    \citep{2014ExA....38..213K} and 3KK \citep{2010SPIE.7735E..3QL} 
    in comparison to the original SDSS filters \citep{2010AJ....139.1628D} and 
    the realization at the PS1 (\cite{2012ApJ...750...99T}). 
    The all-sky catalog produced by the latter is used for calibration of the observed 
    images of 41P. The table lists for each filter the effective wavelength 
    $\lambda_{\M{eff}}$ and the short- ($\lambda_{\M{blue}}$) and long-wavelength 
    $\lambda_{\M{red}}$) for 50\%\ transmission of the filter band passes.} 
  \label{tab2}
\end{table}

Two imaging instruments were used for the comet observations, namely the 3KK
multi-channel imager \citep{2010SPIE.7735E..3QL} and the Wendelstein wide field
imager WWFI \citep{2014ExA....38..213K}. The 3KK instrument allows simultaneous
exposures in three filter bands, namely two in the visible and one in the 
near-infrared. For this analysis of the 41P observations, only the Sloan Digital 
Sky Survey SDSS r and SDSS i images were used (image pixel scale = 0.200 
arcsec/pixel) . Single 
exposures provide a field of view of $7\times 7$ arcmin on the $2048 \times 2048$ 
pixels CCD detectors. The WWFI instrument (pixel scale = 0.199 arcsec/pixel) took 
image series of the comet sequentially through SDSS g, SDSS r, and SDSS i 
filters. Those data provide a field of view of 27.6 x 29.0 arcmin on a 
mosaic of four 4096 x 4096 pixel CCD detectors. Both instruments are mounted on 
the telescope and allowed us to obtain images of 41P with both on several nights. 
The typical nightly observing interval of 41P
extended over about 15 to 60 min; on a few nights the comet was exposed over
longer time intervals or during a second time window. The log file of the 41P
observations is compiled in Table~\ref{tab1}, and information on the filter
properties is listed in Table~\ref{tab2}.

The exposure series of the comet consist of sequences of 60~sec exposures
with jitter offsets in between the individual filter exposures (WWFI) or filter
set exposures (3KK). Non-sidereal tracking was applied to the telescope to
compensate the differential motion of the comet. The exposure time of the
individual comet images was chosen to avoid saturation of the central part of
the coma as well as to avoid smearing of the coma center beyond actual seeing
conditions. In addition to the images of the comet, the usual set of calibration
exposures (bias, sky flat-field) was taken. Since the flux calibration is
performed by means of background stars in the field of view of the science
target, exposures of standard star fields were not collected.

\subsection{Data reduction}\label{sec_2.2}

Data reduction was performed using a pipeline process developed and tuned at
the University Observatory Munich specifically for the Mt. Wendelstein 2m
telescope and the WWFI camera; it is described for the application to comet
observations in \citet{2016MNRAS.462S.376B} and references therein. This
pipeline was adapted to the 3KK images in the visible wavelength range and
applied in a similar way to the comet observations. The pipeline process
corrects for the bias and flat-field properties as well as for the sky background
and for detector artifacts (i.e., `hot' and `cold' pixels, cosmic rays) in the 
images. It also provides zero points (to convert photons
per second to magnitudes) for the exposures using background stars
for the individual images from the Pan-STARRS-3 Pi source catalog (PV3) given 
in AB magnitudes \citep{2013ApJS..205...20M} and corrected to the SDSS gri 
filter system. 
Over 90\% of the WWFI r frames have photometric zero points between 23.2 and 25.2 
mag which were determined using between 200 and 17,000 reference stars; for 3KK 
the number of reference stars is down by a factor of 16 compared to the WWFI case.
For WWFI, a purely photon error for the zero point determination would be smaller 
than 0.01 mag (having at least 200 stars brighter than 17 mag and a minimum sky 
brightness of 17.5 mag/arcsec$^2$). For images with over 10,000 reference stars, 
our large aperture of 10 arcsec may count multiple sources, thereby shifting 
the zero point to a slightly higher value of maximum 0.03 mag (e.g., in the most 
extreme case of 17,000 stars on 2017 July 17). Color terms of $0.035(r-i)$ between 
WWFI and SDSS as measured by \citep{2014ExA....38..213K} play only a minor role, 
having colors in the range $-0.1 < (r-i) < 0.1$.

The data post-processing for the preparation of the analysis of 41P properties
requires mean- and median-averaged composite frames of the comet, which are calculated
from the flux-calibrated image series of a single night - or typically one-hour
 segments in cases of extended or multiple observing windows per
night. Before averaging the image series, we aligned the individual frames
to the central brightness peak in the coma. The centering accuracy is
estimated to be well within 0.1 arcsec (10 to 46 km at the distance of the
comet). From the average composite images a minor sky background level was
subtracted that remained from the averaging process of the individual
sky-removed images. The optimum sky level is determined via the method of the
asymptotic radial profile fitting of the cometary coma in the median-averaged
image. We estimate 
the flux calibration error on a case-by-case basis from the 
nonflatness and bias variations with typical values of less than 1.5\%, from
the sky subtraction and remnant star signal in the coma below 2\%, and from 
the zero points to be better than 1.5\%. Overall, the flux calibration of the
comet signal should be better than 3 to 4\%.

\subsection{Analysis methods}\label{sec_2.3}

For the assessment of the coma activity and in particular for the search of coma
and tail structures in the comet, beyond visual inspection of the
images, we used structure enhancement techniques, such as an adaptive Laplace filter
\citep{1994A&AS..107..101B} and radial normalization
\citep{1986Natur.324..649A}. Both numerical recipes were applied to the
individual images as well as to the nightly or hourly averaged composite frames.

The global coma status is investigated by means of radial coma profiles obtained
from the median averaged composite images of the comet. The radial profiles are
calculated by integrating and then averaging comet flux in concentric 
rings of increasing
radius centered on the brightness peak in the coma. For a quantitative
assessment, the gradient slope in the average radial profile is approximated in
linear fits to the log-log pairs for the flux versus the distance at the comet
using different distance ranges in the coma.

The light scattering and emission behavior of the coma is analyzed through color
maps of the coma which are  produced from nightly median composite images of the SDSS gri
filter exposures. The respective procedure calculates the apparent magnitude in
the median composite filter images and subtracts filter pair images (g-r, g-i,
r-i) after proper centering to the brightness peak in the coma.

We also calculate the model flux distribution of the inner coma region within 8 
arcsec from the central brightness peak in the comet image. This corresponds to 
about 200-900 km distance from the nucleus depending on the observation date. The 
model includes flux contributions of light reflected by the coma and the nucleus, 
and it considers the atmospheric smearing of the signal. This new model allows the 
assessment of the contribution of nucleus-reflected sunlight. In the following, we 
describe the respective analysis steps  in detail.

The flux from the coma is represented by a radial power-law profile that is used
to create a 2D image around the nucleus position. The central
coma pixel (i.e., the coma at the nucleus position) is set to the level of the
closest neighboring distance in the radial profile. In order to mimic the 
object-specific asymmetric light distribution in the coma, the 2D symmetric coma image
is multiplied by a 2D mask file. This mask represents the   pixel-wise ratio
between the observed pixel flux at a certain coma position and the mean coma
flux, as it is determined for the respective nucleus distance in the observed
image. We note that the 2D mask file is a direct output of the normalization method
for the analysis of structures inherent in coma images (for examples, see
Fig.~\ref{fig11} and Fig.~\ref{fig14}). In order to avoid abrupt pixel-to-pixel
variations, we smoothed the coma mask file by a 6x6 pixel averaging
filter. The smoothing process is adequate for the relatively wide and low-contrast 
coma features seen in the coma region beyond about 20 pixels from the
nucleus. However, it may not well represent small-scale variations very close to
the nucleus. The model coma image is normalized to achieve the same mean flux as
measured in the observed image in a ring aperture of a few pixels (in our
case 5 pixels or 1 arcsec in the sky) and with an inner diameter well beyond the
seeing of the exposures (in our case 35.5 pixels or 7.1 arcsec nucleus distance 
in the sky).

The nucleus light is represented by a 2D image with a single pixel at the
appropriate flux level and all other pixels at zero. The degradation of the
image quality due to the atmosphere and the instrument is described by a 2D
radially symmetric point-spread function (PSF). Either a Gaussian or a
Moffat-type distribution \citep{1969A&A.....3..455M} is applied. Here, the
Moffat function is proportional to $(1+(\rho/\alpha)^2)^{-\beta}$ with the
Moffat parameters $\alpha$ and $\beta$ and $\rho$ as radial distance. Both
distribution kernels are normalized to unity over the 2D image size chosen for
the model description.

The input images for the simulation, that is, the coma flux distribution, the
nucleus image, and the PSF image, are each centered to maximum brightness. In
order to improve the sensitivity of the solution for the nucleus, an
oversampling of the pixel size by integer factors of up to 5 in $x$ and $y$ is
introduced. This brings the effective pixel resolution of the model image to a
range below the nominal resolution of the telescope aperture (0.08 arcsec for
the 2m Wendelstein telescope) and allows a better representation of the mean
coma flux distribution close to the nucleus position.

The comet model image in space, that is, before transit through the 
terrestrial atmosphere and the telescope and instrument, is represented by 
the co-addition of the aligned coma and the nucleus images. The comet 
image at the telescope is the result of the convolution of the comet 
model image in space with the PSF image that represents the changes due 
to atmosphere and observing equipment. The resulting image from the 
convolution can be compared with the observed image and appropriate fitting 
parameters can be chosen. For the evaluation of the nucleus flux we apply the 
comparison of the telescope model data to median-averaged composites of 
the observed images. We selected the best solution from a large set 
of forward calculations. Selection criteria were: (1) the minimum sum of
$\chi^2$ deviations in the mean radial 
profiles of the model image and of the observed image in the inner 20 pixels 
(4 arcsec) around the nucleus and (2) a constant ratio (close to unity) of 
the mean radial model profile to the  observed profile  in the radial range 
between 20 and 
40 pixels (4 to 8 arcsec) from the nucleus. We required that criteria (1) and (2)  be
fulfilled at the same time. The pixel range for criterion (1) reflects best the
nucleus distance in the observed image that is most affected by the PSF and by
the nucleus flux. The radial flux profile in the coma is reflected further away
from the nucleus where the PSF and the nucleus flux properties are
unimportant. The fit solution produced a model image with a radial profile very
close to those of the chosen observed image – or in other words the ratio of the
radial profiles of the model and the observed images is close to unity over the
full pixel range (40 pixels). An example case for comet 41P is shown in
Fig.~\ref{fig8} below.

Model parameters for the fitting process are as follows: The exponent of the radial coma
profile (for a geometric expanding coma, the exponent should be 1), the nucleus
brightness, and the PSF parameters, namely the sigma (or equivalent
full width at half maximum (FWHM)) of the seeing disk for the Gaussian
distribution and the FWHM plus the asymptotic exponent for the Moffat case
(representing the $\alpha$ and $\beta$ parameter in the Moffat
distribution). The PSF parameters (Gaussian $\sigma$ or Moffat $\alpha$ and
$\beta$) are determined to the greatest extent possible from the width of the trails of
background stars in the mean composite comet images. The Gaussian $\sigma$ and
Moffat $\alpha$ were found to be well defined by these measurements, and therefore, for
the calculation of the model image, they are kept fixed or very close to the
measured values (rarely up to 0.05 arcsec deviation). It should be noted that the 
fitting process for the inner 20 pixels in the comet image usually gave
better results for the Moffat model (as compared to the Gaussian
approximation). The seeing measured by the FWHM of stars in the comet images was 
between 0.8 and 2 arcsec with a mean value of 1.2 arcsec, and three cases beyond 2 arcsec.

\section{Results}\label{sec_3}

The main objectives of our research are to estimate the size of the nucleus and
to  analyse the coma and tail structures in order to constrain the rotation
properties and the location of active regions on the nucleus of 41P. In preparation 
for these tasks, it was deemed important to gather a description of the general
appearance of comet 41P during the 2017 apparition, including the coma
brightness evolution, the measured colors, and radial gradients in the coma.

\subsection{The cometary coma}\label{sec_3.1}

\begin{figure}[tbp]
  \centering
  \includegraphics[width=\columnwidth]{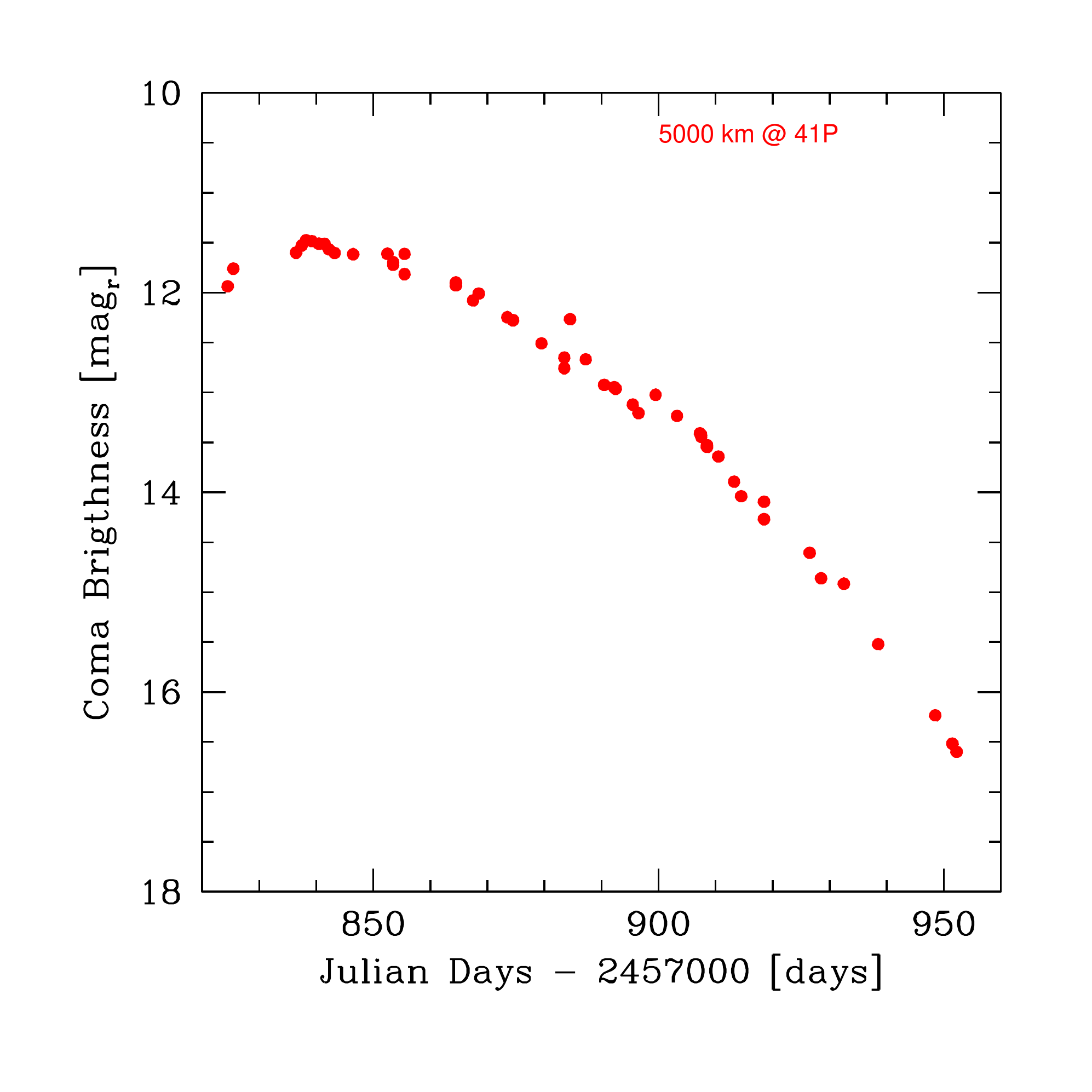}
  \includegraphics[width=\columnwidth]{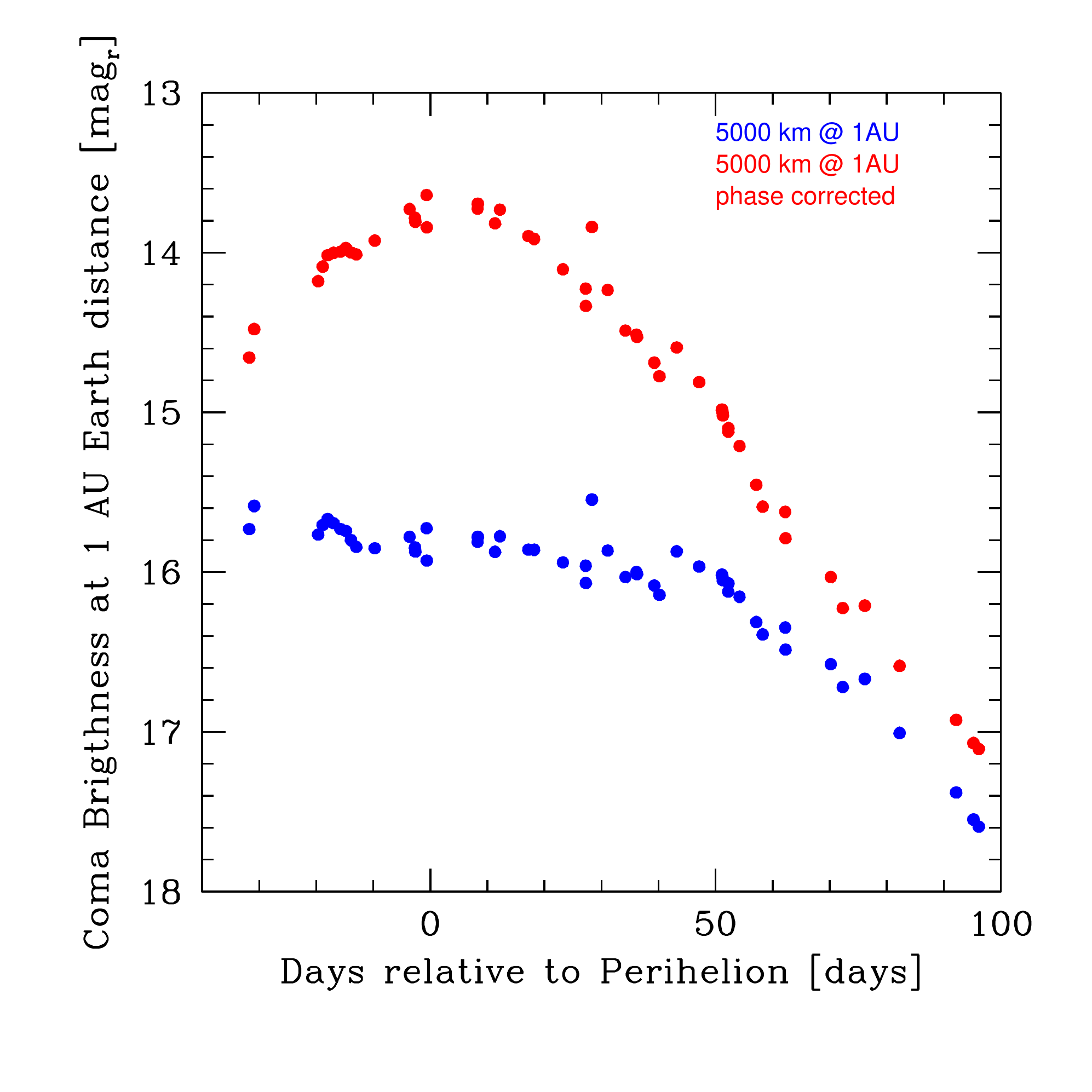}
  \caption{SDSS r filter brightness of the coma of 41P over time. Top panel: 
  Apparent coma brightness versus time (in Julian Date) measured in an aperture 
    corresponding to 5\,000 km at the distance of the comet. Measurement 
    uncertainties are typically of the size of the plotted symbols. Bottom panel: 
    Coma brightness in a 5\,000 km aperture scaled to 1 AU Earth distance 
    (blue dots), and additionally corrected for a linear phase
    function with a slope of 0.04 mag/deg (red dots). Perihelion passage of 41P 
    happened on Julian Date 247856.24959 (2017 April12)}
  \label{fig1}
\end{figure}

\textbf{Coma brightness:} During the observing interval from 2017 March 11 to
2017 July 17, comet 41P showed an extended coma of a few tens of thousands of kilometers in diameter with
slightly oval, asymmetric, and sometimes dented (due to inherent coma
structures) isophote patterns. The total apparent coma brightness in SDSS r
filter varied from about 11.5 to 16.6 mag and was measured in a circular aperture of
constant 5\,000 km diameter at the distance of the comet and centered at the central
brightness peak in the coma (Table~\ref{tab3}). The maximum apparent brightness
was reached about 1-3 weeks before perihelion (Fig.~\ref{fig1}, top panel). The
absolute brightness of the coma (corrected to 1 AU Earth distance and to
zero phase angle using a linear phase angle coefficient of 0.04 mag/deg
\citep{2004come.book..223L}) indicates that 41P may have been brightest
intrinsically around the time of perihelion passage (Fig.~\ref{fig1}, bottom
panel). The overall brightness evolution of the comet shows a few deviations
from a smooth light curve, that is, a short-term maximum (outburst) 28 days after
perihelion passage (2017 May 11), which is discussed in section \ref{sec_3.8}, as well as
short-term dips and peaks of up to 0.1 mag amplitude due to activity variations
of the comet and/or measurement scatter. The light curve of the comet in SDSS i
filter is, in general, quite similar to the one in SDSS r; our data in the SDSS g
filter are too sparse and do not provide a global picture of the light curve
over the 2017 apparition.

The measurements of the comet brightness represent mostly the coma activity in
the dust phase with – likely -- small contributions from coma gases
(C$_2$, C$_3$, NH$_2$, [OI]). Maximum activity around perihelion is therefore not
surprising, as solar illumination and heating of the nucleus for ice
sublimation is highest and also drives the dust activity.
\citet{2018A+A...619A.156M} report the maximum gas production rates in 41P for
the OH, NH, CN, C$_2$, and C$_3$ daughter species before perihelion (within 10-20
days). Comet 41P displayed strong and variable gas and dust jet activity
\citep{2019AJ....157..108S}. Currently, it is unclear to what extend the
occurrence of jets and the increasing rotation period have impacted the
assessment of the global gas activity status of the comet. For the $\M{af}\rho$
activity profile of 41P (see also the paragraph on coma activity below),
which is representative of the dust production of the comet, Moulane et
al. found a maximum 17 days before perihelion. This maximum was followed by a
drop by almost half of the maximum amplitude from 10 days before to about 16
days after perihelion passage of the comet – while the gas production decreased
only by a smaller amount. This $\M{af}\rho$ behavior is in contrast to the
rather smooth brightness evolution of the comet seen in our measurements 
(Fig.~\ref{fig1}, bottom panel), which do not support dust activity variations of
almost a factor two. The $\M{af}\rho$ estimation implies correction of the
measurements to zero phase angle using an assumed phase angle relationship to
the brightness data at unity Earth distance. Different phase angle corrections,
applied to our measurements of 41P, result in a diverse predicted behavior of
the activity profile for zero phase angle (reflected in the $\M{af}\rho$ curves
of 41P in Fig.~\ref{fig7}). Therefore, conclusions on the actual dust activity
profile of the comet have to be handled with care, in particular for corrections
covering a larger phase angle range (see Tab.~\ref{tab1}) like that of 41P.

\begin{figure*}[tbp]
  \centering
  \includegraphics[width=\textwidth]{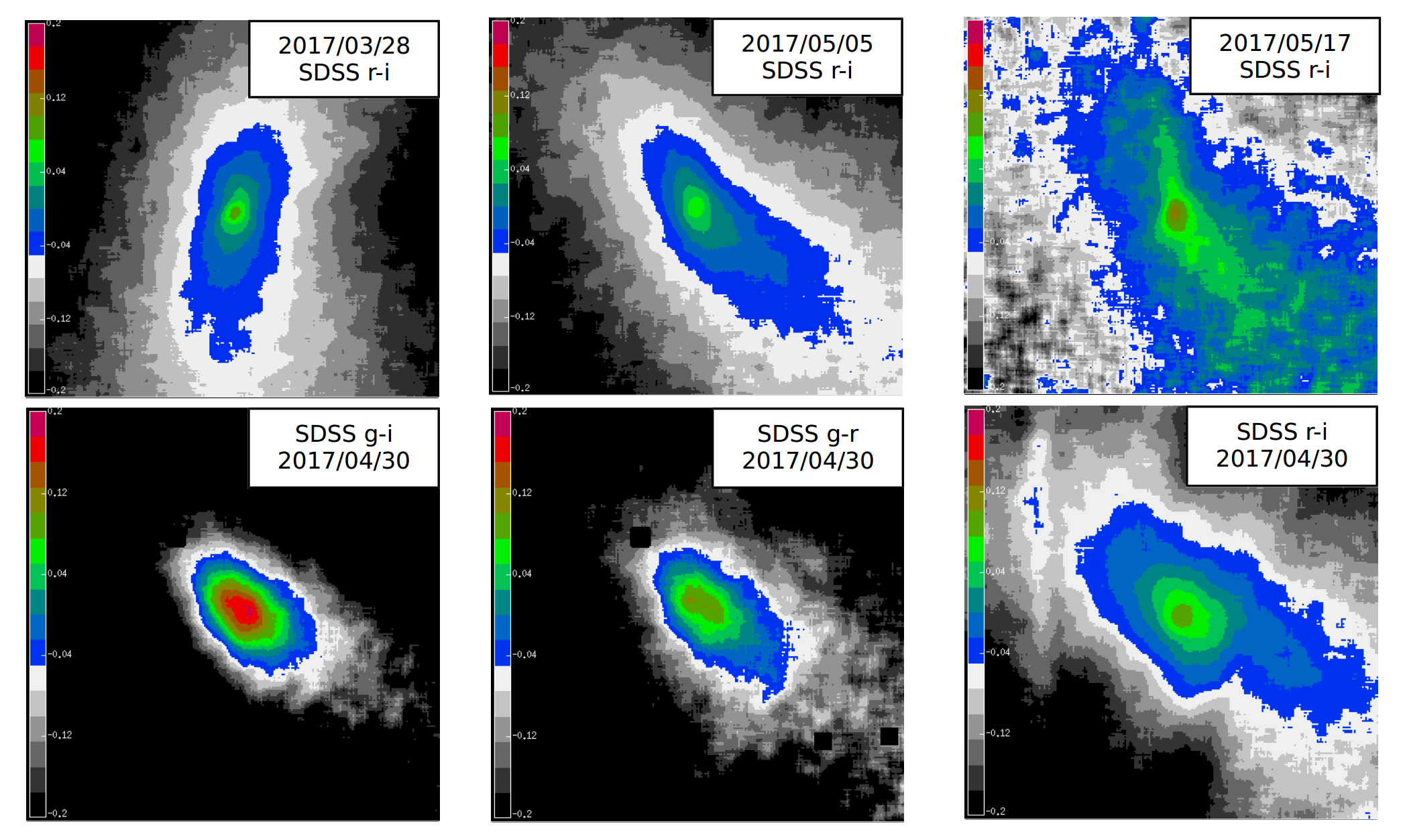}
  \caption{Color maps of the coma of 41P. Top row: Color maps in SDSS r-i 
    (field of view at the distance of the comet in parenthesis, date format 
    in Year/month/day): Left panel on
    2017 March 28 (5\,260 x 5\,260 km), Middle panel: on 2017 May 05 (7\,590 x 7\,590
    km), Right panel: on 2017 May 17 (8\,450 x 8\,450 km). Bottom row: Color maps
    on 2017 April 30 (field of view: 7\,080 x 7\,080 km at the comet): SDSS g-i 
    (left panel), SDSS g-r (middle panel), SDSS r-i
    (right panel). The images are smoothed by a 5x5 pixels box size. The color
    bar on the left of the panels provides the color difference (in mag)
    compared to solar colors (range from -0.2 mag (black) to +0.2 mag (red)). 
    North is up and east is to the left. Sun direction is opposite to 
    the tail orientation in the southern coma hemisphere.}
  \label{fig2}
\end{figure*}

\textbf{Coma color:} The perihelion arc of 41P is covered by SDSS r-i color data
from 2017 March 12 to 2017 June 02. Colors involving the SDSS g filter (SDSS g-r and
SDSS g-i) exist for 7 nights (2017 May 01+19+22+23+25+29 and 2017 June 02). The
color maps of the coma of 41P (Fig.~\ref{fig2}) show an approximately oval shape
of the isocolor lines similar to the isophote pattern of the comet for the respective
observing epoch. The isocolor lines are roundish in the inner part of the coma
(within a few hundred kilometers of the nucleus) and extend further into the tail
direction compared to the solar direction for the more distant part of the
coma. A significant gradient from red color in the coma center to blue color at 
larger distances is noted. In most images, coma structures are only barely 
noticeable in the color 
maps of the comet. However, it is noted that a coma jet is identified in color images 
from 2017 May 17 (see Fig. \ref{fig2}), indicating color properties of this feature that 
are different from the behavior of the main coma. 

\begin{figure}[tbp]
  \centering
  \includegraphics[width=\columnwidth]{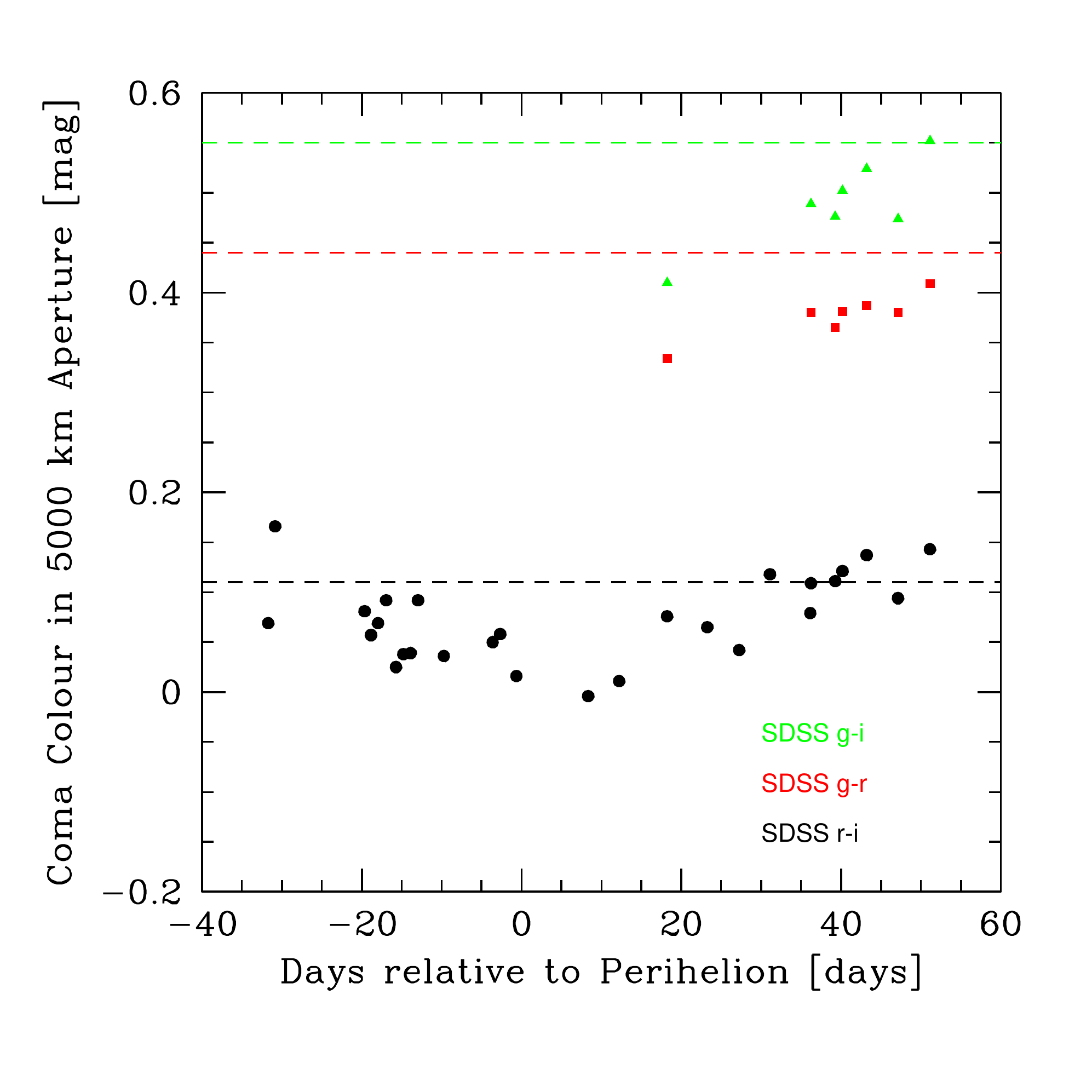}
  \caption{Colors in the coma of comet 41P in SDSS filters vs. time, measured 
    for 5\,000 km aperture diameter at the distance of the comet: SDSS g-i 
    (green triangles), SDSS g-r (red squares), SDSS r-i (black circles). Maximum 
    error of the color 
    values is 0.07mag. The broken lines show the solar colors for the respective 
    filters. }
   \label{fig3}
\end{figure}

\begin{figure}[tbp]
  \centering
  \includegraphics[width=\columnwidth]{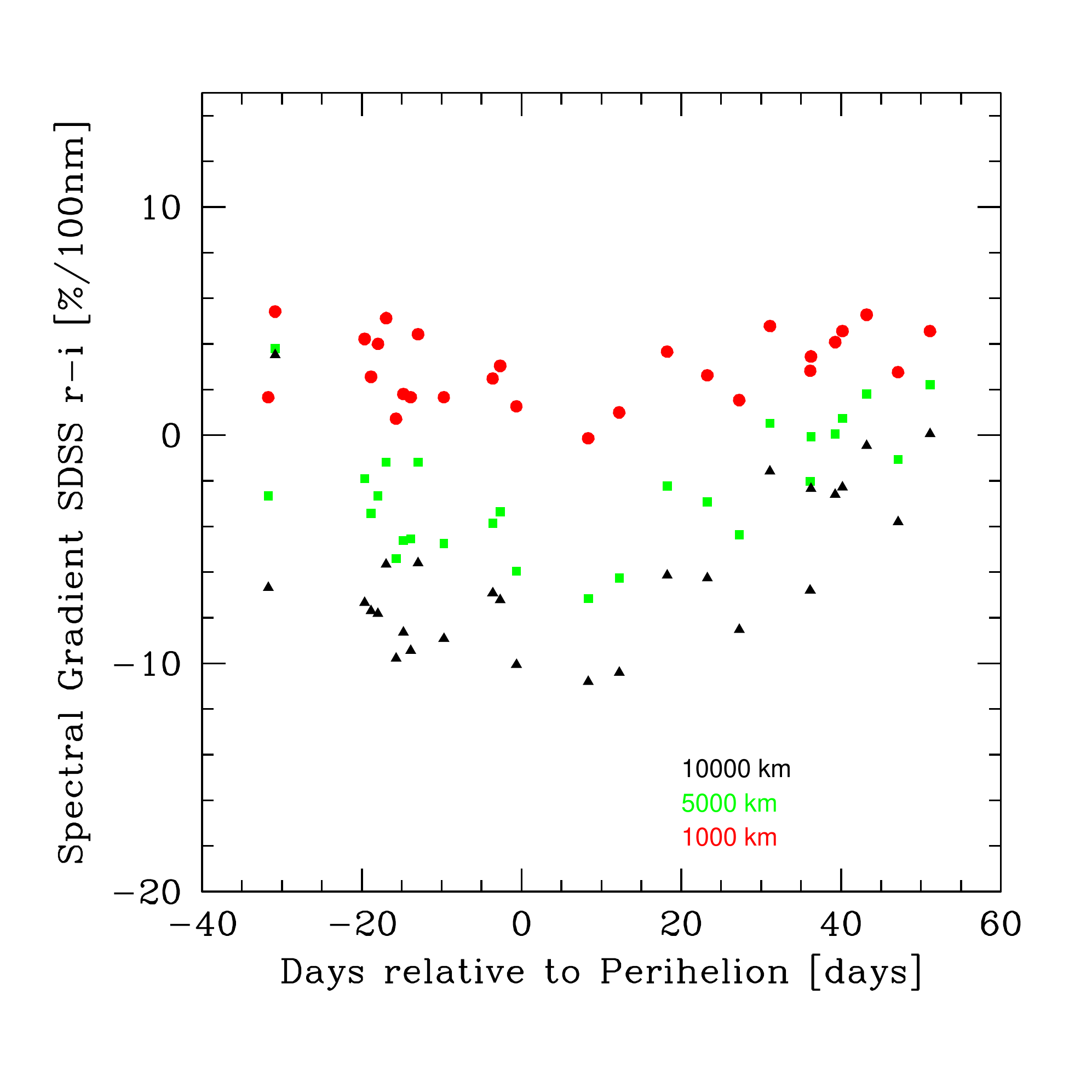}
  \caption{Spectral gradients in the coma (vertical direction) of 41P vs. time 
  (horizontal direction), obtained for the SDSS r-i color for aperture
    diameters of 1\,000 (red circles), 5\,000 (green squares), and 10\,000 km (black
    triangles) at the distance of the comet.}
   \label{fig4}
\end{figure}

Figure~\ref{fig3} quantifies the findings on the colors of 41P using aperture
photometry of different diameters. On different scales, the coma displays  colors that are:
slightly redder (for 500 and 1\,000 km apertures) than 
solar\footnote{solar
  colors calculated from the Sun brightness in the respective filter 
  pairs as given by \cite{2018ApJS..236...47W}} --  SDSS g-i = 0.550 mag, 
  SDSS g-r = 0.440 mag, SDSS r-i = 0.110 mag; at
about solar (for 2\,500 and 5\,000 km apertures), and slightly bluer than solar
(for 10\,000 km aperture). Within about a month before and a month after perihelion, the
SDSS r-i color of the comet shows a trend of bluer slope with increasing coma
aperture. The bluish trend reaches an amplitude of -0.15 mag compared to the Sun
around perihelion passage of the comet (Fig.~\ref{fig3}). This trend is more pronounced 
for the colors involving the SDSS g filter. A similar behavior is seen in the 
ring aperture results that exclude the
nonzero flux contributions from the cometary nucleus. Therefore, we conclude
that the likely source of the bluish colors in 41P around perihelion is light
emission from gaseous species in the coma (for SDSS g: C$_2$ and C$_3$, for SDSS
r: C$_2$, [OI] and NH$_2$, for SDSS i: NH$_2$). Because of the observed trend in
the SDSS colors of the comet around perihelion and the reduction of gas emission
with increasing solar distance, the best values of the intrinsic colors of the
dust in 41P is obtained from the observations on 2017 June 02: SDSS
$\M{g}-\M{r} = 0.41 \pm 0.05$ mag, SDSS $\M{g}-\M{i} = 0.55 \pm 0.06$ mag, and
SDSS $\M{r}-\M{i} = 0.14 \pm 0.07$ mag for the 5\,000 km aperture diameter in
the coma.  The mean spectral gradient varies between -2, 0, and
$+2 \%/100\;\M{nm}$ (error is $\pm 2\% / 100\;\M{nm}$) for SDSS g-r, SDSS g-i,
and SDSS r-i, respectively; at that time the dust of 41P had close to solar
colors and no significant spectral reddening. The spectral gradient for the SDSS
r-i color versus aperture size is presented in Fig.~\ref{fig4}. It is noted that
the color in the central coma part of the 41P color maps, that is, where the gas
contamination is lowest, shows a mean reddening in SDSS r-i of 0.09 ($\pm$0.03)
mag compared to the Sun, which indicates a spectral slope of $6 (\pm 3) \%/100$
nm. This value is close to the reddening found by Rosetta for dust in 
comet 67P/Churyumov-Gerasimenko (7.3 $\pm0.2$ \%/100nm for the 
orange/near-IR filter set; 
\cite{2017MNRAS.469S.195F}). Our measurement results on the coma color of 41P 
illustrated in Figs. \ref{fig2} to \ref{fig4} suggest possible variations of dust 
properties with time and/or distance from the nucleus. For instance, the SDSS r-i 
filter colors and spectral gradients became bluer when the comet was closer to 
perihelion and they were also bluer with increasing distance from the nucleus. 
Both trends might be an effect of the gas contamination of the flux in 
the SDSS r and 
i filters used for the measurements. Taking these trends to be at least in 
part due to the coma dust, one could argue that dust size distributions change 
with distance from the Sun, favoring the enhanced presence of small dust close 
to perihelion, as for instance discussed in the case of comet 67P/Churyumov-Gerasimenko
by \cite{2017MNRAS.469S.443B}. The 
small dust may reflect the light in the Rayleigh regime, producing a trend of bluer 
colors around perihelion. Dust disintegration in the inner coma can have a similar
consequence on the SDSS r-i colors and spectral gradients, resulting in decreasing
reddening with increasing distance from the nucleus. Slopes below unity in the radial profile 
of the coma flux, as described below under `Radial coma profiles', could be 
seen as a supportive argument for the occurrence of dust fragmentation in the inner 
coma of comet 41P.

\citet{2019MNRAS.485.4013L} present color measurements of 41P in broadband V and
R filters. While the mean spectral slope found by these latter authors is in the range of our data
of 41P, the temporal pattern of the spectral gradient shows considerable
variability with amplitude, which is not seen in our results. In particular, Luk'yanyk et al. claim
rapid (on timescales of 1 day) changes in the spectral slope of the comet
(namely for 2017 March 3-4) and interpret this finding as being due to
different dust components released by the comet from different regions on the
nucleus at different times. It is noteworthy that on 2017 March 03, Trappist
measurements of 41P in cometary dust filters show distinctly different gradient
values ($+11 \%/100$nm) compared to the results by Luk'yanyk et al.
($-10 \%/100$nm), although the observations were taken just about 1.5 hours
apart (with different aperture sizes). A similar discrepancy (now for identical
2\,000 km apertures) exists between our data on 2017 March 24 02:25UT
($+4 \%/100$nm in SDSS ri filters) and the results of  Luk'yanyk et al.  measured
about 1.5 hours earlier ($+20 \%/100$nm in Kron-Cousins VR filters). Although
the modeling by Luk'yanyk et al. offers an interesting explanation for
their color variations, it does not seem to explain the smoothly varying color 
gradients we see in 41P. In addition, it is unclear whether their model can 
also explain the rather different
behavior of the radial flux gradients in the coma of 41P as described below or
whether both phenomena have different origins.

The neutral (compared to the Sun) spectral gradient of 41P in the broadband
filters is different from the typical reddening of the solid component seen in
other comets \citep{2004come.book..577K} like for instance also measured in the
Rosetta target comet 67P/Churyumov-Gerasimenko at the Mt. Wendelstein
observatory \citep{2016MNRAS.462S.376B}. Assuming solar colors of the dust in
41P implies that about 20\%\ of the flux in the SDSS g
filter in late April 2017 has to be attributed to contributions from the coma gas.

\begin{figure}[tbp]
  \centering
  \includegraphics[width=\columnwidth]{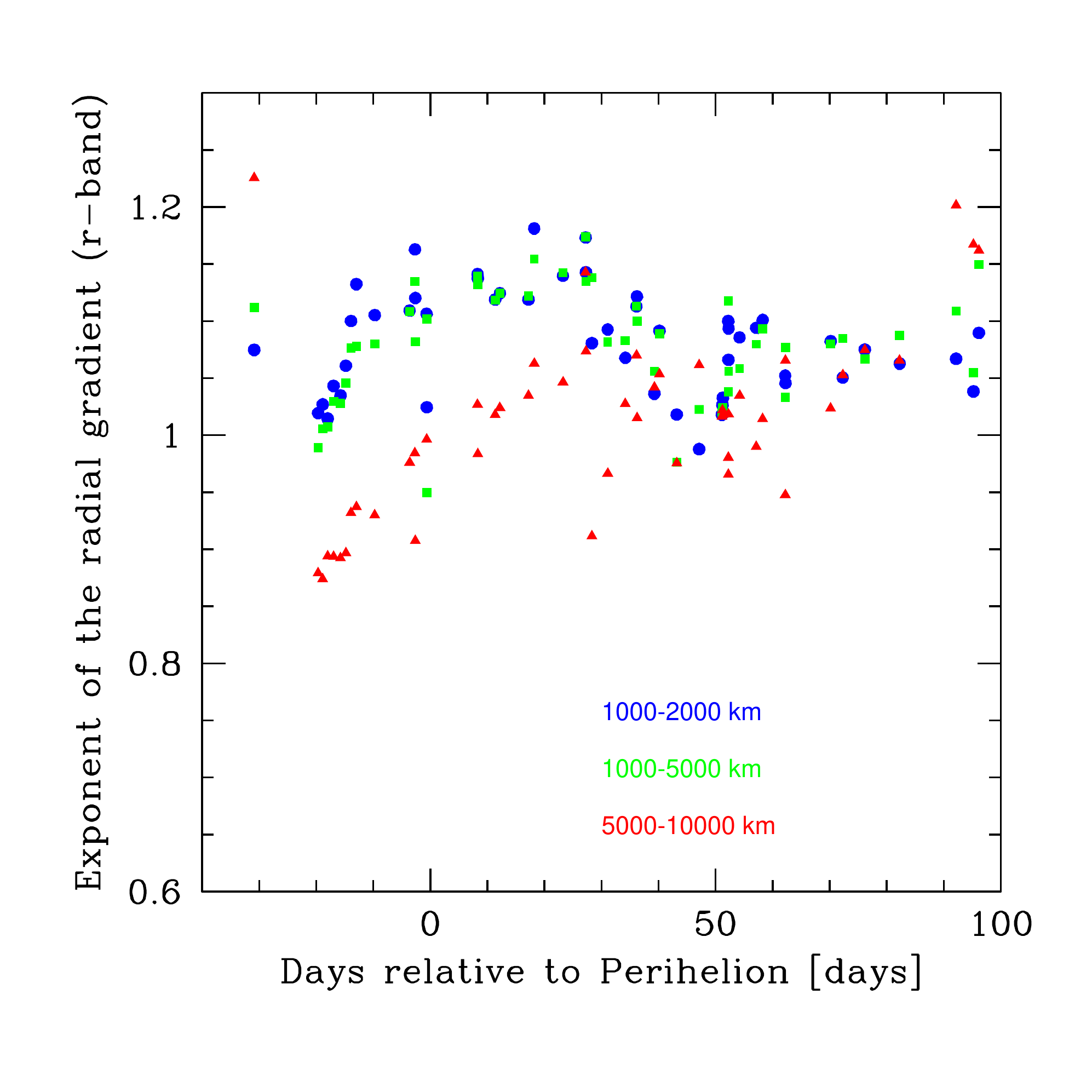}
  \includegraphics[width=\columnwidth]{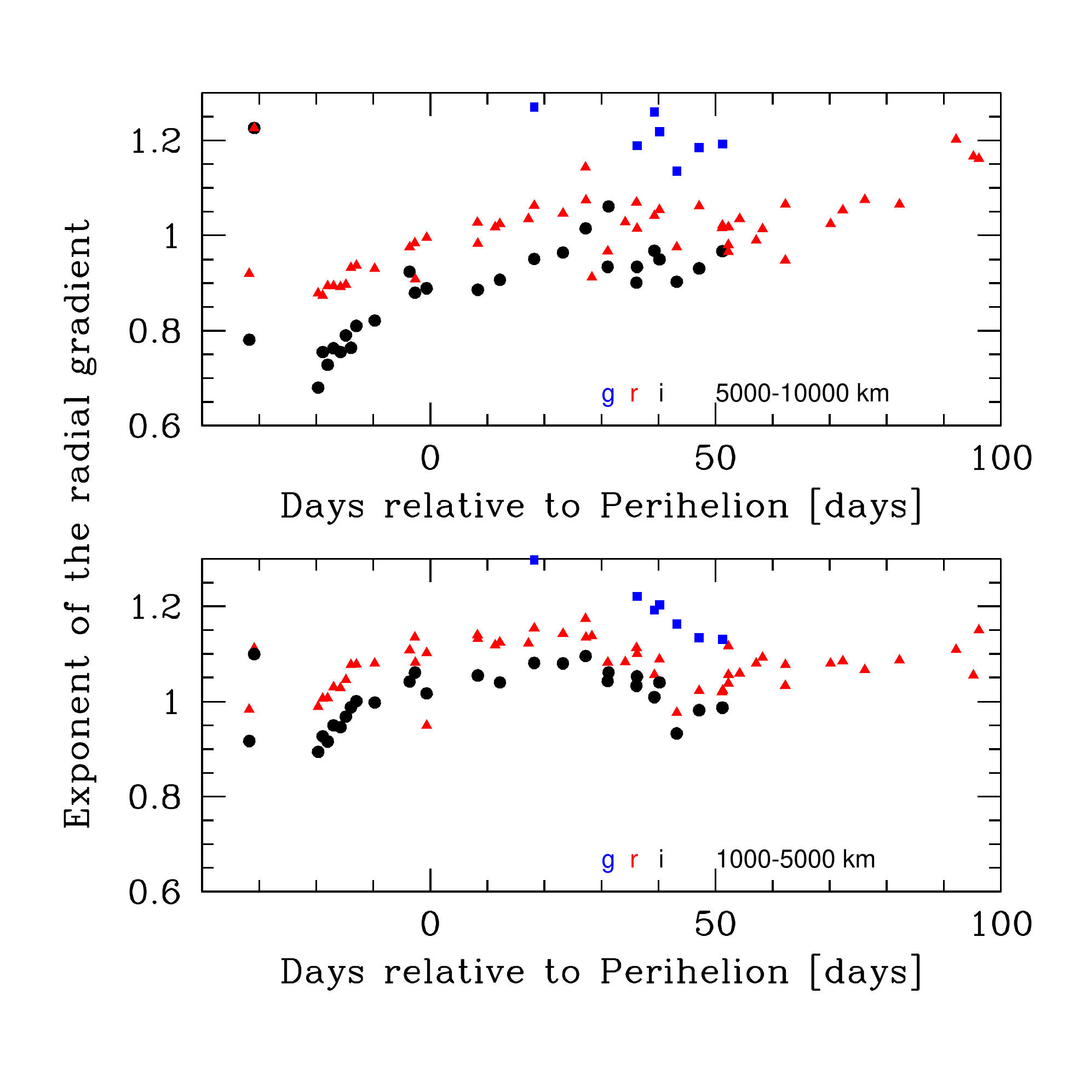}
  \caption{Exponents of the radial coma profile in comet 41P vs.
    time. Top panel: SDSS r filter - the exponents are determined for various 
    distance ranges in the coma, i.e., in the ranges 1\,000-2\,000 (blue),
    1\,000-5\,000 (green), and 5\,000-10\,000 km (red) from the coma center at
    the distance of the comet. Middle and bottom panels: Temporal evolution
    of the radial coma exponent in SDSS g filter (blue squares), in 
    SDSS r filter (red triangles), and in SDSS i filter (black 
    circles), for 5\,000 - 10\,000 km distance
    (middle panel) and for 1\,000 - 5\,000 km distance (bottom panel). The numerical 
    error of the individual results for the exponents is $\pm$0.03.}
  \label{fig5}
\end{figure}

\begin{figure}[tbp]
  \centering
  \includegraphics[width=\columnwidth]{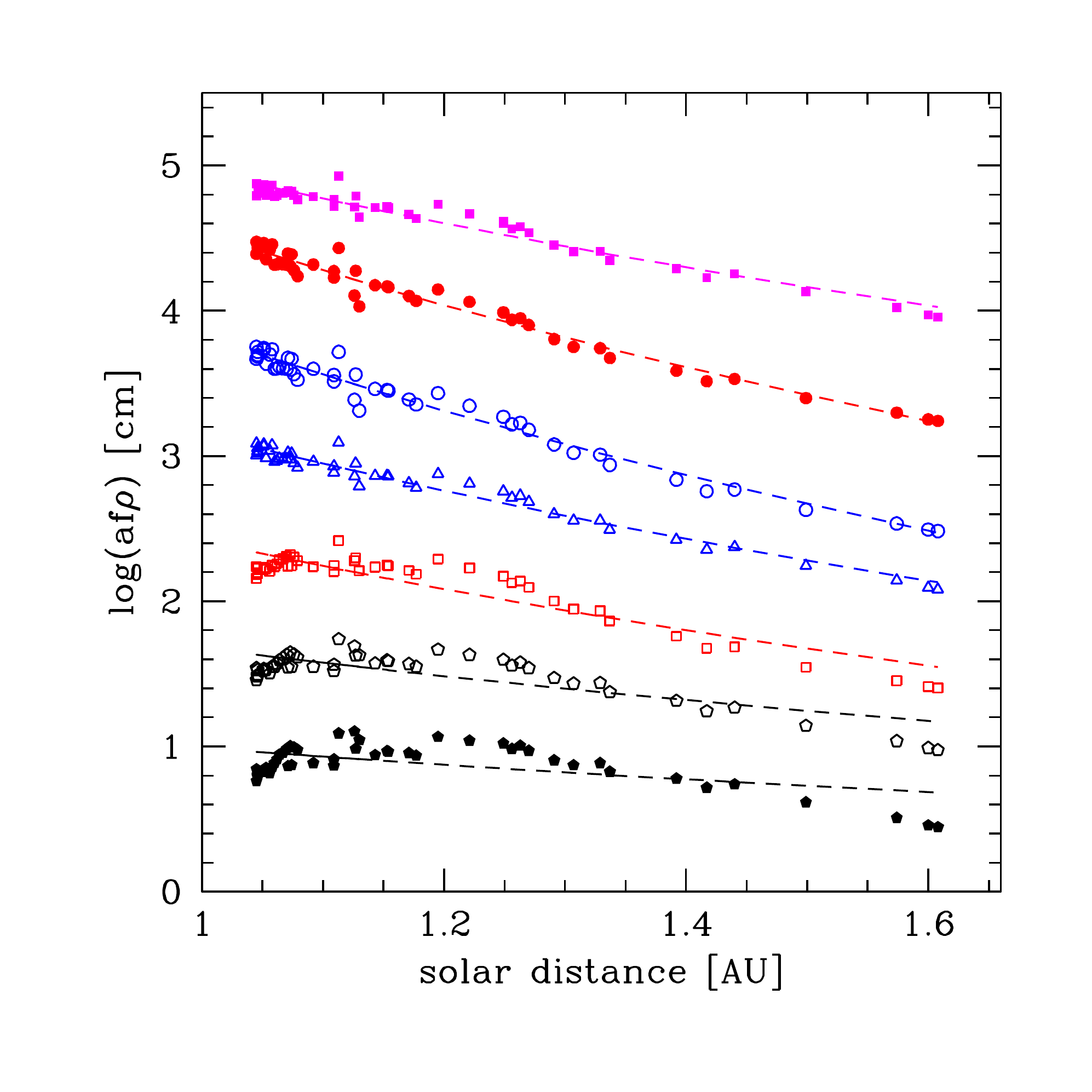}
  \caption{Activity of 41P vs. solar distance. The plot shows the
    $\M{af}\rho$ parameter of the comet estimated from the SDSS r filter
    images through a 5000 km aperture. Different phase-correction 
    models, represented by different colors, symbols, and offsets in 
    log $\M{af}\rho$, were applied to this data set: 
    Rosetta (\cite{2018AJ....156..237M}; filled magenta squares, offset +2.75);
    linear phase angle correction (filled red dots, offset +2.25);
    Bowell A (\cite{1989aste.conf..524B}; open blue dots, offset +1.25);  
    Bowell B (\cite{1989aste.conf..524B}; open blue triangles, offset +1.0); 
    Schleicher 2010 (\cite{2010AJ....140..973S}; open black diamonds, offset +0.0); 
    Mueller (\cite{1999PhDT..........M}; open red squares, offset +0.25); 
    Schleicher 1998 (\cite{1998Icar..132..397S}; filled black diamonds, 
    offset -0.75). The colored lines show the corresponding fits for the 
    cometary activity. Corrections for flux contributions from coma gas 
    and from the nucleus are also applied.}
  \label{fig7}
\end{figure}

\textbf{Radial coma profiles:} Radial coma profiles were extracted from the
median averaged SDSS filter images of 41P. They represent the azimuth-averaged
(over 360 deg) brightness of the comet with distance from the nucleus. For a
geometrically expanding coma (the expected case for the cometary dust in the
inner part of the coma) the flux of the radial profile should increase linearly
with aperture radius. For our 41P observations, the radial exponents are
obtained for various distance ranges (1\,000-2\,000, 1\,000-5\,000, and
5\,000-10\,000 km) in order to assess systematic changes with distance and with
time (see Fig.~\ref{fig5}).

The radial exponent values obtained for distance ranges 1\,000-2\,000 and
1\,000-5\,000 km display a scatter of less than 0.03 for single nights which is
attributed to data reduction artifacts, namely the incomplete removal of
signatures from background objects shining through the cometary coma. The
temporal evolution seen for the SDSS r filter (see Fig.~\ref{fig5}, top panel)
shows an increase of the coma gradients towards perihelion, followed by a plateau up to
40 days post-perihelion; thereafter, it gradually decreases and stabilizes to a
rather constant level until the end of the observing campaign. A similar
temporal behavior is also seen – at least up to about 50 days post-perihelion -
for the radial gradient measured through SDSS i filter, shifted to lower slope
values. We note that the shift between the
coma gradients in SDSS r and SDSS i filters has a decreasing trend between about
25 and 45 days post-perihelion. Only a few nights had coma observations through
SDSS g filters that show higher values of the radial coma slopes and follow the
decreasing gradient trend as seen in the SDSS r filter between about 30 and 55
days post-perihelion (see Fig.~\ref{fig5}, bottom panel).

The increasing and decreasing trends in the radial coma slopes, respectively,
before and after perihelion are attributed to the increasing and decreasing gas
contamination in the SDSS filters during the perihelion passage of the
comet. This interpretation is in line with that of the color variations of the
coma around perihelion. It is noted that the radial exponents of the SDSS g
filter images are in general higher than those for the SDSS r and SDSS i filter
images taken at the same epochs; they also remain steeper at larger radial
distances. This can be explained by the presence of C$_2$ gas in the SDSS g
filter, which has prominent emissions in this filter pass band and also a large
scale length for photo-dissociation compared to the NH$_2$ species that might be
the dominant contaminants in the SDSS ri filters. The radial exponent of the
coma measured in SDSS r filter images remained quasi-constant as of about 60
days post-perihelion and also the coma slopes seen in SDSS g and SDSS i filters
tend to converge towards those of the SDSS r images. Both findings are
indicative of the decreasing influence of the coma gas on the measured fluxes
of the comet. These findings also support the interpretation that the colors of 41P
measured in early June 2017 are representative of the coma dust.

It is noted that the radial exponents of the coma measured in SDSS i filter
before perihelion (as well as those around 50 days after perihelion) are below
unity. The exponents in the radial coma profile found for the very inner part of
the coma (see Tab.~\ref{tab4} described in section \ref{sec_3.2}) 
also indicate
a decrease of the radial exponent from about unity to values of around 0.9 and
below when the comet was approaching perihelion as well as for about a month
after perihelion passage. Thereafter, the exponent values were found to be
mostly between 0.9 and 1.0. This behavior may indicate additional dust
production in the coma, for instance by dust disintegration very close to the
nucleus of 41P \citep[for possible dust fragmentation scenarios
see][]{1990A&A...231..543B}. A similar behavior, though more pronounced in
amplitude and different in its temporal profile, was seen in ground-based images
of comet 67P/Churyumov-Gerasimenko \citep{2016MNRAS.462S.376B}.

\textbf{Coma Activity:} The dust activity of the comet is represented by the
$\M{af}\rho$ parameter of the coma (\citet{1984AJ.....89..579A}). This parameter provides a
quantitative equation that connects the measured coma flux with the projected
aperture radius $\rho$ at the distance of the comet, the so-called filling
factor f of the cometary dust in the light path, and the albedo a of the dust
(which is assumed to be constant for all dust grains):

\begin{equation}
  \M{af}\rho = 4 \cdot \Delta^2  \cdot  \frac{r^2}{\rho}  \cdot \frac{f_{\M{comet}}}{f_{\M{sun}}}
  \label{eq1}
,\end{equation}

where, $\Delta$ and r denote the Earth and Sun distance of the
comet, respectively, and $f_{\M{comet}}$ and $f_{\M{sun}}$ are the fluxes of the
comet and Sun, respectively, in the filter band used.

For the Wendelstein observations of 41P, the $\M{af}\rho$ values around
perihelion contain small contributions from gaseous species in the coma
(possibly up to about 0.15 mag) in SDSS r filter images (see paragraphs on the
coma colors and radial coma profiles). Furthermore, nucleus reflected
sunlight may also contribute to the total aperture brightness of the coma. For 41P, a
nucleus radius of 700 m and an albedo of 0.04 together with a linear phase angle
correction (0.03 mag/deg) is applied. Different phase functions for the dust can be applied
as published in the literature (the parameter values applied here are given in
parenthesis): A linear phase angle correction \citep[][parameter: 0.04
mag/deg]{2004come.book..223L} or slightly more complex ones as for instance used
by \citet[][parameters: 0.01807, -0.000177]{1998Icar..132..397S} and
\citet[][phase corrections taken from table file]{2010AJ....140..973S},
\citet[][parameters: $0.03$, $-0.00025$]{1999PhDT..........M} or the HG-type
phase correction after \citet[][parameters: 3.332, 0.631, 0.986 for Bowell A and
1.862, 1.218, 0.238 for Bowell B]{1989aste.conf..524B}. From Fig. 2 of 
\citet{2018AJ....156..237M}, we obtained approximate values for the phase function 
of cometary dust at comet 67P/Churyumov-Gerasimenko measured by the OSIRIS camera on 
board the Rosetta spacecraft. The data were normalized to unity at zero phase angle 
(extrapolation value 6.5 in the notation of Fig. 2 in \citet{2018AJ....156..237M}) 
and were then used for 
an estimation of $\M{af}\rho$ for comet 41P. Linear interpolation between the 
measurement points
of the Rosetta phase function at 20, 30, 40, 50, and 60 deg is applied to obtain the 
phase function for the phase angle values of our 41P observations.

\begin{table*}
  \centering
  \begin{tabular}{r|rrrr}           
    \hline\hline
    41P                & af$\rho$ Exponent    & af$\rho$ Exponent    & af$\rho$ Exponent    & af$\rho$ Exponent    \\

    Activity           & corrected for        & corrected for        & 
    corrected for        & corrected for        \\
    \hline
    Nucleus brightness & no                   & no                   & yes                  & yes                  \\
    Gas contamination  & no                   & yes                  & no                   & yes                  \\
    \hline
    linear             & $6.4 \pm 0.2 / 0.97$ & $6.7 \pm 0.2 / 0.97$ & $6.2 \pm 0.2 / 0.97$ & $6.4 \pm 0.2 / 0.97$ \\
    Schleicher (1998)  & $1.5 \pm 0.3 / 0.37$ & $1.8 \pm 0.3 / 0.43$ & $1.3 \pm 0.3 / 0.25$ & $1.5 \pm 0.3 / 0.31$ \\
    Schleicher (2010)  & $2.5 \pm 0.2 / 0.67$ & $2.7 \pm 0.3 / 0.72$ & $2.2 \pm 0.3 / 0.60$ & $2.5 \pm 0.3 / 0.62$ \\
    M\"uller             & $4.3 \pm 0.2 / 0.87$ & $4.5 \pm 0.2 / 0.87$ & $4.0 \pm 0.3 / 0.82$ & $4.2 \pm 0.3 / 0.83$ \\
    Bowell A           & $6.7 \pm 0.2 / 0.97$ & $6.9 \pm 0.2 / 0.97$ & $6.4 \pm 0.2 / 0.97$ & $6.6 \pm 0.2 / 0.97$ \\
    Bowell B           & $5.0 \pm 0.1 / 0.97$ & $5.2 \pm 0.1 / 0.97$ & $4.7 \pm 0.1 / 0.97$ & $5.0 \pm 0.1 / 0.96$ \\
    Rosetta            & $4.5 \pm 0.1 / 0.97$ & $4.8 \pm 0.1 / 0.95$ & $4.3 \pm 0.1 / 0.95$ & $4.5 \pm 0.1 / 0.95$ \\
  \hline
  \end{tabular}
  \caption{Exponents for the dust activity of comet 41P versus solar distance
    during the 2017 perihelion passage assuming different phase functions
    (`\texttt{linear}', `\texttt{Schleicher (1998)}', `\texttt{Schleicher
    (2010)}', `\texttt{M\"uller}', `\texttt{Bowell A}', `\texttt{Bowell
    B}' and `\texttt{Rosetta}'). The heliocentric activity profile for an 
    aperture diameter of 5\,000
    km at the comet was used for the fits. The listed data represent the
    exponent value and its error. The third number in each cell indicates the
    coefficient of determination for the linear fit which compares measured and
    estimated $\M{af}\rho$ values (confidence values between 0 and 1).  The
    table columns indicate whether or not corrections for the nucleus brightness
    and coma gas contamination are applied as described in the text.}
  \label{tab5}
\end{table*}

Figure~\ref{fig7} shows the $\M{af}\rho$ data obtained for the coma brightness
measurements of 41P through the SDSS r filter in a 5\,000 km diameter
aperture. The dashed lines in the figure represent fits to the $\M{af}\rho$
values for the individual phase functions used, in all cases assuming that a
single radial exponent describes the evolution of the dust activity of the comet
over the duration of the whole observing campaign. Corrections for gas
contamination (about 15\%\ for gas emissions in the SDSS 
r filter band) and for the nucleus brightness are applied. Table~\ref{tab5} lists
the activity exponents with solar distance obtained from fits to the
$\M{af}\rho$ results for the various phase functions applied, with or
without corrections for gas contamination and reflected light from the
nucleus. The $\M{af}\rho$ fits for the linear, Bowell A/B, and Rosetta 
phase functions have the highest confidence levels (0.95 and higher) and suggest that
a single-exponent parameter (though of different value depending  on the phase
function) is sufficient to describe the $\M{af}\rho$ activity of comet 41P over 
an extended part of its perihelion passage. The Rosetta phase function 
has the lowest radial exponent value, while the linear and Bowell A phase 
functions have the highest exponent values. It is noted that the Bowell phase
function is derived for atmosphereless asteroid-like bodies and applying it
to cometary dust can at least be debated. The linear phase function was
found not to be appropriate for the dust in comet 1P/Halley \citep{1987A&A...187..585M} 
and is likely not a well-suited description of the light scattering by cometary dust over
a wider phase angle range. The $\M{af}\rho$ data using the Schleicher 
and M\"uller phase functions display a larger scatter to the fit lines 
and have considerably lower confidence parameters (see Fig.~\ref{fig7}). 
Here, improvement might be possible with fits in segments of radial distances.
Taking gas production rates, if measured in parallel, as a reference for the dust 
activity of the comet can provide a useful guidance.
However, careful considerations need to be applied for variable ratios in the
gas and dust activity of comets that may change with time and distance from the 
Sun. \citet{2017MNRAS.469S.404B} performed an assessment of the extent to 
which the Rosetta phase function values 
determined at comet 67P/Churymov-Gerasimenko are suitable for the interpretation 
of dust measurements of other comets; they found cases with good and less good 
agreement between measurements and modeling results.

The activity of 41P in the outbound leg of its orbit was found to be steeper
than that of the Rosetta target comet 67P/Churyumov-Gerasimenko
\citep{2016MNRAS.462S.376B,2016A&A...588A..80S}, assuming that the linear,
M\"uller, and Bowell A and B phase corrections can be applied. For the Schleicher phase
function, the activity exponent versus solar distance of 41P is smaller than for
67P, although a single parameter description may be far from optimum for 41P
(and for 67P as well). Given that the phase range of the 41P observations is
larger than for other comets with a measured coma activity, it is not possible
to draw a firm conclusion on the dust activity profile of the comet versus solar
distance. This is because the phase angle corrections have a strong impact on
the $\M{af}\rho$ estimations of the comet and cannot be independently and 
securely determined from our data. Taking the gas activity into account
\citep{2018A+A...619A.156M}, a behavior of the dust activity with a
Schleicher-type phase function appears to be plausible for 41P. However, this
argues for a considerable activity drop of the comet around perihelion while the
phase angle changed only by about 10 deg and a smooth light curve behavior is
observed (Fig.~\ref{fig1}). It is noted that, independent of the phase angle
correction applied, the $\M{af}\rho$ activity of 41P is several factors smaller than
that of 67P \citep[see][]{2016MNRAS.462S.376B}, taking the perihelion distance
of the latter (1.25 AU) as a reference. This difference may reflect the ratio 
of the active surface area of 67P displayed to solar illumination  to that of 41P, because
their equivalent radii differ by a factor of between about three and four (see section
\ref{sec_3.2}).

\subsection{ Nucleus brightness and size}\label{sec_3.2}

\begin{figure*}[tbp]
  \centering
  \includegraphics[width=0.305\textwidth]{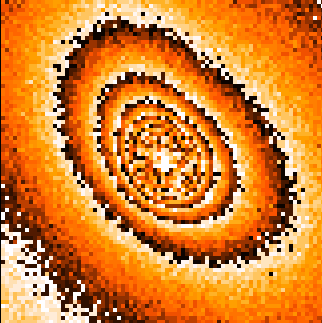}
  \includegraphics[width=0.305\textwidth]{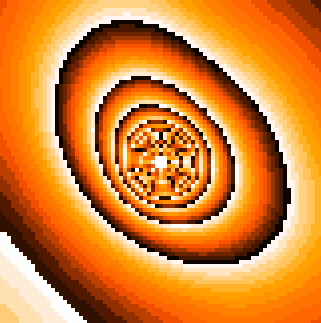}
  \includegraphics[width=0.33\textwidth]{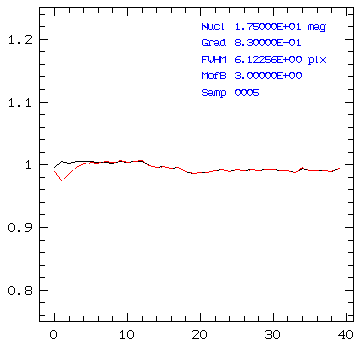}
  \caption{Nucleus brightness fitting in the inner coma of comet 41P. Left
    panel: Innermost 40x40 pixel of the observed comet image on
    2017 May 09. Middle panel: Respective 40x40 pixel model image using the
    Moffat PSF, with the same flux levels as in the observed image. Right panel:
    Ratio of the radial profiles between the simulated and observed
    image. Parameters of the simulated image are: nucleus brightness (Nucl =
    17.5 mag), gradient of the inner coma flux (Grad = 0.83),
    FWHM of the PSF  (6.12 pixels), and Moffat $\beta$
    parameter (MofB = 3.0). The plot shows the ratio of the normal pixel resolution
    of the model image (red line) to that of an oversampled (factor 5) version (black
    line).}
  \label{fig8}
\end{figure*}

As described in section \ref{sec_2.3}, the nucleus size of 41P was estimated via
a fit of the radial brightness profile in the inner coma of the comet, that is, from
the measured light distribution profile within 40 pixels (8 arcsec) around the
central brightness peak in the coma. The best solution was found with a set of
forward calculations with iterative improvements of the fit parameters. The
usage of a Moffat-type PSF 
considerably improved the fit quality for the radial profile over that of 
the Gaussian PSF. The four nominal fit parameters were
treated as follows: The FWHM of the image quality was determined from the width
of star trails in the comet images and it was set to the value as obtained for
the respective observing epoch of 41P. Variations of this fit parameter by up to
0.05 arcsec only very rarely improved the fit solution. The $\beta$ parameter of
the Moffat function could usually not be 
directly determined a priori from the images, which was due to low signal 
levels in the stellar PSF wings and in
part to the presence of the cometary coma light in the field of
view. Hence, $\beta$ was a free fit variable, usually varied in steps of 0.2 to
0.3. As expected, the $\beta$ parameter of the PSF is most sensitive to the
radial profile in the wing range of the PSF; in our cases, typically to the
radial profile from about 1.5 to 3 arcsec from the nucleus (7-20 pixels from the
nucleus). The Moffat $\beta$-values obtained are between 2.5 and 5, with a
median value of 3.2, and the vast majority are between 3 and 4, which is inside the expected
range for a Kolmogorov-type atmosphere. The radial exponent of the coma profile
was typically adjusted via the slope of the radial profile in the distance range of
4 - 8 arcsec (20-40 pixels) from the central brightness peak. The nucleus
brightness contributed mostly to the central part of the radial profile, that is, to
the innermost 2 - 3 arcsec (0-15 pixels from the nucleus). Figure~\ref{fig8}
exemplifies results of the fitting process for the inner coma and the nucleus
brightness of comet 41P for the observations on 2017 May 09.
 
The data set analyzed for the nucleus size estimation comprises 82 
median-averaged images of 41P (7 through SDSS g, 47 through SDSS r, 28 through SDSS i)
from observations of the comet on 42 nights between 2017 March 12 and 2017 July 16
(Tab.~\ref{tab4}). In total, 65 nucleus brightness estimations were achieved for
35 nights (4 for SDSS g, 39 for SDSS r, 22 for SDSS i). Exponents for the radial
profile in the inner coma were obtained for most images. The letters `ne' in 
Tab.~\ref{tab4} indicate observing epochs for which the nucleus brightness estimation 
was unsuccessful. In these cases, the seeing caused
a wider spread of the actual nucleus flux in the PSF such that it was close
to (or below) the sensitivity limit in the radial profile of the respective
night, that is, the nucleus signal was in the noise of the coma profile.

\begin{figure}[tbp]
  \centering
  \includegraphics[width=\columnwidth]{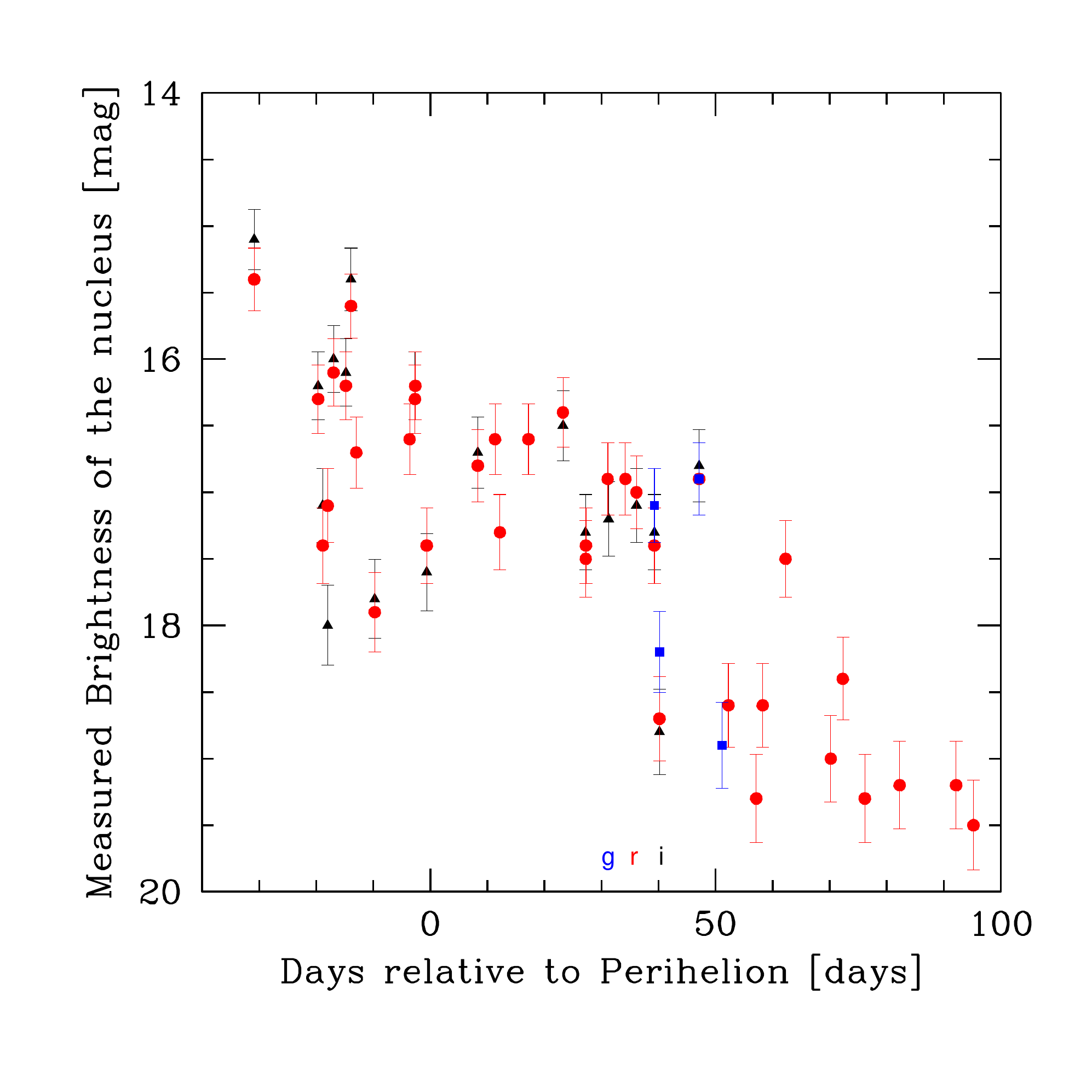}
  \includegraphics[width=\columnwidth]{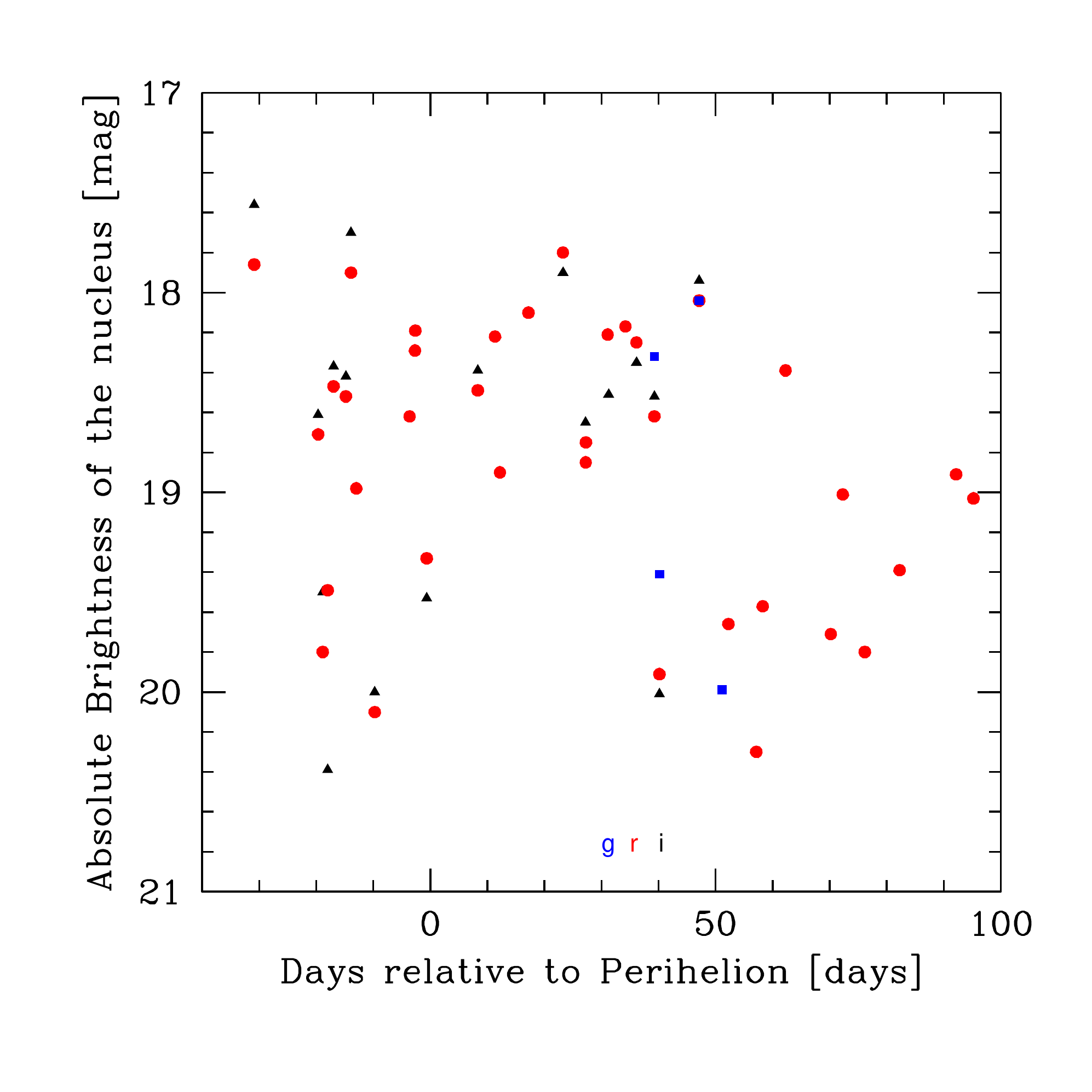}
  \caption{Nucleus brightness of 41P. Top panel: Apparent
    magnitude $m(r,\Delta,\varphi)$ of the nucleus of 41P vs. time to or from
    perihelion as determined through filters SDSS gri. Bottom panel: 
    Absolute magnitude $H(1,1,0)$ of the nucleus of 41P at unity Sun and Earth
    distance vs. time for a linear phase angle correction of 0.03
    mag/deg. For both panels, results in SDSS g filter are plotted as green squares,
    in SDSS r filter as red circles, and in SDSS i filter as black triangles.}
  \label{fig9}
\end{figure}

\textbf{Absolute magnitude of the nucleus:} It is noted that the method used 
to determine the nucleus brightness through the cometary coma
implicitly assumes optically thin conditions for the coma. For 41P we have no
direct proof in support of this assumption. However, the agreement of our results with
alternative size estimations for this nucleus as well as the detection of a
variable nucleus brightness of the comet (see below) can be seen as an indication 
in favor of the optically thin case
for 41P. Attempts to determine the extinction in cometary comae – mostly due to
the dust released by the nucleus – used close appulses of background stars to
cometary nuclei. The results are controversial, both in quality and quantity; for
a discussion see \citet{2012ApJ...760L...2L}. Measurements of these latter authors at
comet 17P/Holmes indicate an optical depth of 0.04 in 1.5 arcsec distance from
the nucleus, from which an optical depth of between 13.3 and 0.19 is concluded for
the nucleus region, depending on the value of the radial exponent in the inner
coma (1 and 0.35, respectively). For comet 17P/Holmes, the dust-production rate
at the time of the observations ($3.5 10^5$ kg/s) was higher by a factor of more
than 3\,000 than the peak dust production rate (110 kg/s) estimated for 41P by
\citet{2018A&A...615A.154P}. This may indicate that optically thin conditions
were prevailing in the inner coma of 41P by the time of our observations of the
comet.

The results achieved for the nucleus brightness and for the gradient exponent of
the inner coma of 41P are compiled in Tab.~\ref{tab4}. From the apparent nucleus
brightness (Fig.~\ref{fig9} top panel) the absolute nucleus brightness
$H(1,1,0)$ is calculated for 1 AU Sun ($r$) and Earth ($\Delta$) distances and 
zero-degree phase angle ($\varphi$), assuming a linear phase angle function with a
constant correction factor ($k$) of 0.03 mag/deg in all filter bands:

\begin{equation}
  H(1,1,0) = m(r,\Delta,\varphi) - 2.5 \cdot \log_{10}(r \cdot \Delta) - k \cdot \varphi
  \label{eq2}
,\end{equation}

where $m(r,\Delta,\varphi)$ represents the estimated nucleus
magnitude at Sun and Earth distance $r$ and $\Delta$, respectively, and phase
angle $\varphi$ of the comet. We note that in the literature on
photometric size estimations, phase correction factors of 0.02
\citep{2013A&A...557A..33S} to 0.04 mag/deg \citep[][and references
therein]{2004come.book..223L} are typically applied for cometary nuclei. More
recently, \cite{2017MNRAS.469S.404B} provided a graphical representation (see 
Fig. 9 in their paper) of the pre- and post-perihelion phase function for the nucleus 
of the Rosetta 
target comet 67P/Churyumov-Gerasimenko which allowed us to obtain an 
approximate average phase correction factor of 0.043 mag/deg (for phase angles 
between 0 and 70 deg). Using our $m(r,\Delta,\varphi)$ values in Tab.~\ref{tab4}
and assuming a phase correction factor of 0.03 mag/deg for 41P, 
the $H(1,1,0)$ magnitudes of the nucleus for the SDSS r filter vary between 17.8 mag 
and 20.3 mag, for the SDSS i filter between 17.6 mag and 20.4 mag, and for SDSS g 
between 18.0 mag and 20.0 mag (see Tab.~\ref{tab4} and Fig.~\ref{fig9}, bottom panel).

\textbf{Nucleus size:} The equivalent radius $R$ (in m) for the nucleus of 41P can 
be estimated using the following equation.

\begin{equation}
  R = 149.6 \cdot 10^9 \;\frac{10^{0.2\left(M-H(1,1,0)\right)}}{A^{1/2}} \quad [\M{m}]
  \label{eq3}
,\end{equation}

where $M$ is the magnitude of the Sun in the respective filter band at 1AU and $A$ is the
geometric albedo of the nucleus. For comets, a typical albedo of 4\%\ can 
be assumed  \citep{2004come.book..223L}, whereas for the Rosetta comet 67P/Churyumov-Gerasimenko 
an albedo of 6.5\%\ is measured for a wavelength of 649 nm \citep{2015A&A...583A..30F}. 
For $M$ the `AB' magnitudes for the SDSS gri 
filters by \citet{2018ApJS..236...47W} can be used ($-26.47$ mag in SDSS g,
$-26.93$ mag in SDSS r, $-27.05$ mag in SDSS i). The equivalent radii determined
for 41P vary between about 800 m and 220 m (Tab.~\ref{tab4} assuming $k$ = 0.03 
mag/deg and A = 0.04).

\begin{table}
  \centering
  \begin{tabular}{r|rr}           
    \hline\hline
    41P      & $H(1,1,0)$    & Radius        \\
    Nucleus  & (mag)         & (km)          \\
    \hline
    SDSS g   & $18.9\pm 0.3$ & 640 (560-730) \\
    SDSS r   & $18.7\pm 0.3$ & 560 (480-640) \\
    SDSS i   & $18.5\pm 0.3$ & 590 (510-680) \\
    \hline
  \end{tabular}
  \caption{Absolute brightness $H(1,1,0)$ and equivalent radius $R$ obtained
    from the SDSS filter photometry of the nucleus of comet 41P. The values for
    $H(1,1,0)$ in the table represent the median of the $H(1,1,0)$ results for
    the respective set of epochs of each filter. The radius column lists the $R$
    values corresponding to the median and the radius range corresponding to the
    respective $H(1,1,0)$ uncertainty. The equivalent radii are calculated
    according to eq.~(\ref{eq3}) assuming a geometric albedo of $0.04$ and a 
    phase correction factor of $0.03$ mag/deg, and
    applying solar AB magnitudes (at 1AU) of $-26.47$, $-26.93$, $-27.05$ mag
    for SDSS gri filters, respectively. The mean values of the $H(1,1,0)$ SDSS
    g, r, i filter magnitudes of 41P are $18.9 \pm 0.9$ mag, $18.8 \pm 0.7$ mag,
    $18.6 \pm 0.8$ mag, respectively; their large scatter reflects the
    brightness variations of the 41P nucleus.}
  \label{tab6}
\end{table}

The median values of the absolute brightness of the nucleus in the SDSS gri
filters (Tab.~\ref{tab6}) lead to values of the equivalent radii of 41P of between
560 and 640 m (for geometric albedo 0.04 and phase correction factor 0.03 
mag/deg). These radius values of 41P are numerically below that  given 
by \citeauthor{2000A&AS..146...73T} ((\citeyear{2000A&AS..146...73T}; 
radius = 700 m), but they appear to be compatible considering the 
uncertainties on both sides. We would like to mention that from our 2017 May 09 data we 
obtain an equivalent radius of about 520 to 560 m for the nucleus of 41P, a value 
that is compatible with the lower limit radius of 450 m as measured from radar 
observations of the comet on the same day \citep{2017DPS....4941424H}. Applying 
the Rosetta values for the albedo (0.065) and phase function (0.043 mag/deg) to our 
measurements for the nucleus of 41P results in median H(1,1,0) brightness of 18.3, 18.0, 
and 17.7 mag for SDSS gri filters, respectively, corresponding to about 660, 600, and
650 m for the equivalent radius of the nucleus. 

\textbf{Axis ratio of the nucleus:} According to \citet{1990A&A...231..548Z}, 
the light curve amplitude of asteroids, or of bodies with irregular shape and
light scattering depending on structural parameters on microscopic and
macroscopic scales in general, depends on the phase angle at the time of observation. This
dependency was further analyzed by \citet{2006A&A...454..367G} and both 
groups of authors arrive at a relationship between the light curve amplitude $\Delta$m and
the phase angle $\varphi$ (for $0 < \varphi < 40$ deg) of

\begin{equation}
  \Delta m(\varphi) = (1 + \kappa \cdot \varphi) \cdot \Delta m(0)
  \label{eq4}
,\end{equation}

where $\Delta m(\varphi)$ and $\Delta m(0)$ are the light curve amplitude at
phase angles $\varphi$ and 0 deg, respectively. The $\kappa$ parameter should
fall in the range 0.012 to 0.030, depending on the applicable light scattering
properties. It is noted that eq. (\ref{eq4}) implies that variability is 
obtained for constant observing geometry of the object.

Comet 41P has a relatively slow rotation period that was increasing 
considerably from about 1 day to more than 2 days in the course of our observing 
campaign (see Section \ref{sec_1} and the discussion in the paragraph on `Nucleus 
rotation' below). In order to sample the amplitude of the brightness variability 
of 41P, we scanned our nucleus brightness estimations for variability in a 
time frame of 7 days. This way the impacts on the nucleus brightness due to the 
changing observing geometry with time should be no more than 0.2 mag, that is, on 
the order of the uncertainty on the nucleus brightness estimations obtained from 
the observations. We found maximum variability of the brightness of the nucleus  of 2.3 mag 
in SDSS r filter for the time interval 2017 March 26 to 2017 April 03. 

The phase angle values for the maximum variability of the nucleus in SDSS r 
($60\pm4$ deg) fall outside of the range for which eq.~(\ref{eq4}) should be
valid. Nevertheless, we take this relationship and its known parameters as a
zero-order approximation for an estimation of the axes ratio of the
comet. Here, $\Delta m(0)$ provides a lower limit for the main axis ratio $a/b$ of
the nucleus (approximated as being of ellipsoidal shape with axes $a>b=c$) according
to

\begin{equation}
  a/b \ge 10^{0.4 \cdot \Delta m(0)}
  \label{eq5}
.\end{equation}

Taking the above-mentioned mean $\Delta m(\varphi)$ value for the SDSS r filter as
a reference, the nucleus of 41P appears to have a minimum axes ratio $a/b$ of
2.1 (for $k = 0.030\;\M{deg}^{-1}$) to 3.4 (for $k = 0.012\;\M{deg}^{-1}$). This
result indicates a prolate or oblate nucleus shape for comet 41P with a value
for the principal axes ratio that is in the upper range of or slightly beyond what has been
found for cometary nuclei so far \citep[see][who give a/b ratios between 1.67 and
2.5 for four comets]{2004come.book..223L,2015Sci...347a1044S}. We note that
among small asteroids (diameter $>$ 150 m), light curve amplitudes of 1.5 mag and
higher (equivalent to similar axis ratios as determined for 41P) seem 
to be rare \citep[e.g., two cases in Fig.~\ref{fig5} of][]{2002aste.book..113P}. On
the other side, the likely interstellar object 1I/2017 U1 Oumuamua was found to
have a very elongated shape with an axis ratio of between 7 and 10
\citep{2017Natur.552..378M}.

\textbf{Nucleus rotation:} \citet{2018Natur.553..186B} present observational
results of 41P, namely time-variable coma structures and the UV light curve of the
inner coma, that allow interpretation of the nucleus rotation period, which
changed from about 20h in early March 2017 to around 46-60h in early May
2017. This finding is supported by results from \citet{2018A+A...619A.156M}
using time-variable gas jet structures seen in the coma within about a month
of perihelion passage of the comet. \citet{2019AJ....157..108S} obtained a
quadratic approximation over time for the rotation period P(t) of 41P from an
analysis of gas jet features in the cometary coma. These latter authors found
$P(t) = 0.8755 - 0.0006892 \cdot t + 0.0003262 \cdot t^2$ ($t = $ time since
2017 March 01 0 UT). This relationship should describe the period increase during
the time interval used for the jet analysis, that is, between about mid-March and
early May 2017.

Our data (see Tab. \ref{tab4} and Fig. \ref{fig9}) show that the 
apparent and absolute brightness of the nucleus of 41P varied on timescales 
of a day to a few days. The amplitude of the brightness variability was larger 
before and during perihelion than towards the end of our observing campaign, 
i.e., beyond about 50 days post-perihelion. Our attempts to fit the variability in the  
brightness of the nucleus with a constant or changing (linear and quadratic functions 
were applied) rotation period gave inconclusive results, most likely because the 
changing light scattering scenario (illumination and viewing geometry, body 
shape, and light scattering properties of the nucleus of 41P) was not considered.

\subsection{The dust and gas tail}\label{sec_3.6}

\begin{figure*}[tbp]
  \centering
  \includegraphics[width=0.52\textwidth]{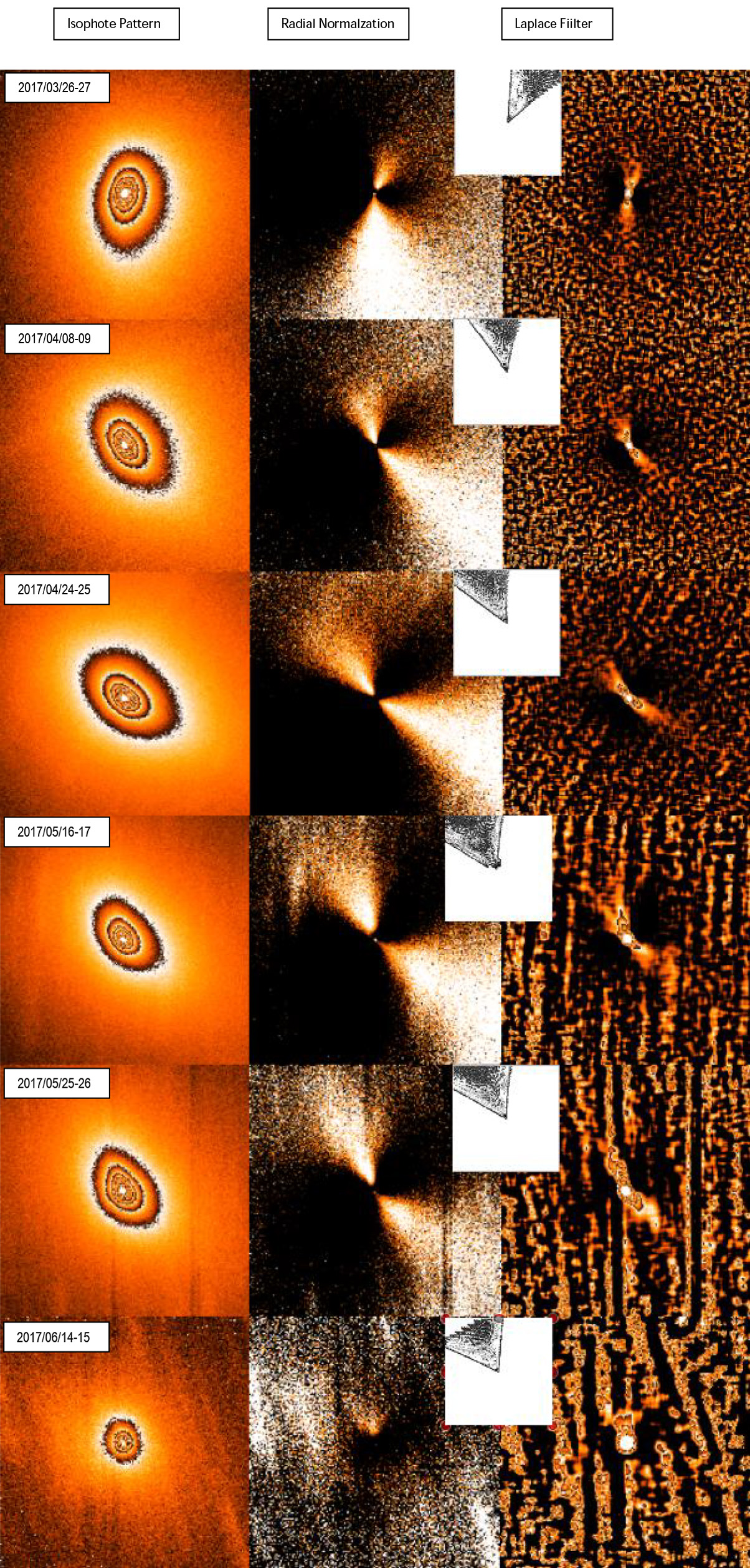}
  \caption{Temporal evolution and geometric structure of coma {\it fan A} in
    comet 41P. This image series shows coma and {\it fan A} therein from top to
    bottom on nights 2017 March 26-27, 2017 April 08-09, 2017 April 24-25, 2017 May 16-17,
    2017 May 25-26, and 2017 June 14-15 (date format in the panels is year/month/day). 
    All images were taken through the SDSS r
    filter and have an extension of 16\,000km at the comet in X and Y direction
    with the nucleus position in the center of the frames. North is up and East
    is to the left. Left column: isophote pattern,
    middle panel: radially normalized version, right panel: Laplace filtered 
    version. For each night, we present median averaged images of all
    exposures. {\it Fan A} appears in the northern
    hemisphere of the coma. Noise 
    patterns of the size of the typical pixel
    window used appear as irregular `salt-and-pepper islands' in the Laplace
    filtered result. Since the Laplace filter is sensitive to curvature in the
    pixel flux distribution of the image, the appearance of the fan differs
    slightly from that in the radially normalized image. The wide structure in
    the south marks the anti-sunward tail orientation of the comet. 
    Image artifacts due to
    incompletely removed star trails or charge overflow patterns appear as
    linear streaks in the images. Between the middle and right panel, small
    inserts illustrate the results of the fan modeling with the
    \citet{1992Ap&SS.191....1B} model. These inserts show the fan geometry for
    the solution of the rotation axis and active area location that we consider
    to best describe the observed fan structure in the coma of 41P.  }
  \label{fig11}
\end{figure*}
 
The dust tail of 41P is noticeable in the calibrated images by the southward
extension of the isophotes and by a broad linear structure in the radially
normalized and Laplace-filtered images of the comet
(Figs.~\ref{fig11},~\ref{fig13},~\ref{fig14}). The tail orientation represented
by its near-nucleus position angle (Tab.~\ref{tab7}) moved from approximately the southern
direction in March 2017 towards the southwest in April and May 2017 and as of the
end of May 2017 back over southern directions towards east-southeast in mid
July 2017. The relatively fast change in the tail position angle as of mid June
2017 is related to the opposition passage of the comet when the comet changed
from the morning to the evening sky and the tail flipped orientation in the
opposite direction towards the Sun. In Tab.~\ref{tab7}, the Sun direction with
respect to the nucleus, represented under column label `\texttt{PA r vector}',
indicates the direction where the recently emitted (within days to 2 weeks) and
radiation-pressure-sensitive dust is found in the tail region of the comet. Old
and radiation-pressure-insensitive dust released months to years before the
observing epoch is found closer to the position angle of the projected velocity
vector (column `PA v vector' in Tab.~\ref{tab7}).

Since the near-nucleus position angle of the tail structure in 41P aligns
(within about 5 deg) with the anti-solar direction (see column `PA r – tail' in
Tab.~\ref{tab7}) while it is far from the direction of the old and heavy
dust (see column `PA v – tail' in Tab.~\ref{tab7}), we conclude that the broad
linear structure in the southern hemisphere represents radiation-pressure-sensitive 
dust that was only recently emitted from the cometary nucleus. Around
opposition on 2017 July 03, no signature of the dust tail is visible in our
images. At that time, the view of the tail from Earth was mostly in the radial
direction (slightly below the orbital plane of the comet), and it became projected 
behind the coma of the comet that was also seen mostly front-on.  No
indication of a trail-like feature is found in our images of the comet, in
particular not at the beginning or towards the end of the observing campaign,
when Earth was not far above the orbital plane of the comet. These findings
may indicate that heavy dust may not be abundant in that comet. However, quantitative
constraint of this statement requires proper dust dynamics and light
scattering modeling.

Signatures of the plasma tail of comet 41P are not apparent in the images. These
can appear in approximately anti-solar direction in the form of tail rays and
narrow, straight, and/or kinked features with embedded isolated substructures
\citep[see][]{1994A&AS..107..101B}. Such signatures should display fast (within hours)
changes in position and geometry and should be most visible in the SDSS g and r
filter images of the comet (due to emission bands of CO$_2^+$, CO$^+$ and 
H$_2O^+$ molecules).

\subsection{Structures in the cometary coma, active regions and rotation axis of 
the nucleus}\label{sec_3.7}

\begin{figure}[tbp]
  \centering
  \includegraphics[width=\columnwidth]{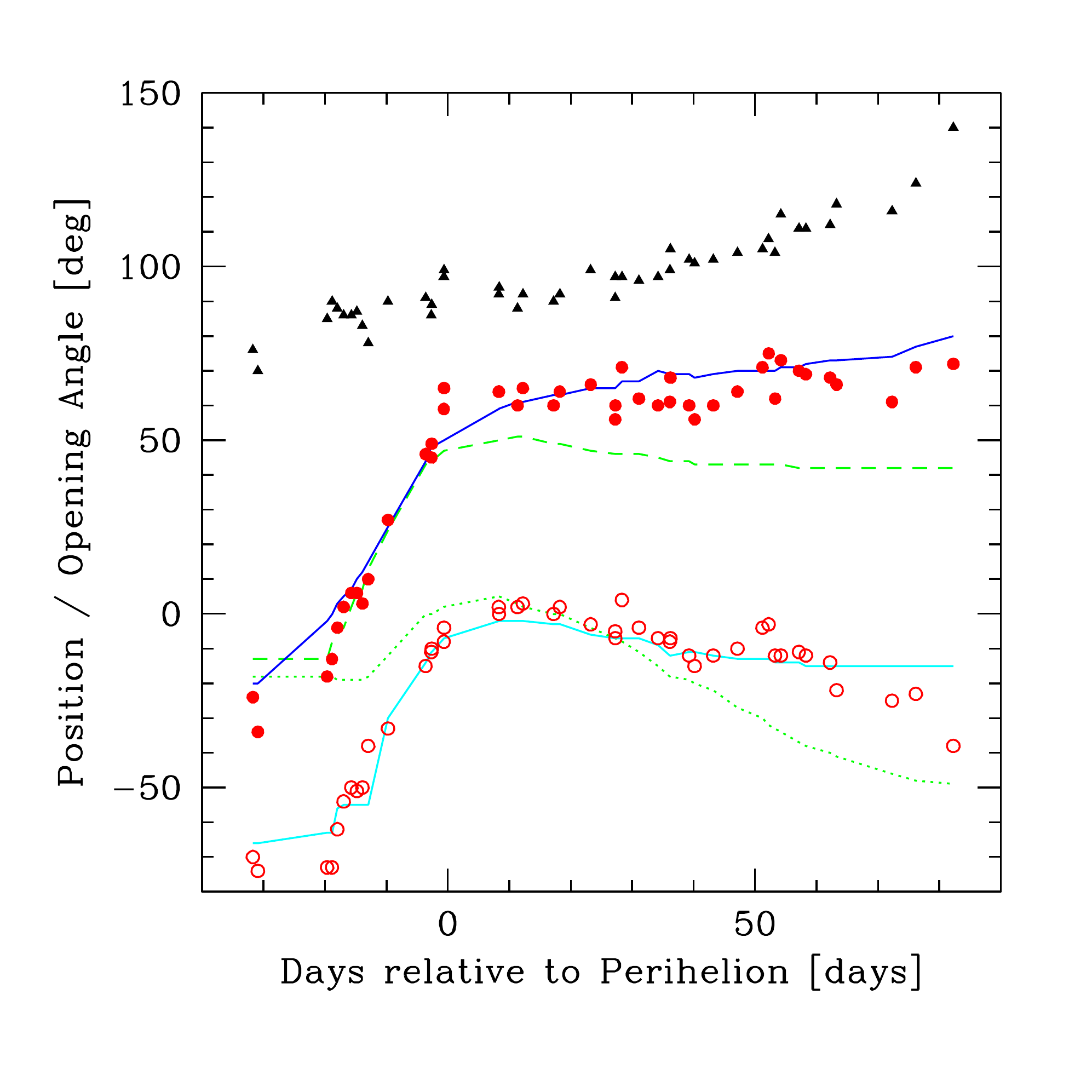}
  \caption{Position angles and opening angle of coma {\it fan A} seen in comet
    41P vs. time to or from perihelion. The plot shows the temporal evolution
    of the observed position angle for the western (`\texttt{Awest}' = red open 
    circle) and eastern
    (`\texttt{Aeast}' = red filled circle) edge of {\it fan A} (for `\texttt{Awest}' 
    and `\texttt{Aeast}' see Tab.~\ref{tab8}). Lines show the PAs of the fan edges 
    obtained from model simulations, i.e., our model: eastern edge (continuous 
    blue line), western edge (continuous cyan line; model of 
    \citet{2018A&A...615A.154P}), eastern edge (broken green line), and western edge 
    (dotted green line). The observed PAs were measured at 2\,000 km projected nucleus 
    distance at flux levels 5\%\ above mean coma background at that distance. 
    The opening angle results (black triangles) are derived from the PAs of the 
    observed fan edges and are plotted with an offset of +30 deg for clarity. }
  \label{fig12}
\end{figure}

\begin{figure*}[tbp]
  \centering
  \includegraphics[width=0.8\textwidth]{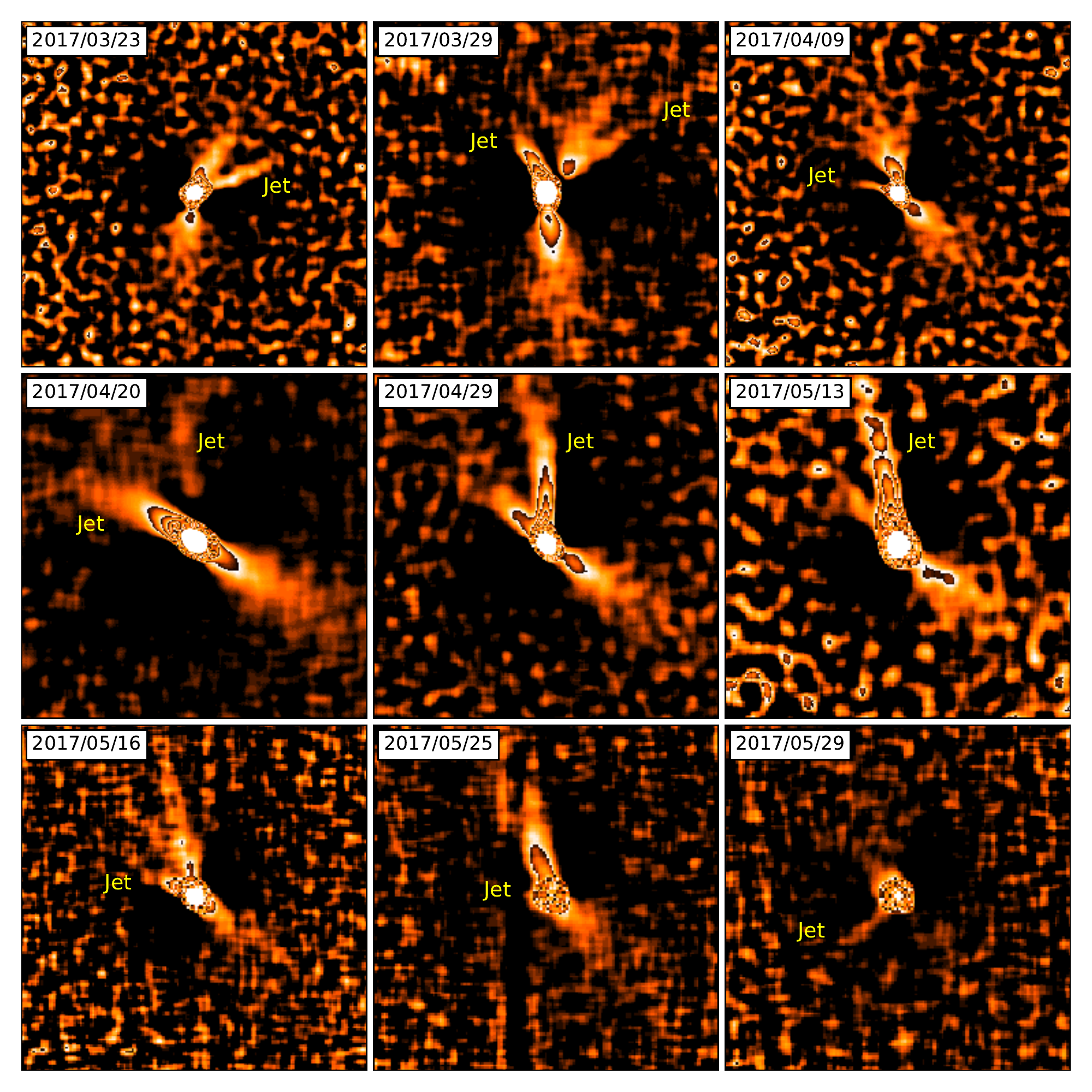}
  \caption{Jet structures in comet 41P outside of the {\it fan A} region. The
    Laplace filtered images show jet features on: 2017 March 23,
    2017 March 29, 2017 April 09 (top row);  2017 April 20, 2017 April 29, 2017 May 13 (middle row),
    and 2017 May 16, 2017 May 25, 2017 May 29 (bottom row). The date format in 
    the panels is year/month/day. In the panels,  only the
    jet(s) are marked ({\it fan A} and artifacts from background star trails or
    charge overflows are not marked in the images for clarity reasons). All
    images are taken through SDSS r filter. The field of view is 3\,000 x
    3\,000 km at the distance of the comet. North is up, east is to the left.
    Sun direction is opposite to the tail orientation in the southern coma 
    hemisphere.}
  \label{fig13}
\end{figure*}

To search for and analyze morphological structures in the coma of 41P
we used the enhanced (Laplace-filtered and radially normalized; see
section \ref{sec_2}) versions  of the individual and combined (median and mean averaged)
images of the comet.

From the images of the more than four-month observing campaign of 41P, we
identified the repeated occurrence of a coma fan, labeled `\texttt{Fan A}' in
Tab.~\ref{tab8}, and of several jet structures, labeled feature EJ and WJ in
Tab.~\ref{tab9}. Since the coma features are seen in the SDSS gri filters that
cover a wide wavelength range, we assume that they are due to the dust
distribution in the coma, although gas contribution to the measured coma light
may be present (see sections \ref{sec_3.2}). In support of the interpretation as
dust signatures, we note that the features analyzed here show in part a rather
collimated nature and are best modeled by expansion speeds of far below the speeds of 100
m/s typical for cometary dust.
 
We analyzed the nature of the respective coma features using dust coma
simulations performed with the cometary coma and tail modeling software
developed and described by \citet{1992Ap&SS.191....1B}. This Monte-Carlo-type
software allows calculation of the brightness distribution of the cometary
dust coma and tail via a Finson-Probstein approach \citep{1968ApJ...154..327F}. 
Coma structures can be simulated through dust released from individual
active areas on the rotating nucleus. Apart from the physical parameters of the dust
(emission velocity, radiation pressure sensitivity, activity profile daily and
with solar distance), important key parameters of the model are the attitude of
the rotation axis and the location of active regions on the nucleus. The
rotation axis is represented by the obliquity angle towards the orbit pole
direction of the comet and the rotation angle of the meridian at the time of
perihelion. An active region is defined by its nuclear longitude and
latitude. The model assumes simple rotation of a spherical nucleus. As the
rotation period of 41P changed considerably over time \citep{2019AJ....157..108S}, 
the results of the model software are restricted to the orientation of the rotation 
axis and the latitudes of active areas on the nucleus. Longitude estimations for 
these regions were not possible in the case of 41P.

The comparison of the simulation results is performed by finding a best possible match
of the near-nucleus position angles of a given coma feature in the
available images. For the coma fan, the opening angle of the feature,
measured in radially normalized images of the comet, is also used. The opening angle
geometry is determined from position angle measurements of the structure at
2\,000 km projected distance from the nucleus where it is 5\%\ above mean
coma flux level at this distance. Beyond that, the overall geometry and
appearance of the simulated coma structure are also compared with the observed
feature to check overall consistency. The matching process makes use of
simulation results that systematically iterate the parameters of the rotation
axis and of the location of an active region on the nucleus. The best-matching
result implies an uncertainty of better than 10 deg for the orientation of the
rotation axis and of close to 5 deg for the obtained information on the location
of active areas. Position-angle measurements of imaged coma features have
estimated uncertainties of 3 deg, except for very weak and/or low-contrast
features where this may be two to three times higher.

\textbf{Fan A} (Figs.~\ref{fig11}+\ref{fig12}, Tab.~\ref{tab8}): This conical
structure in the sunward coma of 41P is seen at all available observing epochs
except on nights 2017 March 10-11, 2017 June 21, and 2017 July 13+16+17. The
nondetection of the fan in March 2017 is most likely a consequence of the
reduced sensitivity of the 3KK detector (in the images of this night only the
very center of the coma is visible), while that on the nights in June and July
2017 might be due to the reduced brightness contrast of the fan and enhanced
disturbances from the very crowded star background in the images. {\it Fan A}
appeared (see Fig.~\ref{fig11}) in the northwestern coma quadrant in March
2017. It swung over North into the northeastern coma part in early April 2017,
where it remained until the first week of May 2017, when it started to drift
back towards the northern coma section. During the observing interval a
continuous increase of the fan opening angle from about 40 to 110 deg is noted
(Tab.~\ref{tab8} and Fig.~\ref{fig12}). {\it Fan A} extended to maximum
projected distances of up to 8\,000 km from the nucleus in March to May 2017 and
up to 4\,000 km in June 2017. Maximum brightness excess in the fan cone was
25-30\%\ above the mean coma background in March to May 2017; thereafter, it
reduced first to 20\%\ and then to 10\%. 
{\it Fan A} displayed time variable fluctuations in
its intrinsic brightness profile, that is, linearly extended fan sections of maximum
brightness varied in position angle over time, as did the overall shading of the
brightness distribution in the fan. This variable appearance might be
due in part to jet structures originating from local areas on the rotating nucleus
for individual nights projected into the position angle range of {\it fan A}.

The forward modeling of {\it fan A} showed immediately that for the slow
rotation period of 41P in March and April 2017 as determined by
\citet{2018Natur.553..186B}, \citet{2018A+A...619A.156M}, and
\citet{2019AJ....157..108S}, localized regions of enhanced activity over a few
degrees extension on the nuclear surface result in narrow collimated jets that
do not smear out to a fan feature of quasi-continuous and wide angular
extension. Hence, {\it fan A} must come from a rather extended source region of
enhanced nuclear activity. The simulation results that best match the
measured geometry of the observed coma fan (Tab.~\ref{tab8}) were obtained for
a rotation axis orientation with 52 deg obliquity towards the orbital plane of
the comet and about 50 deg rotation angle for the perihelion meridian of the
nucleus. The active region extends beyond about +50 deg latitude, and it may
cover the whole range up to +90 deg latitude and all longitudes. This scenario
is highly compatible with the appearance of a rather homogeneous brightness
distribution within the fan as well as over time. Individual localized spots of
even further enhanced activity may produce time-variable jet features and
brightness variations (shading) that are noticed inside the fan area. It is noted
that from the appearance of {\it fan A}, it is not possible to firmly
distinguish the sense of rotation of the nucleus. This implies that the
mirrored rotation axis with 128 deg obliquity and 230 deg rotation angle for the
perihelion meridian also produce acceptable fan shapes and geometries. The sense of
rotation of the nucleus is again addressed in the last paragraph of this 
section as well as in section \ref{sec_3.8}.

The best-matching solution explains the overall geometry of coma {\it fan A} as
a large active region on the nucleus and a rotation axis attitude, which both
remained constant in time (within our uncertainty limits), at least from 
mid-March to mid-June 2017. Solutions for enhanced rotation of the nucleus might
also be possible but were not explored here. It is noted that the constant
rotation axis solution presented above has difficulties in reproducing the wide
opening angles of {\it fan A} measured as of mid-June 2017 (see Tab.~\ref{tab8}
and Fig.~\ref{fig12}). Although the fan contrast towards the end of our
observing campaign is at a low to marginal level, the possible deviations
between measured and modeled fan geometry might be an indication that at that
time the rotation axis of 41P may have been in an enhanced state, as was
speculated by \citet{2019AJ....157..108S}.

From the analysis of the Trappist observations of 41P,
\citet{2018A&A...615A.154P} found a rotation axis orientation with an obliquity
of 25 ($\pm$ 15) deg and a rotation angle of the meridian at perihelion of 5
($\pm$ 3) deg. These latter authors also argue for the presence of a large-scale region of
enhanced activity extending beyond 45 ($\pm$ 10) deg north on the
nucleus. Using this information for simulations of the appearance and geometry
of a resulting fan at the epochs of the Mt. Wendelstein campaign on 41P confirms
the overall trend in the orientation of {\it fan A}. However, it is noted (see
Fig.~\ref{fig12}) that the width of {\it fan A} tends to be narrower and fan
edges are shifted towards the west compared to the ones seen in our images. Whether
these deviations can be compensated by a time-variable orientation of the
rotation axis of the comet or by the temporal and spatial evolution of the
activity profile in the wide active region on the nuclear surface remains
unexplored here.
 
\textbf{Jet structures} (Fig.~\ref{fig13}, Tab.~\ref{tab9}): A number of
narrow linear features are seen in Laplace-filtered images of the coma of 41P
(for examples see Fig.~\ref{fig13}). These occur preferentially in the sunward coma
hemisphere with variable intensity and spatial extension, and are best
identified outside of the {\it fan A} region in the coma, that is, eastward (larger
position angles) and westward (smaller position angles) of the fan
structure. The appearance of these linear jet features is episodic as can be
seen from Tab.~\ref{tab9}. In the table, a quasi-alternating appearance of the
features over time is indicated: they are seen westward of {\it fan A} between
20 and 9 days before perihelion, eastward of {\it fan A} between 3 days before
and 8 days post-perihelion,  westward again between 8 and 31 days post-perihelion,  
and once more eastward of {\it fan A} between 28
and 51 days post-perihelion. Narrow linear features are not
identified in our 41P images thereafter, that is, after 2017 June 02. This may be 
due in part to the fainter flux level of the comet and lower contrast in the coma.

The collimated appearance of the jet-like features, which is similar in the different
filters used during the comet observations, is in favor of a dust-dominated
nature. For 41P, this interpretation can be firmly supported by dust coma
simulations: local regions of enhanced dust emission on a slowly rotating
nucleus produce collimated quasi-linear jets. Contrary to the extended source
region of {\it fan A}, the jets in the coma of 41P originate from small areas on
the nucleus surface localized in a narrow range (not more than a few degrees
extension) in longitude and latitude. The slow rotation of the nucleus results
in quasi-constant near-nucleus position angles of the jets that are compatible
with the constant appearance of the jets over the nightly observing
intervals. The prominent occurrence of the dust jets on the sunward side implies
a `sun-focused' emission profile of the dust.

The simultaneous appearance of two dust jets, one on the eastward and one on
the westward side of {\it fan A} on 2017 March 29 and 2017 April 20
(Fig.~\ref{fig13}), requires the presence of at least two source regions that
are widely separated on the nucleus of the comet. \citet{2019AJ....157..108S} 
identified two gas jets, labeled J1 and J2, originating from two local
areas that have a wide separation in longitude on the nucleus. Using the J1 and
J2 identification as given in Fig.~\ref{fig8} of \citet{2019AJ....157..108S},
we tentatively associate the dust jets in our images with the same
respective source region as the gas jets, as follows: gas jet J2 with the westward
dust jet in our images of nights 2017 March 29 and 2017 April 20 and gas jet J1 with the
eastern dust jets of nights 2017 March 29 and 2017 April 20.

We applied our solution for the rotation axis attitude of 41P, which was obtained
from the simulation of the coma {\it fan A} in order to constrain the location
of the source regions of the dust jets identified in our images eastward and
westward of the fan area. The numerical simulations of the dust jets indicate
that the source regions of the eastern and western dust jets seen in March and
April 2017 should have latitudes of +35 deg or less. The dust jets seen in May
2017 on the eastern side of the coma require close-to-equatorial source regions
in order to allow for the southeastern position angles measured in our
images. Our simulations also show that the source regions of gas jets J1 and J2
– and the associated dust jets in our images – require near equatorial
latitudes in order to appear on the anti-sunward side of the nucleus.

It is noted that the dust jet east of the coma fan on 2017 March 29 shows a slight
anti-clockwise jet curvature, as does the dust jet towards the east on 2017 April 09
(Fig.~\ref{fig13}) – and the gas jets in the images of
\citet{2019AJ....157..108S} during this time interval. Our dust coma simulation
clearly indicates that the anti-clockwise curvature requires clockwise rotation
of the nucleus. This finding in turn identifies the rotation axis with 52 deg
obliquity and 50 deg rotation angle of the meridian at perihelion as the preferred
solution compared to the rotation axis pointing in the opposite direction. The
simulations also show that the eastward-oriented dust jets represent the morning
side on the nucleus, while westward features originate from the afternoon to
evening side.

\subsection{The outburst on 2017 May 10-11}\label{sec_3.8}

\begin{figure*}[tbp]
  \centering
  \includegraphics[width=0.3\textwidth]{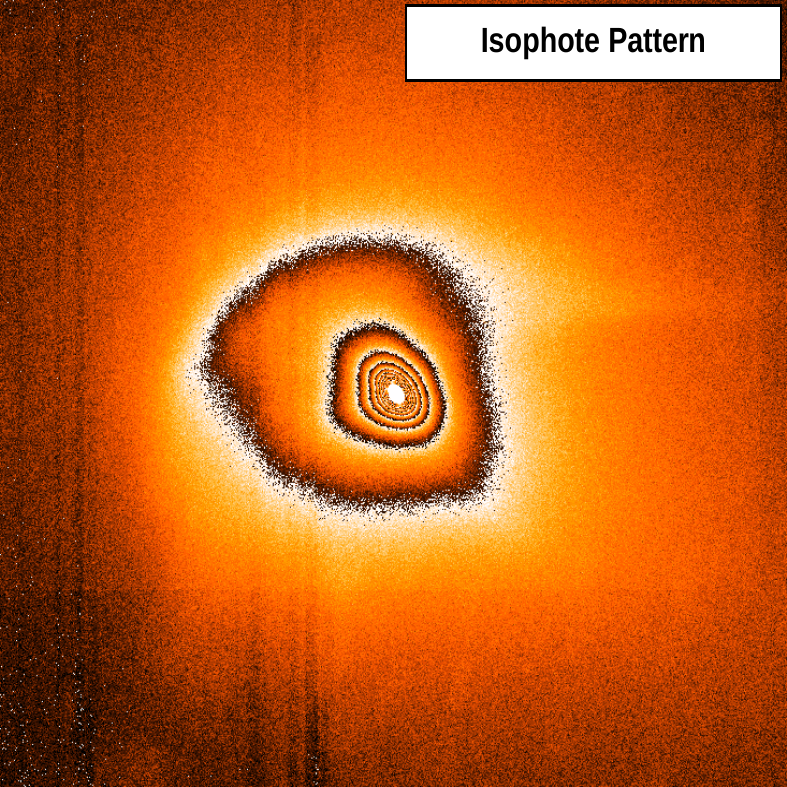}
  \includegraphics[width=0.3\textwidth]{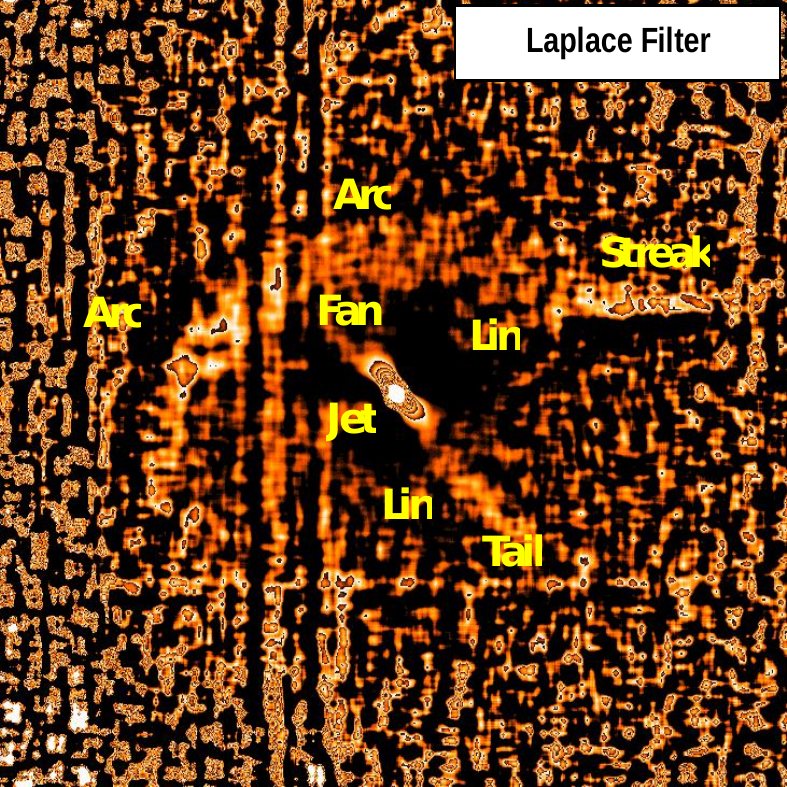}
  \includegraphics[width=0.3\textwidth]{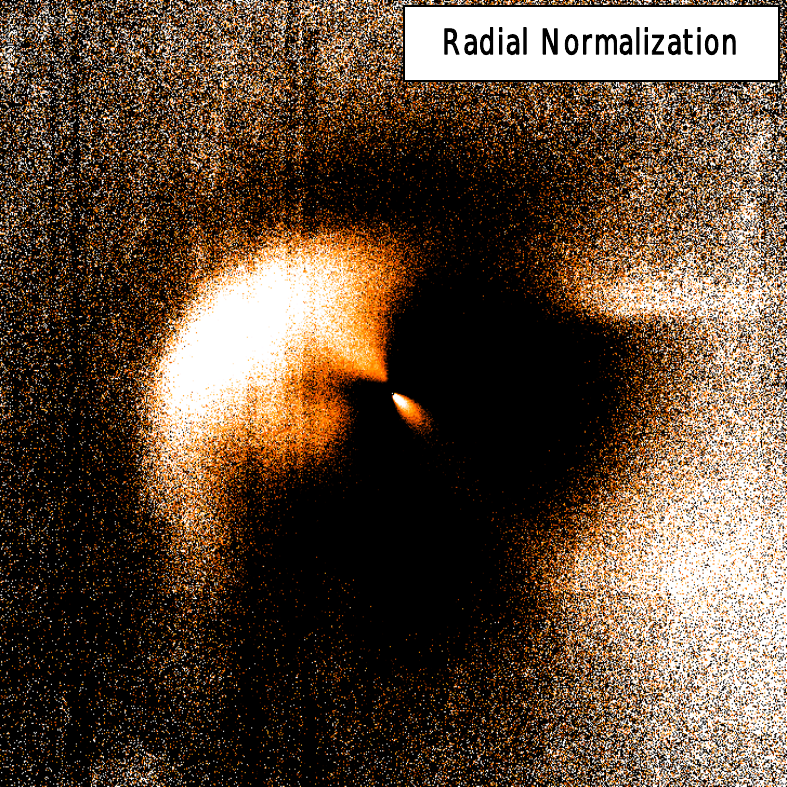}
  \caption{Outburst of comet 41P observed on 2017 May 11. We show 
  the isophote pattern of the comet in the SDSS r filter (left panel), the 
  respective Laplace-filtered result (middle panel), and the radially normalized 
  pattern (right panel). The images cover 22400 x 22400 km of the coma at the 
  distance of the comet. North is up, east is to the left. In the Laplace-filtered 
  image, various coma structures are marked:  the arc, the jet, 
  the lineaments (Lin) from the outburst, the fan, and the dust tail from the 
  normal activity of the comet. Other features are noise patterns or originate 
  from charge overflows.}
  \label{fig14}
\end{figure*}

SDSS r filter photometry of the coma of 41P shows an approximately 0.6 mag
brightness increase ($\sim$ 70 \%) on 2017 May 11 1-2 UT (Tab.~\ref{tab3})
as compared to the measurements about 25 hours earlier (2017 May 10 00:30UT) and
43 hours later (2017 May 13 20UT). Similar brightness variations of the comet are
not seen during other time intervals covered by our observing campaign on
41P. The brightness increase is accompanied by unique structures seen in the
coma of the comet on the night of  2017 May 11 only (Fig.~\ref{fig14}): The Laplace-filtered
coma image displays an arc of 2\,100 km in width and spanning over about 195 deg (from PA
$\sim$ -65 over PA = 0 to PA $\sim$ 130 deg). The arc reaches a maximum intensity
ratio of about 1.5 above the mean coma background in the east-northeast
region of the coma and a minimum ratio approximately equal to the mean coma level in the
north-northwest region of the coma. The arc is accompanied by a straight
jet-like feature of about 25 deg opening angle at a position angle of 95 – 120 deg
that extends from the coma center to about 4\,500 km into the eastern coma. In the 
Laplace-filtered image a narrow linear feature is seen at a marginal level 
that extends from about the eastern part to the western part of the arc in the 
tailward side of the coma. Moreover, at least one narrow streak is seen extending
from $\M{PA} \sim -55\;\M{to}\; -75\;\M{deg}$ at 2\,500 to 11\,500 km projected
distance from the coma center, respectively. In addition to the outburst-related coma
structures, the {\it fan A} and the dust tail can still be identified, which are seen 
essentially undisturbed in the images of that night.

By the time of the observations on 2017 May 11 1-2 UT, the outburst may have
lasted for a maximum of about 1 day which is close to half the rotation cycle of the
comet measured by \citet{2019AJ....157..108S} about 10 days earlier. Also, the
more than 180 deg coverage of the arc supports the interpretation that the
outburst activity continued for about half a rotation of the nucleus. The
narrow linear features and the straight jet indicate that the outburst signal in
the SDSS r filter images of this night is due to dust-reflected sunlight,
although strictly speaking gas activity cannot be completely excluded. The maximum
projected distance of the arc from the nucleus (about 7300 km) is compatible
with a dust expansion speed of about 80 m/s (for a 24h outburst period). The 
radial gradient of the inner coma (inside about 1500 km nucleus distance) shows 
an exponent (0.98, see Tab. \ref{tab4}) close to unity, which is indicative 
of an almost geometric expansion of the dust in this distance range. It is noted that
the radial exponents of the inner coma show significantly smaller values for
the observing epochs immediately preceding and succeeding the outburst date.  

Our dust coma modeling using a rotation axis solution obtained from a fan
analysis and assuming a rotation period of about 50 hours allows the following
conclusions: The rotation motion of the nucleus is from east to west in the
images, with the former being the morning side and the latter the evening side of
the cometary activity. Since outburst signatures are seen in the sunward and
tailward side of the coma center, the outburst activity extended over a wider
range in latitude. The eastern jet of  25 deg  in width implies an extension of the
activity from about 30 deg north to about 5 deg south on the nucleus (assuming 
the preferred rotation axis solution from Sect. \ref{sec_3.7}). It is
noted that the outburst may have started on the evening side of the nucleus, 
lasting until at least the source region had rotated to the morning
side of the comet. The wide extension of the outburst in latitude covers at
least more than 400 m of terrain on the (600 m radius) nucleus of 41P and is therefore not
representative for the local active regions from which the dust jets originate.

Possible scenarios for the outburst are: (a) a nucleus splitting event, (b) an
up-welling of a dust cloud from a landslide on the nucleus, and (c) a singular
activity outburst from an extended region on the nucleus.

\textbf{`Nucleus splitting' scenario:} The potential nucleus splitting of 41P on
2017 May 10 did not produce any noticeable relics
in our images, such as daughter nuclei. The separation velocities of subnuclei are typically on the
order of 1 m/s, as is estimated for the daughter nuclei of comet splitting
events \citep{2004come.book..301B}. Therefore, the separation distance of split
nuclei in 41P should have been 80 and 250 km during the two observing epochs
following the assumed splitting date 2017 May 10. Inspection of the images of
these two nights did not reveal any indication of the presence of a daughter
nucleus in 41P. The same negative finding applies for later observing epochs of
the comet until the end of our campaign. It is also noted that the overall coma
brightness returned to the normal level 3 days after the outburst, a behavior that
is considered atypical for nucleus splitting events reported in the literature
\citep{2004come.book..301B}. Therefore, apart from the signatures related to the
outburst, direct or further indirect evidence for a nucleus splitting of 41P on
2017 May 10 is absent in our data of the comet as well as in the reports of the
2017 apparition published so far.

\textbf{`Dust cloud from a landslide' scenario:} The total amount of dust
$M_{\M{Dust}}$ released during the outburst between 2017 May 10 and 2017 May 11
(i.e., within about $\Delta$t = 24h) can be approximated using the measured
brightness increase $\Delta m = 0.6$ mag compared to the dust-production rate
$Q_{\M{Dust}}$ of the comet (about 45 kg/s) given by \citet{2018A&A...615A.154P}
for early May 2018: 
$M_{\M{Dust}} \sim Q_{\M{Dust}} \cdot \Delta t \cdot \left(10^{-0.4 \Delta m}
  -1\right) \sim 2900$ tons (approximately three-quarters of the dust mass released by the nucleus
per day) or roughly  5700 m$^3$ of dust for an assumed mean bulk density of
0.5 g/cm$^3$ for the nucleus solids as found in comet 67P/Churyumov-Gerasimenko
\citep{2016Natur.530...63P}. A landslide scenario for nonrepetitive outbursts
on this comet was introduced by \citet{2016MNRAS.462S.184V} and further analyzed
by \citet{2017NatAs...1E..92P}. This scenario involves the collapse of cliff structures on
the cometary nucleus, an event that was found to be short in time for 
67P/Churyumov-Gerasimenko (order
10 min) and involving the release of approximately a few percent 
of the mass of the collapsing material. The dust leaves the surface in a dust 
cloud of wide opening angle. The cliff
collapse may be triggered due to a weakening of the cliff strength by thermal
stresses and in the end may leave fresh volatile ice-rich material from
deeper layers on the cliff surface. As a consequence, the collapsed region may
display enhanced activity over a much longer time interval than the duration of
the actual cliff collapse. The ejected mass of cometary material from the Aswan
cliff collapse over about 70 m \footnote{It is noted that a landslide structure 
in the Hathor cliff of 67P/Churyumov-Gerasimenko extends over several hundred meters.} 
in length, one of the stronger ones seen on 67P/Churyumov-Gerasimenko,
amounts to 500 - 1\,000 tons and should represent only a small percentage of the mass
that was sliding down the cliff wall \citep{2017NatAs...1E..92P}. The ejected
mass of the 41P outburst was greater than that of the Aswan outburst on
67P/Churyumov-Gerasimenko by a factor of between three and six and would 
therefore indicate a more massive cliff collapse, 
possibly also
indicated by the wide latitude extension of the active outburst region. The
outburst of 41P may have started on the afternoon side of the nucleus, and it
involved an extended surface region, as is also observed on 
67P/Churyumov-Gerasimenko. However, the duration of the outburst of 41P is 
much longer than found in the case of 67P/Churyumov-Gerasimenko. It
also displayed a variable intensity profile over the 24h duration that was
imaged in our data: The initial outburst phase was followed by a decay in
activity and thereafter by a considerable brightness increase towards and on
the morning side of the coma. The latter might be due to the enhanced
sublimation of volatile ices that were freshly  exposed to the Sun
after the cliff collapse. It is noted that in the second half of May 2017, short
jet features appeared in the morning side of the coma in the position angle
range representing the latitude range of the outburst region on 2017 May 10 (see
Tab.~\ref{tab9}).
 
\textbf{Outburst from an extended region on the nucleus:} The scenario of a
singular outburst from an extended region on the nucleus involves a thermal
trigger for a sudden gas release from subsurface layers, caused by either a heat
wave from solar illumination reaching a massive pocket of very volatile ice
(CO$_2$, CO) or  subsurface gas that has not been able to diffuse into the coma
with time, has built up significant gas pressure, and at some moment exceeding
the tensile strength of the enclosing cometary material. The available results
from the observations do not allow us to distinguish the gas-triggered outburst
from the cliff-collapse scenario. However, we note that the region affected
by the outburst seems to be very extended in latitude, that is, over more than 400 m
in length, which is more than half of the mean radius of the nucleus, that is, the
subsurface gas reservoir must have been very extended.

\section{Conclusions}\label{sec_4}

During its last perihelion passage at a distance of 0.14 AU from Earth, comet 41P
was imaged in broadband SDSS gri filters between 2017 March 11 and 2017 July17 from
the Mt. Wendelstein observatory with the 2m Fraunhofer telescope over a wide
phase angle range (14-69 deg). The apparent coma brightness peaked around
perihelion (2017 April 12). Determination of the dust activity profile of 41P and of
its maximum is not possible without the adoption of proper phase angle
corrections. Depending on the chosen phase function, we found a single power
law of the solar distance profile with a maximum activity around perihelion and
a steep radial exponent or a segmented profile with its maximum a few weeks before
perihelion and a shallower decrease with increasing distance from the Sun. The
measured filter flux of the comet is dominated by dust-reflected
sunlight. However, during the time period from mid-March to the end of 
May 2017 the filter flux was contaminated
by light from emissions of coma gases. In the very inner coma ($<$ 1\,000 km
from the nucleus), the power exponent of the radial profile was mostly below
unity, indicating ongoing fragmentation of dust. The overall color and spectral
gradient of the cometary dust was found to be solar ($\sim -2$ to $+2 \%/100$ nm).
Trends towards bluer SDSS r-i coma colors with time to and from perihelion 
and with increasing distance from the nucleus are noted and are interpreted as a 
possible signature of changing dust size distribution with time and within the
cometary coma, respectively. Signatures in the anti-solar direction in the images
indicate the presence of a dust tail made of grains that were only recently 
(within days to a few weeks) released from the nucleus. No dust trail of older 
and heavier dust grains and no signatures of an ion tail are found in the images.

The size of the nucleus of 41P was determined directly from the light
distribution in the central brightness peak of the coma. The value of about 600
m for the equivalent radius of the nucleus confirms an earlier size
estimation. Significant variability of the nucleus brightness suggests a minimum
axis ratio of more than two. This ratio value is higher than what has been measured for other
comets, and marks the upper end of ratios seen in other small bodies observed
in the Solar System, except for 1I//2017 U1 Oumuamua, which is potentially an extra-solar
object. 
The rotation axis of the nucleus was determined
from the continuous fan structure in the northern coma hemisphere. It remained
approximately constant from mid-March to early June 2017 (obliquity = 52 deg, meridian
at perihelion passage = 50 deg, uncertainty $\sim 10$ deg). 
Deviations of the fan geometry (i.e., wide opening angle) may indicate
changes in the rotation motion of 41P as of about June 2017. The fan itself
originates from a wider region on the nucleus beyond about 50 deg north up to
polar latitudes. Two prominent linear and a few narrow jet features in the coma
can be traced to local areas at moderate northern to equatorial latitudes, with
the former two being identified on two epochs as being associated with the J1 and
J2 gas jets seen by \citet{2019AJ....157..108S}. On 2017 May 10-11, a 0.6 mag
outburst is observed in 41P that is responsible for various structures (arc,
wide jet, lineaments) in the coma of the comet. The outburst produced about
3\,000 tons of dust within a day. Together with the curvature of some dust jets
seen earlier in the coma, this outburst provided the opportunity to constrain 
the rotation direction
(prograde) and the rotation period of the nucleus ($>2$ days). The outburst
originated from a region that extended over more than 400 m (between 30 deg north
and 5 deg south) on the nucleus and may have been produced by a major landslide
on the nucleus or the explosion of an extended subsurface gas pocket.

The analysis of our observational results for comet 41P has shown that a good
understanding of the light scattering of the dust population is essential for 
proper interpretation of the dust activity versus solar distance, in
particular if the comet passes through a wide phase angle range during the
observing interval. As the comet came close to Earth, we were able to test a 
new approach for the
estimation of the nucleus brightness, shining through the inner coma. This
 delivered quantitative results on the dimensions of the
body. By comparing the observed geometry of dust structures (fan, jet, arc,
lineaments) in the cometary coma with that of numerical simulations of these
features, we were able to assess the rotation axis and the locations of active regions
on the nucleus. An upgrade of the simulation tools considering
enhanced rotation modes (including a changing rotation period) will allow for further
improvement and consolidation of the results for 41P.

\section{Acknowledgements  }

We thank Dr. Klaus Beisser for the provision of the simulation software for the
cometary dust coma and tail. We wish to acknowledge the important contribution
to this work resulting from the Pan-STARRS1 Survey catalog. This research was
supported by the cluster of excellence 'Origin and structure of the universe' of
the Deutsche Forschungsgemeinschaft DFG. This paper contains data obtained at 
the Wendelstein Observatory of the Ludwig-Maximilian University Munich.

\bibliographystyle{aa}
\bibliography{aa37663-20}

\begin{thebibliography}{56}
\expandafter\ifx\csname natexlab\endcsname\relax\def\natexlab#1{#1}\fi

\bibitem[{{A'Hearn} {et~al.}(1986){A'Hearn}, {Hoban}, {Birch}, {Bowers},
  {Martin}, \& {Klinglesmith}}]{1986Natur.324..649A}
{A'Hearn}, M.~F., {Hoban}, S., {Birch}, P.~V., {et~al.} 1986, \nat, 324, 649

\bibitem[{{A'Hearn} {et~al.}(1984){A'Hearn}, {Schleicher}, {Millis}, {Feldman},
  \& {Thompson}}]{1984AJ.....89..579A}
{A'Hearn}, M.~F., {Schleicher}, D.~G., {Millis}, R.~L., {Feldman}, P.~D., \&
  {Thompson}, D.~T. 1984, \aj, 89, 579

\bibitem[{{Beisser} \& {Drechsel}(1992)}]{1992Ap&SS.191....1B}
{Beisser}, K. \& {Drechsel}, H. 1992, Earth, Moon, and Planets, 191, 1

\bibitem[{{Belyaev}(1986)}]{1986cspc.book.....B}
{Belyaev}, N.~A. 1986, {Catalogue of short-period comets} ({Astron. Inst.
  Slovak Academy of Sciences})

\bibitem[{{Bertini} {et~al.}(2017){Bertini}, {La Forgia}, {Tubiana},
  {G{\"u}ttler}, {Fulle}, {Moreno}, {Frattin}, {Kovacs}, {Pajola}, {Sierks},
  {Barbieri}, {Lamy}, {Rodrigo}, {Koschny}, {Rickman}, {Keller}, {Agarwal},
  {A'Hearn}, {Barucci}, {Bertaux}, {Bodewits}, {Cremonese}, {Da Deppo},
  {Davidsson}, {Debei}, {De Cecco}, {Drolshagen}, {Ferrari}, {Ferri},
  {Fornasier}, {Gicquel}, {Groussin}, {Gutierrez}, {Hasselmann}, {Hviid}, {Ip},
  {Jorda}, {Knollenberg}, {Kramm}, {K{\"u}hrt}, {K{\"u}ppers}, {Lara},
  {Lazzarin}, {Lin}, {Moreno}, {Lucchetti}, {Marzari}, {Massironi}, {Mottola},
  {Naletto}, {Oklay}, {Ott}, {Penasa}, {Thomas}, \&
  {Vincent}}]{2017MNRAS.469S.404B}
{Bertini}, I., {La Forgia}, F., {Tubiana}, C., {et~al.} 2017, \mnras, 469, S404

\bibitem[{{Bockel{\'e}e-Morvan} {et~al.}(2017){Bockel{\'e}e-Morvan}, {Rinaldi},
  {Erard}, {Leyrat}, {Capaccioni}, {Drossart}, {Filacchione}, {Migliorini},
  {Quirico}, {Mottola}, {Tozzi}, {Arnold}, {Biver}, {Combes}, {Crovisier},
  {Longobardo}, {Blecka}, \& {Capria}}]{2017MNRAS.469S.443B}
{Bockel{\'e}e-Morvan}, D., {Rinaldi}, G., {Erard}, S., {et~al.} 2017, \mnras,
  469, S443

\bibitem[{{Bodewits} {et~al.}(2018){Bodewits}, {Farnham}, {Kelley}, \&
  {Knight}}]{2018Natur.553..186B}
{Bodewits}, D., {Farnham}, T.~L., {Kelley}, M. S.~P., \& {Knight}, M.~M. 2018,
  \nat, 553, 186

\bibitem[{{Boehnhardt}(2004)}]{2004come.book..301B}
{Boehnhardt}, H. 2004, in Comets II, ed. M.~C. {Festou}, H.~U. {Keller}, \&
  H.~A. {Weaver} (University of Arizona Press, Tucson), 301

\bibitem[{{Boehnhardt} \& {Birkle}(1994)}]{1994A&AS..107..101B}
{Boehnhardt}, H. \& {Birkle}, K. 1994, \aaps, 107, 101

\bibitem[{{Boehnhardt} {et~al.}(1990){Boehnhardt}, {Fechtig}, \&
  {Vanysek}}]{1990A&A...231..543B}
{Boehnhardt}, H., {Fechtig}, H., \& {Vanysek}, V. 1990, \aap, 231, 543

\bibitem[{{Boehnhardt} {et~al.}(2016){Boehnhardt}, {Riffeser}, {Kluge}, {Ries},
  {Schmidt}, \& {Hopp}}]{2016MNRAS.462S.376B}
{Boehnhardt}, H., {Riffeser}, A., {Kluge}, M., {et~al.} 2016, \mnras, 462, S376

\bibitem[{{Bowell} {et~al.}(1989){Bowell}, {Hapke}, {Domingue}, {Lumme},
  {Peltoniemi}, \& {Harris}}]{1989aste.conf..524B}
{Bowell}, E., {Hapke}, B., {Domingue}, D., {et~al.} 1989, in Asteroids II, ed.
  R.~P. {Binzel}, T.~{Gehrels}, \& M.~S. {Matthews} (University of Arizona
  Press, Tucson), 524

\bibitem[{{Doi} {et~al.}(2010){Doi}, {Tanaka}, {Fukugita}, {Gunn}, {Yasuda},
  {Ivezi{\'c}}, {Brinkmann}, {de Haars}, {Kleinman}, {Krzesinski}, \& {French
  Leger}}]{2010AJ....139.1628D}
{Doi}, M., {Tanaka}, M., {Fukugita}, M., {et~al.} 2010, \aj, 139, 1628

\bibitem[{{Fern{\'a}ndez} \& {Sosa}(2015)}]{2015P&SS..118...14F}
{Fern{\'a}ndez}, J.~A. \& {Sosa}, A. 2015, \planss, 118, 14

\bibitem[{{Finson} \& {Probstein}(1968)}]{1968ApJ...154..327F}
{Finson}, M.~J. \& {Probstein}, R.~F. 1968, \apj, 154, 327

\bibitem[{{Fornasier} {et~al.}(2015){Fornasier}, {Hasselmann}, {Barucci},
  {Feller}, {Besse}, {Leyrat}, {Lara}, {Gutierrez}, {Oklay}, {Tubiana},
  {Scholten}, {Sierks}, {Barbieri}, {Lamy}, {Rodrigo}, {Koschny}, {Rickman},
  {Keller}, {Agarwal}, {A'Hearn}, {Bertaux}, {Bertini}, {Cremonese}, {Da
  Deppo}, {Davidsson}, {Debei}, {De Cecco}, {Fulle}, {Groussin}, {G{\"u}ttler},
  {Hviid}, {Ip}, {Jorda}, {Knollenberg}, {Kovacs}, {Kramm}, {K{\"u}hrt},
  {K{\"u}ppers}, {La Forgia}, {Lazzarin}, {Lopez Moreno}, {Marzari}, {Matz},
  {Michalik}, {Moreno}, {Mottola}, {Naletto}, {Pajola}, {Pommerol}, {Preusker},
  {Shi}, {Snodgrass}, {Thomas}, \& {Vincent}}]{2015A&A...583A..30F}
{Fornasier}, S., {Hasselmann}, P.~H., {Barucci}, M.~A., {et~al.} 2015, \aap,
  583, A30

\bibitem[{{Frattin} {et~al.}(2017){Frattin}, {Cremonese}, {Simioni}, {Bertini},
  {Lazzarin}, {Ott}, {Drolshagen}, {La Forgia}, {Sierks}, {Barbieri}, {Lamy},
  {Rodrigo}, {Koschny}, {Rickman}, {Keller}, {Agarwal}, {A'Hearn}, {Barucci},
  {Bertaux}, {Da Deppo}, {Davidsson}, {Debei}, {De Cecco}, {Deller}, {Ferrari},
  {Ferri}, {Fornasier}, {Fulle}, {Gicquel}, {Groussin}, {Gutierrez},
  {G{\"u}ttler}, {Hofmann}, {Hviid}, {Ip}, {Jorda}, {Knollenberg}, {Kramm},
  {K{\"u}hrt}, {K{\"u}ppers}, {Lara}, {Lopez Moreno}, {Lucchetti}, {Marzari},
  {Massironi}, {Mottola}, {Naletto}, {Oklay}, {Pajola}, {Penasa}, {Shi},
  {Thomas}, {Tubiana}, \& {Vincent}}]{2017MNRAS.469S.195F}
{Frattin}, E., {Cremonese}, G., {Simioni}, E., {et~al.} 2017, \mnras, 469, S195

\bibitem[{{Guti{\'e}rrez} {et~al.}(2006){Guti{\'e}rrez}, {Davidsson}, {Ortiz},
  {Rodrigo}, \& {Vidal-Nu{\~n}ez}}]{2006A&A...454..367G}
{Guti{\'e}rrez}, P.~J., {Davidsson}, B.~J.~R., {Ortiz}, J.~L., {Rodrigo}, R.,
  \& {Vidal-Nu{\~n}ez}, M.~J. 2006, \aap, 454, 367

\bibitem[{{Howell} {et~al.}(2017){Howell}, {Lejoly}, {Taylor},
  {Rivera-Valentin}, {Zambrano-Marin}, {Giorgini}, {Nolan}, {Samarasinha},
  {Mueller}, {Aponte-Hernandez}, {Saran Bhiravarasu}, {Rodriguez
  Sanchez-Vahamonde}, \& {Harris}}]{2017DPS....4941424H}
{Howell}, E.~S., {Lejoly}, C., {Taylor}, P.~A., {et~al.} 2017, in AAS/Division
  for Planetary Sciences Meeting Abstracts \#49, 414.24

\bibitem[{{Hughes}(1991)}]{1991kap..book.....H}
{Hughes}, D.~W. 1991, {Comets in the Post-Halley Era}, ed. R.~L. {Newburn Jr.},
  M.~{Neugebauer}, \& R.~{Rahe} ({Kluwer Acad. Publ.}), 825

\bibitem[{{Kolokolova} {et~al.}(2004){Kolokolova}, {Hanner},
  {Levasseur-Regourd}, \& {Gustafson}}]{2004come.book..577K}
{Kolokolova}, L., {Hanner}, M.~S., {Levasseur-Regourd}, A.~C., \& {Gustafson},
  B. {\r{A}}.~S. 2004, in Comets II, ed. M.~C. {Festou}, H.~U. {Keller}, \&
  H.~A. {Weaver} (University of Arizona Press, Tucson), 577

\bibitem[{{Kosyra} {et~al.}(2014){Kosyra}, {G{\"o}ssl}, {Hopp}, {Lang-Bardl},
  {Riffeser}, {Bender}, \& {Seitz}}]{2014ExA....38..213K}
{Kosyra}, R., {G{\"o}ssl}, C., {Hopp}, U., {et~al.} 2014, Experimental
  Astronomy, 38, 213

\bibitem[{{Kresak}(1991)}]{1991kap..book.....K}
{Kresak}, L. 1991, {Comets in the Post-Halley Era}, ed. R.~L. {Newburn Jr.},
  M.~{Neugebauer}, \& R.~{Rahe} ({Kluwer Acad. Publ.}), 607

\bibitem[{{Lacerda} \& {Jewitt}(2012)}]{2012ApJ...760L...2L}
{Lacerda}, P. \& {Jewitt}, D. 2012, \apjl, 760, L2

\bibitem[{{Lamy} {et~al.}(2004){Lamy}, {Toth}, {Fernandez}, \&
  {Weaver}}]{2004come.book..223L}
{Lamy}, P.~L., {Toth}, I., {Fernandez}, Y.~R., \& {Weaver}, H.~A. 2004, in
  Comets II, ed. M.~C. {Festou}, H.~U. {Keller}, \& H.~A. {Weaver} (University
  of Arizona Press, Tucson), 223

\bibitem[{{Lang-Bardl} {et~al.}(2010){Lang-Bardl}, {Hodapp}, {Jacobson},
  {Bender}, {G{\"o}ssl}, {Fabricius}, {Grupp}, {Hopp}, \&
  {Mitsch}}]{2010SPIE.7735E..3QL}
{Lang-Bardl}, F., {Hodapp}, K., {Jacobson}, S., {et~al.} 2010, in Proceedings
  of the SPIE, Volume 7735, 3Q

\bibitem[{{Lovell} {et~al.}(2017){Lovell}, {Amason}, {Howell},
  {Presler-Marshall}, {Reid}, {Cunningham}, {Frayer}, {Lockman}, \&
  {Church}}]{2017DPS....4942004L}
{Lovell}, A.~J., {Amason}, C., {Howell}, E.~S., {et~al.} 2017, in AAS/Division
  for Planetary Sciences Meeting Abstracts \#49, 420.04

\bibitem[{{Luk'yanyk} {et~al.}(2019){Luk'yanyk}, {Zubko}, {Hus{\'a}rik},
  {Ivanova}, {Svore{\r{A}}}, {Kochergin}, {Baransky}, \&
  {Videen}}]{2019MNRAS.485.4013L}
{Luk'yanyk}, I., {Zubko}, E., {Hus{\'a}rik}, M., {et~al.} 2019, \mnras, 485,
  4013

\bibitem[{{Magnier} {et~al.}(2013){Magnier}, {Schlafly}, {Finkbeiner}, {Juric},
  {Tonry}, {Burgett}, {Chambers}, {Flewelling}, {Kaiser}, {Kudritzki},
  {Morgan}, {Price}, {Sweeney}, \& {Stubbs}}]{2013ApJS..205...20M}
{Magnier}, E.~A., {Schlafly}, E., {Finkbeiner}, D., {et~al.} 2013, \apjs, 205,
  20

\bibitem[{{McKay} {et~al.}(2018){McKay}, {DiSanti}, {Cochran}, {Dello Russo},
  {Bonev}, {Vervack}, {Gibb}, {Roth}, \& {Kawakita}}]{2018AAS...23114413M}
{McKay}, A., {DiSanti}, M., {Cochran}, A., {et~al.} 2018, in American
  Astronomical Society Meeting Abstracts \#231, 144.13

\bibitem[{{McKay} {et~al.}(2017){McKay}, {DiSanti}, {Cochran}, {Dello Russo},
  {Bonev}, {Vervack}, {Gibb}, {Roth}, \& {Kawakita}}]{2017DPS....4930509M}
{McKay}, A., {DiSanti}, M.~A., {Cochran}, A.~L., {et~al.} 2017, in AAS/Division
  for Planetary Sciences Meeting Abstracts \#49, 305.09

\bibitem[{{Meech} \& {Jewitt}(1987)}]{1987A&A...187..585M}
{Meech}, K.~J. \& {Jewitt}, D.~C. 1987, \aap, 187, 585

\bibitem[{{Meech} {et~al.}(2017){Meech}, {Weryk}, {Micheli}, {Kleyna},
  {Hainaut}, {Jedicke}, {Wainscoat}, {Chambers}, {Keane}, {Petric}, {Denneau},
  {Magnier}, {Berger}, {Huber}, {Flewelling}, {Waters}, {Schunova-Lilly}, \&
  {Chastel}}]{2017Natur.552..378M}
{Meech}, K.~J., {Weryk}, R., {Micheli}, M., {et~al.} 2017, \nat, 552, 378

\bibitem[{{Moffat}(1969)}]{1969A&A.....3..455M}
{Moffat}, A.~F.~J. 1969, \aap, 3, 455

\bibitem[{{Moreno} {et~al.}(2018){Moreno}, {Guirado}, {Mu{\~n}oz}, {Bertini},
  {Tubiana}, {G{\"u}ttler}, {Fulle}, {Rotundi}, {Della Corte}, {Ivanovski},
  {Rinaldi}, {Bockel{\'e}e-Morvan}, {Zakharov}, {Agarwal}, {Mottola}, {Toth},
  {Frattin}, {Lara}, {Guti{\'e}rrez}, {Lin}, {Kolokolova}, {Sierks}, {Naletto},
  {Lamy}, {Rodrigo}, {Koschny}, {Davidsson}, {Barucci}, {Bertaux}, {Bodewits},
  {Cremonese}, {Da Deppo}, {Debei}, {De Cecco}, {Deller}, {Fornasier}, {Ip},
  {Keller}, {Lazzarin}, {L{\'o}pez-Moreno}, {Marzari}, \&
  {Shi}}]{2018AJ....156..237M}
{Moreno}, F., {Guirado}, D., {Mu{\~n}oz}, O., {et~al.} 2018, \aj, 156, 237

\bibitem[{{Moulane} {et~al.}(2018){Moulane}, {Jehin}, {Opitom}, {Pozuelos},
  {Manfroid}, {Benkhaldoun}, {Daassou}, \& {Gillon}}]{2018A+A...619A.156M}
{Moulane}, Y., {Jehin}, E., {Opitom}, C., {et~al.} 2018, \aap, 619, A156

\bibitem[{{M\"uller}(1999)}]{1999PhDT..........M}
{M\"uller}, M. 1999, Ph.~D.~Thesis (University of Heidelberg)

\bibitem[{{Pajola} {et~al.}(2017){Pajola}, {H{\"o}fner}, {Vincent}, {Oklay},
  {Scholten}, {Preusker}, {Mottola}, {Naletto}, {Fornasier}, {Lowry}, {Feller},
  {Hasselmann}, {G{\"u}ttler}, {Tubiana}, {Sierks}, {Barbieri}, {Lamy},
  {Rodrigo}, {Koschny}, {Rickman}, {Keller}, {Agarwal}, {A'Hearn}, {Barucci},
  {Bertaux}, {Bertini}, {Besse}, {Boudreault}, {Cremonese}, {da Deppo},
  {Davidsson}, {Debei}, {de Cecco}, {Deller}, {Deshapriya}, {El-Maarry},
  {Ferrari}, {Ferri}, {Fulle}, {Groussin}, {Gutierrez}, {Hofmann}, {Hviid},
  {Ip}, {Jorda}, {Knollenberg}, {Kovacs}, {Kramm}, {K{\"u}hrt}, {K{\"u}ppers},
  {Lara}, {Lin}, {Lazzarin}, {Lucchetti}, {Lopez Moreno}, {Marzari},
  {Massironi}, {Michalik}, {Penasa}, {Pommerol}, {Simioni}, {Thomas}, {Toth},
  \& {Baratti}}]{2017NatAs...1E..92P}
{Pajola}, M., {H{\"o}fner}, S., {Vincent}, J.~B., {et~al.} 2017, Nature
  Astronomy, 1, 0092

\bibitem[{{P{\"a}tzold} {et~al.}(2016){P{\"a}tzold}, {Andert}, {Hahn}, {Asmar},
  {Barriot}, {Bird}, {H{\"a}usler}, {Peter}, {Tellmann}, {Gr{\"u}n},
  {Weissman}, {Sierks}, {Jorda}, {Gaskell}, {Preusker}, \&
  {Scholten}}]{2016Natur.530...63P}
{P{\"a}tzold}, M., {Andert}, T., {Hahn}, M., {et~al.} 2016, \nat, 530, 63

\bibitem[{{Pozuelos} {et~al.}(2018){Pozuelos}, {Jehin}, {Moulane}, {Opitom},
  {Manfroid}, {Benkhaldoun}, \& {Gillon}}]{2018A&A...615A.154P}
{Pozuelos}, F.~J., {Jehin}, E., {Moulane}, Y., {et~al.} 2018, \aap, 615, A154

\bibitem[{{Pravec} {et~al.}(2002){Pravec}, {Harris}, \&
  {Michalowski}}]{2002aste.book..113P}
{Pravec}, P., {Harris}, A.~W., \& {Michalowski}, T. 2002, in Asteroids III, ed.
  W.~F. {Bottke Jr.}, A.~{Cellino}, P.~{Paolicchi}, \& R.~P. {Binzel}
  (University of Arizona Press, Tucson), 113--122

\bibitem[{{Rumyantsev} {et~al.}(2019){Rumyantsev}, {Kiselev}, \&
  {Ivanova}}]{2019SoSyR..53...91R}
{Rumyantsev}, V.~V., {Kiselev}, N.~N., \& {Ivanova}, A.~V. 2019, Solar System
  Research, 53, 91

\bibitem[{{Schleicher}(2010)}]{2010AJ....140..973S}
{Schleicher}, D.~G. 2010, {Composite Dust Phase Function for Comets - Details}
  ({https://asteroid.lowell.edu/comet/dustphase\_details.html})

\bibitem[{{Schleicher} {et~al.}(2019){Schleicher}, {Knight}, {Eisner}, \&
  {Thirouin}}]{2019AJ....157..108S}
{Schleicher}, D.~G., {Knight}, M.~M., {Eisner}, N.~L., \& {Thirouin}, A. 2019,
  \aj, 157, 108

\bibitem[{{Schleicher} {et~al.}(1998){Schleicher}, {Millis}, \&
  {Birch}}]{1998Icar..132..397S}
{Schleicher}, D.~G., {Millis}, R.~L., \& {Birch}, P.~V. 1998, \icarus, 132, 397

\bibitem[{{Sekanina}(1991)}]{1991kap..book.....S}
{Sekanina}, Z. 1991, {Comets in the Post-Halley Era}, ed. R.~L. {Newburn Jr.},
  M.~{Neugebauer}, \& R.~{Rahe} ({Kluwer Acad. Publ.}), 769

\bibitem[{{Sierks} {et~al.}(2015){Sierks}, {Barbieri}, {Lamy}, {Rodrigo},
  {Koschny}, {Rickman}, {Keller}, {Agarwal}, {A'Hearn}, {Angrilli}, {Auger},
  {Barucci}, {Bertaux}, {Bertini}, {Besse}, {Bodewits}, {Capanna}, {Cremonese},
  {Da Deppo}, {Davidsson}, {Debei}, {De Cecco}, {Ferri}, {Fornasier}, {Fulle},
  {Gaskell}, {Giacomini}, {Groussin}, {Gutierrez-Marques}, {Guti{\'e}rrez},
  {G{\"u}ttler}, {Hoekzema}, {Hviid}, {Ip}, {Jorda}, {Knollenberg}, {Kovacs},
  {Kramm}, {K{\"u}hrt}, {K{\"u}ppers}, {La Forgia}, {Lara}, {Lazzarin},
  {Leyrat}, {Lopez Moreno}, {Magrin}, {Marchi}, {Marzari}, {Massironi},
  {Michalik}, {Moissl}, {Mottola}, {Naletto}, {Oklay}, {Pajola}, {Pertile},
  {Preusker}, {Sabau}, {Scholten}, {Snodgrass}, {Thomas}, {Tubiana}, {Vincent},
  {Wenzel}, {Zaccariotto}, \& {P{\"a}tzold}}]{2015Sci...347a1044S}
{Sierks}, H., {Barbieri}, C., {Lamy}, P.~L., {et~al.} 2015, Science, 347,
  aaa1044

\bibitem[{{Snodgrass} {et~al.}(2016){Snodgrass}, {Jehin}, {Manfroid}, {Opitom},
  {Fitzsimmons}, {Tozzi}, {Faggi}, {Yang}, {Knight}, {Conn}, {Lister},
  {Hainaut}, {Bramich}, {Lowry}, {Rozek}, {Tubiana}, \&
  {Guilbert-Lepoutre}}]{2016A&A...588A..80S}
{Snodgrass}, C., {Jehin}, E., {Manfroid}, J., {et~al.} 2016, \aap, 588, A80

\bibitem[{{Snodgrass} {et~al.}(2013){Snodgrass}, {Tubiana}, {Bramich}, {Meech},
  {Boehnhardt}, \& {Barrera}}]{2013A&A...557A..33S}
{Snodgrass}, C., {Tubiana}, C., {Bramich}, D.~M., {et~al.} 2013, \aap, 557, A33

\bibitem[{{Tancredi} {et~al.}(2000){Tancredi}, {Fern{\'a}ndez}, {Rickman}, \&
  {Licandro}}]{2000A&AS..146...73T}
{Tancredi}, G., {Fern{\'a}ndez}, J.~A., {Rickman}, H., \& {Licandro}, J. 2000,
  \aaps, 146, 73

\bibitem[{{Tonry} {et~al.}(2012){Tonry}, {Stubbs}, {Lykke}, {Doherty},
  {Shivvers}, {Burgett}, {Chambers}, {Hodapp}, {Kaiser}, {Kudritzki},
  {Magnier}, {Morgan}, {Price}, \& {Wainscoat}}]{2012ApJ...750...99T}
{Tonry}, J.~L., {Stubbs}, C.~W., {Lykke}, K.~R., {et~al.} 2012, \apj, 750, 99

\bibitem[{{Vincent} {et~al.}(2016){Vincent}, {A'Hearn}, {Lin}, {El-Maarry},
  {Pajola}, {Sierks}, {Barbieri}, {Lamy}, {Rodrigo}, {Koschny}, {Rickman},
  {Keller}, {Agarwal}, {Barucci}, {Bertaux}, {Bertini}, {Besse}, {Bodewits},
  {Cremonese}, {Da Deppo}, {Davidsson}, {Debei}, {De Cecco}, {Deller},
  {Fornasier}, {Fulle}, {Gicquel}, {Groussin}, {Guti{\'e}rrez},
  {Guti{\'e}rrez-Marquez}, {G{\"u}ttler}, {H{\"o}fner}, {Hofmann}, {Hviid},
  {Ip}, {Jorda}, {Knollenberg}, {Kovacs}, {Kramm}, {K{\"u}hrt}, {K{\"u}ppers},
  {Lara}, {Lazzarin}, {Lopez Moreno}, {Marzari}, {Massironi}, {Mottola},
  {Naletto}, {Oklay}, {Preusker}, {Scholten}, {Shi}, {Thomas}, {Toth}, \&
  {Tubiana}}]{2016MNRAS.462S.184V}
{Vincent}, J.~B., {A'Hearn}, M.~F., {Lin}, Z.~Y., {et~al.} 2016, \mnras, 462,
  S184

\bibitem[{{Vsekhsvyatskii}(1964)}]{1964pcc..book.....V}
{Vsekhsvyatskii}, S.~K. 1964, {Physical characteristics of comets} ({Israel
  Program for Scientific Translations Ltd.})

\bibitem[{{Whipple} \& {Huebner}(1976)}]{1976ARA&A..14..143W}
{Whipple}, F.~L. \& {Huebner}, W.~F. 1976, \araa, 14, 143

\bibitem[{{Willmer}(2018)}]{2018ApJS..236...47W}
{Willmer}, C. N.~A. 2018, \apjs, 236, 47

\bibitem[{{Zappala} {et~al.}(1990){Zappala}, {Cellino}, {Barucci},
  {Fulchignoni}, \& {Lupishko}}]{1990A&A...231..548Z}
{Zappala}, V., {Cellino}, A., {Barucci}, A.~M., {Fulchignoni}, M., \&
  {Lupishko}, D.~F. 1990, \aap, 231, 548

\end{thebibliography}

\appendix
\section{}

\begin{table*}
  \centering
  \tiny
  \setlength{\tabcolsep}{3pt}
  \begin{tabular}{r|rrrr|r|r|rrr|rrr|rrr|rr}           
    \hline\hline
    Time to/from       &      &          &           &      &       &      & SDSS r &     &     & SDSS i  &     &     & SDSS g &     &     & TRP  & TRP  \\
    Perihelion   & $r$  & $\Delta$ & $\varphi$ & PR   & Inst. & Date & Time   & $N$ & TEX & Time    & $N$ & TEX & Time   & $N$ & TEX & Min  & Max  \\
    (days) & (AU) & (AU)     & (deg)     & (km) &       & (UT) & (UT)   &     & (s) & (UT)    &     & (s) & (UT)   &     & (s) & (\%) & (\%) \\
         \hline
    -32.73 & 1.135 & 0.178 & 34.32 & 25.81 &  3KK & 2017/03/11 &  00:06 - 00:56 &   6 &  360 &             &     &   &  &  &  & na & na \\
    -31.71 & 1.130 & 0.174 & 35.76 & 25.33 &  3KK & 2017/03/12 &  00:44 - 01:05 &  12 &  720 & 00:44 - 01:05 &  10 & 600 &  &  &  & 68 & 78 \\
    -30.90 & 1.126 & 0.172 & 36.93 & 24.95 &  3KK & 2017/03/12 &  18:15 - 22:25 & 104 & 6240 & 18:15 - 22:25 & 104 & 6240 &  &  &  & 42 & 84 \\
    -19.64 & 1.079 & 0.147 & 52.89 & 21.35 &  3KK & 2017/03/24 &  02:25 - 02:53 &  13 &  780 & 02:25 - 02:53 &   2 & 120 &  &  &  & 29 & 56 \\
    -18.84 & 1.076 & 0.146 & 53.95 & 21.21 &  3KK & 2017/03/24 &  21:39 - 22:09 &  13 &  780 & 21:39 - 22:09 &  13 & 780 &  &  &  & 43 & 67 \\
    -17.97 & 1.073 & 0.145 & 55.08 & 21.07 &  3KK & 2017/03/25 &  18:29 - 18:58 &  11 &  660 & 18:29 - 18:58 &   6 & 360 &  &  &  & 51 & 59 \\
    -16.95 & 1.070 & 0.144 & 56.39 & 20.94 &  3KK & 2017/03/26 &  19:05 - 19:30 &  60 &  780 & 19:05 - 19:21 &   7 & 420 &  &  &  & 58 & 62 \\
    -15.71 & 1.067 & 0.143 & 57.90 & 20.80 &  3KK & 2017/03/28 &  00:36 - 01:06 &  13 &  780 & 00:36 - 01:06 &  10 & 600 &  &  &  & 72 & 78 \\
    -14.79 & 1.064 & 0.143 & 59.01 & 20.73 &  3KK & 2017/03/28 &  22:53 - 23:23 &  12 &  720 & 22:53 - 23:23 &  10 & 600 &  &  &  & 62 & 68 \\
    -13.88 & 1.062 & 0.142 & 60.03 & 20.67 &  3KK & 2017/03/29 &  20:33 - 21:02 &  13 &  780 & 20:33 - 21:02 &  12 & 720 &  &  &  & 52 & 59 \\
    -12.95 & 1.060 & 0.142 & 61.05 & 20.64 &  3KK & 2017/03/30 &  18:56 - 19:20 &  13 &  780 & 18:56 - 19:20 &  13 & 780 &  &  &  & 65 & 72 \\
     -9.73 & 1.053 & 0.142 & 64.21 & 20.66 &  3KK & 2017/04/03 &  00:03 - 01:07 &  25 & 1500 & 00:03 - 01:07 &  17 & 1020 &  &  &  & 37 & 59 \\
     -3.61 & 1.046 & 0.147 & 68.40 & 21.31 &  3KK & 2017/04/09 &  02:53 - 03:44 &  22 & 1320 & 02:53 - 03:42 &  21 & 1260 &  &  &  & 57 & 69 \\
     -2.67 & 1.046 & 0.148 & 68.83 & 21.47 &  3KK & 2017/04/10 &  01:46 - 02:16 &  13 &  780 & 01:46 - 02:16 &  13 & 780 &  &  &  & 67 & 76 \\
     -2.61 & 1.046 & 0.148 & 68.85 & 21.38 & WWFI & 2017/04/10 &  03:15 - 03:24 &   5 &  300 &       -       &     &   &  &  &  & 62 & 82 \\
     -0.64 & 1.045 & 0.151 & 69.52 & 21.87 &  3KK & 2017/04/12 &  02:19 - 02:47 &  13 &  780 & 02:19 - 02:47 &  13 & 780 &  &  &  & 54 & 62 \\
     -0.61 & 1.045 & 0.151 & 69.53 & 21.77 & WWFI & 2017/04/12 &  03:16 - 03:20 &   2 &  120 &       -       &     &   &  &  &  & 29 & 49 \\
      8.30 & 1.051 & 0.167 & 69.52 & 24.20 & WWFI & 2017/04/21 &  01:04 - 01:29 &  12 &  720 &       -       &     &   &  &  &  & 82 & 86 \\
      8.34 & 1.051 & 0.167 & 69.51 & 24.33 &  3KK & 2017/04/21 &  01:41 - 02:49 &  26 & 1560 & 01:41 - 02:49 &  24 & 1440 &  &  &  & 65 & 69 \\
     11.36 & 1.056 & 0.174 & 68.51 & 25.22 & WWFI & 2017/04/24 &  02:18 - 02:39 &  11 &  660 &       -       &     &   &  &  &  & 72 & 84 \\
     12.22 & 1.058 & 0.177 & 68.15 & 25.65 &  3KK & 2017/04/24 &  22:42 - 23:40 &  25 & 1500 & 22:42 - 23:40 &  24 & 1440 &  &  &  & 59 & 70 \\
     17.22 & 1.071 & 0.189 & 65.46 & 27.37 & WWFI & 2017/04/29 &  23:00 - 23:25 &  13 &  780 &       -       &     &   &  &  &  & 74 & 92 \\
     18.22 & 1.074 & 0.192 & 64.83 & 27.76 & WWFI & 2017/04/30 &  23:02 - 23:23 &  13 &  780 & 22:34 - 22:59 &  13 & 780 & 00:59 - 01:05 & 2 & 120 & 26 & 72 \\
     23.26 & 1.092 & 0.206 & 61.16 & 29.93 &  3KK & 2017/05/05 &  23:36 - 00:43 &  30 & 1800 & 23:36 - 00:43 &  27 & 1620 &  &  &  & 67 & 79 \\
     27.24 & 1.109 & 0.218 & 57.85 & 31.62 &  3KK & 2017/05/09 &  23:36 - 00:05 &  13 &  780 & 23:36 - 00:05 &  13 & 780 &  &  &  & 71 & 77 \\
     27.28 & 1.109 & 0.218 & 57.83 & 31.47 & WWFI & 2017/05/10 &  00:24 - 00:49 &  13 &  780 &       -       &     &   &  &  &  & 92 & 95 \\
     28.33 & 1.113 & 0.221 & 56.90 & 31.93 & WWFI & 2017/05/11 &  01:13 - 02:29 &  18 & 1080 &       -       &     &   &  &  &  & 59 & 63 \\
     31.10 & 1.127 & 0.229 & 54.39 & 33.31 &  3KK & 2017/05/13 &  20:11 - 20:33 &  10 &  600 & 20:11 - 20:39 &  11 & 660 &  &  &  & 22 & 64 \\
     31.26 & 1.128 & 0.230 & 54.24 & 33.22 & WWFI & 2017/05/14 &                &     &      & 00:07 - 00:14 &   4 & 240 &  &  &  & 82 & 89 \\
     34.20 & 1.143 & 0.239 & 51.46 & 34.55 & WWFI & 2017/05/16 &  22:43 - 23:01 &  12 &  720 &       -       &     &   &  &  &  & 73 & 79 \\
     36.14 & 1.153 & 0.245 & 49.58 & 35.64 &  3KK & 2017/05/18 &  21:05 - 21:33 &  13 &  780 & 21:05 - 21:33 &  10 & 600 &  &  &  & 15 & 54 \\
     36.23 & 1.154 & 0.246 & 49.49 & 35.50 & WWFI & 2017/05/18 &  23:26 - 23:45 &   3 &  180 & 23:46 - 00:05 &  13 & 780 & 00:06 - 00:24 & 12 & 720 & 58 & 86 \\
     39.27 & 1.171 & 0.256 & 46.48 & 36.97 & WWFI & 2017/05/22 &  00:18 - 00:37 &  13 &  780 & 00:38 - 00:57 &  13 & 780 & 00:58 - 01:17 & 13 & 780 & 82 & 92 \\
     40.17 & 1.177 & 0.259 & 45.58 & 37.42 & WWFI & 2017/05/22 &  21:51 - 22:10 &  13 &  780 & 22:11 - 22:30 &  13 & 780 & 22:30 - 22:49 & 13 & 780 & 68 & 87 \\
     43.20 & 1.095 & 0.270 & 42.51 & 38.97 & WWFI & 2017/05/25 &  22:33 - 22:52 &  12 &  720 & 22:52 - 23:10 &  10 & 600 & 23:12 - 23:18 & 5 & 300 & 65 & 87 \\
     47.13 & 1.221 & 0.284 & 38.51 & 41.12 & WWFI & 2017/05/29 &  20:54 - 21:12 &  13 &  780 & 21:13 - 21:18 &   3 & 180 & 21:33 - 21:52 & 13 & 780 & 44 & 71 \\
     51.21 & 1.249 & 0.301 & 34.39 & 43.54 & WWFI & 2017/06/02 &  20:55 - 01:01 &  25 & 1500 & 21:14 - 21:33 &  25 & 1500 & 21:34 - 21:53 & 13 & 780 & 68 & 87 \\
     52.22 & 1.256 & 0.306 & 33.37 & 44.18 & WWFI & 2017/06/03 &  21:41 - 01:00 &  33 & 1980 &             &     &  &  &  &  & 50 & 78 \\
     53.26 & 1.263 & 0.310 & 32.36 & 44.83 & WWFI & 2017/06/05 &  00:07 - 00:14 &   5 &  300 &             &     &  &  &  &  & 24 & 27 \\
     54.21 & 1.270 & 0.314 & 31.43 & 45.46 & WWFI & 2017/06/05 &  22:44 - 23:18 &  21 & 1260 &             &     &  &  &  &  & 40 & 67 \\
     57.14 & 1.291 & 0.328 & 28.63 & 47.47 & WWFI & 2017/06/08 &  20:24 - 22:54 &  43 & 2580 &             &     &  &  &  &  & 85 & 95 \\
     58.27 & 1.307 & 0.339 & 26.67 & 49.01 & WWFI & 2017/06/11 &  00:08 - 00:47 &  27 & 1620 &             &     &  &  &  &  & 93 & 94 \\
     62.21 & 1.329 & 0.355 & 24.10 & 51.27 & WWFI & 2017/06/13 &  21:32 - 23:47 &  20 & 1200 &             &     &  &  &  &  & 55 & 74 \\
     63.27 & 1.337 & 0.361 & 23.21 & 52.13 & WWFI & 2017/06/15 &  00:16 - 00:35 &  13 &  780 &             &     &  &  &  &  & 74 & 83 \\
     70.19 & 1.392 & 0.403 & 18.20 & 58.31 & WWFI & 2017/06/21 &  22:24 - 22:54 &  17 & 1020 &             &     &  &  &  &  & 75 & 82 \\
     72.29 & 1.417 & 0.425 & 16.50 & 61.41 & WWFI & 2017/06/25 &  00:43 - 01:20 &  21 & 1260 &             &     &  &  &  &  & 65 & 80 \\
     76.16 & 1.440 & 0.446 & 15.29 & 64.48 & WWFI & 2017/06/27 &  21:19 - 22:20 &  26 & 1560 &             &     &  &  &  &  & 33 & 75 \\
     82.27 & 1.499 & 0.505 & 13.99 & 72.95 & WWFI & 2017/07/05 &  00:05 - 00:47 &  23 & 1380 &             &     &  &  &  &  & 87 & 89 \\
     92.16 & 1.574 & 0.590 & 15.15 & 85.30 & WWFI & 2017/07/13 &  21:39 - 22:05 &  14 &  840 &             &     &  &  &  &  & 73 & 82 \\
     95.20 & 1.600 & 0.622 & 15.95 & 89.97 & WWFI & 2017/07/16 &  22:45 - 22:52 &   5 &  300 &             &     &  &  &  &  & 35 & 86 \\
     96.14 & 1.608 & 0.633 & 16.22 & 91.47 & WWFI & 2017/07/17 &  21:09 - 21:46 &  25 & 1500 &             &     &  &  &  &  & 71 & 84 \\
    \hline
  \end{tabular}
  \caption{Log file of the observations of comet 41P at the Mt. Wendelstein 2m
    Fraunhofer telescope. The table rows list for each observing epoch: Time
    to/from perihelion (in days), Sun distance $r$ (in AU), Earth distance
    $\Delta$ (in AU), phase angle $\varphi$ (in deg), pixel resolution PR (in
    km at the distance of the comet), instrument Inst. (3KK or WWFI) used for
    the observations of the comet, civil date (in UT; format year/month/day), 
    and thereafter per filter
    SDSS r, SDSS i, SDSS g the time interval (in UT), the number N of images
    available and the total integration time TEX (in s) of the respective filter
    sequence. The last two columns give values for the minimum/maximum
    transparency TRP Min/Max (in percent) in the field of view of the comet
    during the nightly exposure series compared to best possible transparency
    for the site (na = not applicable due to unreliable flux calibration of the
    instrument). Julian Date of perihelion passage of 41P is 2457856.24959 (2017 April 12).}
  \label{tab1}
\end{table*}

\begin{table*}
  \centering
  \tiny
  \setlength{\tabcolsep}{4pt}  
  \begin{tabular}{r|rrr|rrr|rrr}           
    \hline\hline
    Time        &  SDSS r &        &         & SDSS i  &         &         &  SDSS g &        &         \\
    to/from     &  1000km & 5000km & 10000km & 1000km  & 5000km  & 10000km &  1000km & 5000km & 10000km \\
    Perihelion  &  Bright.& Bright.& Bright. & Bright. & Bright. & Bright. &  Bright. & Bright. & Bright. \\
    (days)      &  (mag)  & (mag)  & (mag)   & (mag)   & (mag)   & (mag)   &  (mag) & (mag) & (mag) \\
    \hline
    -32.73  & 13.94 & 11.50 & 10.17 &       &       &       &        &       &  \\
    -31.71  & 13.62 & 11.94 & 11.21 & 13.49 & 11.87 & 11.21 &        &       &  \\
    -30.90  & 13.54 & 11.44 & 10.38 & 13.37 & 11.23 & 10.15 &        &       &  \\
    -19.64  & 13.33 & 11.59 & 10.87 & 13.16 & 11.51 & 10.88 &        &       &  \\
    -18.84  & 13.30 & 11.53 & 10.78 & 13.15 & 11.47 & 10.80 &        &       &  \\
    -17.97  & 13.20 & 11.48 & 10.73 & 13.04 & 11.41 & 10.75 &        &       &  \\
    -16.95  & 13.26 & 11.49 & 10.73 & 13.08 & 11.39 & 10.71 &        &       &  \\
    -15.71  & 13.30 & 11.51 & 10.75 & 13.18 & 11.48 & 10.80 &        &       &  \\
    -14.79  & 13.32 & 11.51 & 10.75 & 13.18 & 11.47 & 10.78 &        &       &  \\
    -13.88  & 13.43 & 11.56 & 10.78 & 13.29 & 11.52 & 10.82 &        &       &  \\
    -12.95  & 13.59 & 11.60 & 10.83 & 13.41 & 11.51 & 10.80 &        &       &  \\
     -9.73  & 13.54 & 11.60 & 10.82 & 13.40 & 11.57 & 10.85 &        &       &  \\
     -3.61  & 13.48 & 11.61 & 10.79 & 13.34 & 11.56 & 10.79 &        &       &  \\
     -2.67  & 13.70 & 11.69 & 10.86 & 13.55 & 11.63 & 10.87 &        &       &  \\
     -2.61  & 13.66 & 11.71 & 10.93 &       &       &       &        &       &  \\
     -0.64  & 13.51 & 11.61 & 10.79 & 13.38 & 11.60 & 10.85 &        &       &  \\
     -0.61  & 13.61 & 11.81 & 11.13 &       &       &       &        &       &  \\
      8.30  & 13.88 & 11.93 & 11.08 &       &       &       &        &       &  \\
      8.34  & 13.84 & 11.90 & 11.05 & 13.74 & 11.90 & 11.12 &        &       &  \\
     11.36  & 14.01 & 12.08 & 11.24 &       &       &       &        &       &  \\
     12.22  & 13.96 & 12.01 & 11.17 & 13.84 & 12.00 & 11.23 &        &       &  \\
     17.22  & 14.19 & 12.24 & 11.39 &       &       &       &        &       &  \\
     18.22  & 14.40 & 12.27 & 11.42 & 14.24 & 12.20 & 11.41 &  14.91 & 12.61 & 11.63 \\
     23.26  & 14.47 & 12.51 & 11.65 & 14.33 & 12.44 & 11.64 &        &  &  \\
     27.24  & 14.68 & 12.65 & 11.76 & 14.54 & 12.61 & 11.78 &        &  &  \\
     27.28  & 14.73 & 12.75 & 11.90 &       &       &       &        &  &  \\
     28.33  & 14.16 & 12.27 & 11.37 &       &       &       &        &  &  \\
     31.10  & 14.58 & 12.67 & 11.86 & 14.41 & 12.55 & 11.77 &        &  &  \\
     31.26  &       &       &       & 14.41 & 12.55 & 11.75 &        &  &  \\
     34.20  & 14.80 & 12.92 & 12.09 &       &       &       &        &  &  \\
     36.14  & 14.89 & 12.95 & 12.10 & 14.73 & 12.87 & 12.10 &        &  &  \\
     36.23  & 14.98 & 12.96 & 12.14 & 14.82 & 12.85 & 12.06 &  15.51 & 13.34 & 12.41 \\
     39.27  & 14.89 & 13.10 & 12.31 & 14.73 & 12.99 & 12.24 &  15.40 & 13.47 & 12.55 \\
     40.17  & 15.17 & 13.20 & 12.39 & 14.99 & 13.08 & 12.31 &  15.67 & 13.58 & 12.67 \\
     43.20  & 14.84 & 13.02 & 12.30 & 14.66 & 12.89 & 12.20 &  15.35 & 13.41 & 12.50 \\
     47.13  & 14.96 & 13.24 & 12.45 & 14.81 & 13.14 & 12.40 &  15.48 & 13.61 & 12.73 \\
     51.21  & 15.29 & 13.42 & 12.64 & 15.09 & 13.27 & 12.52 &  15.79 & 13.82 & 12.93 \\
     52.22  & 15.41 & 13.53 & 12.67 &       &       &       &        &  &  \\
     53.26  & 15.58 & 13.64 & 12.86 &       &       &       &        &  &  \\
     54.21  & 15.41 & 13.53 & 12.67 &       &       &       &        &  &  \\
     57.14  & 15.87 & 13.87 & 13.08 &       &       &       &        &  &  \\
     58.27  & 15.98 & 14.04 & 13.22 &       &       &       &        &  &  \\
     62.21  & 16.16 & 14.30 & 13.52 &       &       &       &        &  &  \\
     63.27  & 16.04 & 14.27 & 13.45 &       &       &       &        &  &  \\
     70.19  & 16.61 & 14.60 & 13.79 &       &       &       &        &  &  \\
     72.29  & 16.71 & 14.83 & 14.00 &       &       &       &        &  &  \\
     76.16  & 16.87 & 14.91 & 14.11 &       &       &       &        &  &  \\
     82.27  & 17.36 & 15.52 & 14.68 &       &       &       &        &  &  \\
     92.16  & 18.04 & 16.23 & 15.35 &       &       &       &        &  &  \\
     95.20  & 18.34 & 16.52 & 15.72 &       &       &       &        &  &  \\
     96.14  & 18.46 & 16.60 & 15.67 &       &       &       &        &  &  \\
    \hline
  \end{tabular}
  \caption{Aperture photometry of comet 41P. The table lists (for the observing
    epochs of Tab.~\ref{tab1}) the brightness of the comet in SDSS r, SDSS i,
    SDSS g filters, measured in circular apertures of 1\,000, 5\,000, 10\,000 km
    diameter (at the distance of the comet) centered on the central brightness
    peak in the cometary coma. The typical maximum uncertainty is $\pm$0.03 mag
    for brightness of up to 16 mag, and  $\pm$0.04 mag for fainter magnitudes.}
  \label{tab3}
\end{table*}

\begin{table*}
  \centering
  \tiny
  \setlength{\tabcolsep}{2.5pt} 
  \begin{tabular}{r|r|rrrr|rrrr|rrrr}           
    \hline\hline
    Time         & Inst. & SDSS r   &            &            &            & SDSS i   &            &            &            & SDSS g   &            &            &            \\
    to/from      &       & Coma     & Nucleus    & Nucleus    & Nucleus    & Coma     & Nucleus    & Nucleus    & Nucleus    & Coma     & Nucleus    & Nucleus    & Nucleus    \\
    Perihelion   &       & Gradient & Estimated  & Absolute   & Equivalent & Gradient & Estimated  & Absolute   & Equivalent & Gradient & Estimated  & Absolute   & Equivalent \\
    (days)       &       &          & Brightness & Brightness & Radius     &          & Brightness & Brightness & Radius     &          & Brightness & Brightness & Radius     \\
                 &       &          & (mag)      & (mag)      & (km)       &          & (mag)      & (mag)      & (km)       &          & (mag)      & (mag)      & (km)       \\
    \hline
    -31.71 &  3KK &  1.00 & 15.7 & 18.2 & 0.72 &  1.02 & 15.5 & 18.0 & 0.78 &       &      &      &      \\
    -30.90 &  3KK &  0.90 & 15.4 & 17.9 & 0.82 &  0.91 & 15.1 & 17.6 & 0.93 &       &      &      &      \\
    -19.64 &  3KK &  0.98 & 16.3 & 18.7 & 0.56 &  1.00 & 16.2 & 18.6 & 0.57 &       &      &      &      \\
    -18.84 &  3KK &  0.99 & 17.4 & 19.8 & 0.34 &  1.03 & 17.1 & 19.5 & 0.38 &       &      &      &      \\
    -17.97 &  3KK &  1.04 & 17.1 & 19.5 & 0.39 &  1.11 & 18.0 & 20.4 & 0.25 &       &      &      &      \\
    -16.95 &  3KK &  0.99 & 16.1 & 18.5 & 0.62 &  1.05 & 16.0 & 18.4 & 0.64 &       &      &      &      \\
    -15.71 &  3KK &  1.03 & ne   & ne   & ne   &  1.00 & ne   & ne   & ne   &       &      &      &      \\
    -14.79 &  3KK &  0.97 & 16.2 & 18.5 & 0.61 &  1.02 & 16.1 & 18.4 & 0.63 &       &      &      &      \\
    -13.88 &  3KK &  0.87 & 15.6 & 17.9 & 0.81 &  0.93 & 15.4 & 17.7 & 0.87 &       &      &      &      \\
    -12.95 &  3KK &  0.86 & 16.7 & 19.0 & 0.49 &  0.89 & 16.7 & 19.0 & 0.48 &       &      &      &      \\
     -9.73 &  3KK &  0.92 & 17.9 & 20.1 & 0.29 &  0.97 & 17.8 & 20.0 & 0.30 &       &      &      &      \\
     -3.61 &  3KK &  0.93 & 16.6 & 18.6 & 0.58 &  0.98 & 16.6 & 18.6 & 0.57 &       &      &      &      \\
     -2.67 &  3KK &  0.83 & 16.3 & 18.3 & 0.68 &  0.90 & 16.2 & 18.2 & 0.70 &       &      &      &      \\
     -2.61 & WWFI &  0.85 & 16.2 & 18.2 & 0.71 &       &      &      &      &       &      &      &      \\
     -0.64 &  3KK &  0.93 & 17.4 & 19.3 & 0.42 &  1.00 & 17.6 & 19.5 & 0.38 &       &      &      &      \\
     -0.61 & WWFI &  0.96 & 17.4 & 19.3 & 0.42 &       &      &      &      &       &      &      &      \\
      8.30 & WWFI &  0.88 & 16.8 & 18.5 & 0.62 &       &      &      &      &       &      &      &      \\
      8.34 &  3KK &  0.90 & 16.8 & 18.5 & 0.62 &  0.94 & 16.7 & 18.4 & 0.64 &       &      &      &      \\
     11.36 & WWFI &  0.88 & 16.6 & 18.2 & 0.70 &       &      &      &      &       &      &      &      \\
     12.22 &  3KK &  0.93 & 17.3 & 18.9 & 0.51 &  1.00 & ne   & ne   & ne   &       &      &      &      \\
     17.22 & WWFI &  0.87 & 16.6 & 18.1 & 0.74 &       &      &      &      &       &      &      &      \\
     18.22 & WWFI &  0.90 & ne   & ne   & ne   &  0.90 & ne   & ne   & ne   &  0.75 & ne   & ne   & ne   \\
     23.26 &  3KK &  0.83 & 16.4 & 17.8 & 0.84 &  0.89 & 16.5 & 17.9 & 0.80 &       &      &      &      \\
     27.24 &  3KK &  0.83 & 17.5 & 18.9 & 0.52 &  0.89 & 17.3 & 18.7 & 0.56 &       &      &      &      \\
     27.28 & WWFI &  0.85 & 17.4 & 18.8 & 0.55 &       &      &      &      &       &      &      &      \\
     28.33 & WWFI &  0.98 & ne   & ne   & ne   &       &      &      &      &       &      &      &      \\
     31.10 &  3KK &  0.90 & 16.9 & 18.2 & 0.70 &  0.94 & 16.9 & 18.2 & 0.69 &       &      &      &      \\
     31.26 & WWFI &       &      &      &      &  0.95 & 17.2 & 18.5 & 0.60 &       &      &      &      \\
     34.20 & WWFI &  0.90 & 16.9 & 18.2 & 0.71 &       &      &      &      &       &      &      &      \\
     36.14 &  3KK &  0.89 & 17.0 & 18.3 & 0.69 &  0.94 & 17.1 & 18.4 & 0.65 &       &      &      &      \\
     36.23 & WWFI &  0.90 & ne   & ne   & ne   &  0.98 & ne   & ne   & ne   &  0.82 & ne   & ne   & ne   \\
     39.27 & WWFI &  0.96 & 17.4 & 18.6 & 0.58 &  1.00 & 17.3 & 18.5 & 0.60 &  0.82 & 17.1 & 18.3 & 0.77 \\
     40.17 & WWFI &  0.93 & 18.7 & 19.9 & 0.32 &  0.97 & 18.8 & 20.0 & 0.30 &  0.82 & 18.2 & 19.4 & 0.47 \\
     43.20 & WWFI &  1.07 & ne   & ne   & ne   &  1.14 & ne   & ne   & ne   &  0.93 & ne   & ne   & ne   \\
     47.13 & WWFI &  0.97 & 16.9 & 18.0 & 0.76 &  1.02 & 16.8 & 17.9 & 0.78 &  0.85 & 16.9 & 18.0 & 0.88 \\
     51.21 & WWFI &  1.01 & ne   & ne   & ne   &  1.07 & ne   & ne   & ne   &  0.92 & 18.9 & 20.0 & 0.36 \\
     52.22 & WWFI &  1.03 & ne   & ne   & ne   &       &      &      &      &       &      &      &      \\
     53.26 & WWFI &  0.82 & 18.6 & 19.7 & 0.36 &       &      &      &      &       &      &      &      \\
     54.21 & WWFI &  0.96 & ne   & ne   & ne   &       &      &      &      &       &      &      &      \\
     57.14 & WWFI &  0.93 & 19.3 & 20.3 & 0.27 &       &      &      &      &       &      &      &      \\
     58.27 & WWFI &  0.90 & 18.6 & 19.6 & 0.38 &       &      &      &      &       &      &      &      \\
     62.21 & WWFI &  1.00 & ne   & ne   & ne   &       &      &      &      &       &      &      &      \\
     63.27 & WWFI &  0.90 & 17.5 & 18.4 & 0.65 &       &      &      &      &       &      &      &      \\
     70.19 & WWFI &  0.93 & 19.0 & 19.7 & 0.35 &       &      &      &      &       &      &      &      \\
     72.29 & WWFI &  0.91 & 18.4 & 19.0 & 0.49 &       &      &      &      &       &      &      &      \\
     76.16 & WWFI &  0.97 & 19.3 & 19.8 & 0.34 &       &      &      &      &       &      &      &      \\
     82.27 & WWFI &  0.90 & 19.2 & 19.4 & 0.41 &       &      &      &      &       &      &      &      \\
     92.16 & WWFI &  0.86 & 19.2 & 18.9 & 0.51 &       &      &      &      &       &      &      &      \\
     95.20 & WWFI &  0.92 & 19.5 & 19.0 & 0.48 &       &      &      &      &       &      &      &      \\
    \hline
  \end{tabular}
  \caption{Estimated and absolute nucleus brightness (m(r,$\Delta$,$\varphi$) and
    $H(1,1,0)$), equivalent radius (R), and gradient of the near-nucleus coma of
    comet 41P. The date is given in days to/from perihelion. The table also
    includes the instrument and filter names. `Here, `ne' stands for not estimated. The
    uncertainty in the nucleus brightness estimations is about 0.2 mag for
    apparent magnitudes of 17-18 mag or brighter, and  is 0.3 mag for fainter
    flux levels of the nucleus. The uncertainty of the exponent of the radial
    flux profile in the inner coma is typically better than 0.02. The images 
    obtained on epochs -32.73 days (2017 March 11) and 96.14 days (2017 July 17) 
    did not allow useful results to be derived.}
  \label{tab4}
\end{table*}

\begin{table*}
  \centering
  \tiny
  \begin{tabular}{r|r|r|rr|rr}           
    \hline\hline
    Time to/from & Inst.  & PA    & PA       & PA       & PA       & PA       \\
    Perihelion   &        & Tail  & r vector & v vector & r - Tail & v - Tail \\
    (days)       &        & (deg) & (deg)    & (deg)    & (deg)    & (deg)    \\
    \hline                         
   -32.73  &  3KK & 150 & 148.2 & 263.8 & -1.8  & 113.8 \\
   -31.71  &  3KK & 152 & 149.2 & 263.1 & -2.8  & 111.1 \\
   -30.90  &  3KK & 155 & 150.0 & 262.7 & -5.0  & 107.7 \\
   -19.64  &  3KK & 170 & 167.1 & 264.0 & -2.9  &  94.0 \\
   -18.84  &  3KK & 168 & 169.0 & 265.1 & 1.0   &  97.1 \\
   -17.97  &  3KK & 168 & 171.3 & 266.5 & 3.3   &  98.5 \\
   -16.95  &  3KK & 170 & 174.2 & 268.4 & 4.2   &  98.4 \\
   -15.71  &  3KK & 175 & 177.8 & 271.0 & 2.8   &  96.0 \\
   -14.79  &  3KK & 177 & 180.9 & 273.4 & 3.9   &  96.4 \\
   -13.88  &  3KK & 187 & 184.1 & 276.0 & -2.9  &  89.0 \\
   -12.95  &  3KK & 185 & 187.5 & 278.7 & 2.5   &  93.7 \\
    -9.73  &  3KK & 197 & 199.8 & 289.5 & 2.8   &  92.5 \\
    -3.61  &  3KK & 215 & 220.4 & 308.7 & 5.4   &  93.7 \\
    -2.67  &  3KK & 225 & 222.9 & 311.1 & -2.1  &  86.1 \\
    -2.61  & WWFI & 220 & 222.9 & 311.1 & 2.9   &  91.1 \\
    -0.64  &  3KK & 228 & 227.0 & 315.2 & -1.0  &  87.2 \\
    -0.61  & WWFI & 225 & 227.0 & 315.2 & 2.0   &  90.2 \\
     8.30  &  3KK & 235 & 235.5 & 323.8 & 0.5   &  88.8 \\
     8.34  & WWFI & 235 & 235.5 & 323.8 & 0.5   &  88.8 \\
    11.36  & WWFI & 230 & 236.0 & 324.2 & 6.0   &  94.2 \\
    12.22  &  3KK & 240 & 236.0 & 324.2 & -4.0  &  84.2 \\
    17.22  & WWFI & 230 & 235.3 & 323.0 & 5.3   &  93.0 \\
    18.22  & WWFI & 230 & 235.1 & 322.6 & 5.1   &  92.6 \\
    23.26  &  3KK & 235 & 233.3 & 319.8 & -1.7  &  84.8 \\
    27.24  &  3KK & 235 & 231.6 & 317.0 & -3.4  &  82.0 \\
    27.28  & WWFI & 235 & 231.5 & 317.0 & -3.5  &  82.0 \\
    28.33  & WWFI & 230 & 231.1 & 316.2 & 1.1   &  86.2 \\
    31.10  &  3KK & 230 & 229.7 & 313.9 & -0.3  &  83.9 \\
    31.26  & WWFI & 231 & 229.7 & 313.9 & -1.3  &  82.9 \\  
    34.20  & WWFI & 230 & 228.0 & 311.2 & -2.0  &  81.2 \\
    36.14  &  3KK & 230 & 226.9 & 309.4 & -3.1  &  79.4 \\
    36.23  & WWFI & 230 & 226.8 & 309.3 & -3.2  &  79.3 \\
    39.27  & WWFI & 222 & 225.0 & 306.6 & 3.0   &  84.6 \\
    40.17  & WWFI & 225 & 224.4 & 305.7 & -0.6  &  80.7 \\
    43.20  & WWFI & 225 & 222.4 & 303.0 & -2.6  &  78.0 \\
    47.13  & WWFI & 225 & 219.5 & 299.5 & -5.5  &  74.5 \\
    51.21  & WWFI & 220 & 216.2 & 296.1 & -3.8  &  76.1 \\
    52.22  & WWFI & 220 & 215.2 & 295.2 & -4.8  &  75.2 \\
    53.26  & WWFI & 220 & 214.2 & 294.4 & -5.8  &  74.4 \\
    54.21  & WWFI & 210 & 213.3 & 293.6 & 3.3   &  83.6 \\
    57.14  & WWFI & 210 & 210.1 & 291.5 & 0.1   &  81.5 \\
    58.27  & WWFI & 205 & 207.5 & 290.0 & 2.5   &  85.0 \\
    62.21  & WWFI & 205 & 203.4 & 288.1 & -1.6  &  83.1 \\
    63.27  & WWFI & 210 & 201.7 & 287.5 & -8.3  &  77.5 \\
    70.19  & WWFI & 190 & 188.1 & 283.9 & -1.9  &  93.9 \\
    72.29  & WWFI & 190 & 180.3 & 282.6 & -9.7  &  92.6 \\
    76.16  & WWFI & 180 & 172.1 & 281.6 & -7.9  & 101.6 \\
    82.27  & WWFI & ne  & 149.7 & 279.5 & ne    &    ne \\
    92.16  & WWFI & ne  & 125.7 & 277.6 & ne    &    ne \\
    95.20  & WWFI & 135 & 119.6 & 277.2 & -15.4 & 142.2 \\
    96.14  & WWFI & 135 & 117.9 & 277.0 & -17.1 & 142.0 \\
    \hline
  \end{tabular}
  \caption{Position angle of the dust tail of 41P. The table lists the observing
    epoch (in days to/from perihelion), the instrument used (Inst.), the
    position angle PA (counted east over north) of the tail at the nucleus
    position in the image (PA Tail), the position angle of the extended radius
    vector of the comet projected in the sky (`\texttt{PA r vector}'), the
    `\texttt{PA}' of the velocity vector of the comet projected in the sky
    (`\texttt{PA v vector}'), the difference between `\texttt{PA Tail}' and
    `\texttt{PA r vector}', the difference between `\texttt{PA Tail}' and
    `\texttt{PA v vector}'. `\texttt{ne}' stands for not estimated. 
    `\texttt{PA Tail}' is given for the SDSS r filter images except for 
    epoch 31.26 days for which it is obtained for SDSS i filter. The
    uncertainty of `\texttt{PA Tail}' measurements is $\pm 3$ deg. }
  \label{tab7}
\end{table*}

\begin{table*}
  \centering
  \tiny
  \begin{tabular}{r|rr|rr|rrccc}           
    \hline\hline
    Time & Fan A & Fan A & Fan A & Fan A & Fan A & Fan A & Fan A & Fan A & Fan A \\
    to/from & PA & PA & PA & PA & PA & Opening & Curva- & Inten- & Exten- \\
    Perihelion & Aeast & Awest & Aeast & Awest & Abright & Angle & ture & sity & sion \\
            & obs & obs & sim & sim & obs & obs & obs & obs & obs \\
    (days) & (deg) & (deg) & (deg) & (deg) & (deg) & (deg) &   &   & (km) \\
        \hline
%    -32.73  &    &     &    &     &    &    &   &     &    \\
    -31.71 & -24 & -70 & -20 & -66 & -30 & 46 & nm & nm & nm \\
    -30.90 & -34 & -74 & -20 & -66 & -27 & 40 & nm & nm & nm \\
    -19.64 & -18 & -73 & -2 & -63 & -35 & 55 & s & 1.2 & 5000 \\
    -18.84 & -13 & -73 & 0 & -63 & -34 & 60 & s & 1.2 & 5000 \\
    -17.97 & -4 & -62 & 3 & -56 & -19 & 58 & s & 1.25 & 4000 \\
    -16.95 &  2 & -54 & 5 & -55 & -18 & 56 & s & 1.3 & 5000 \\
    -15.71 &  6 & -50 & 7 & -55 & -23 & 56 & s & 1.25 & 5000 \\
    -14.79 &  6 & -51 & 10 & -55 & -9 & 57 & s & 1.3 & 4000 \\
    -13.88 &  3 & -50 & 12 & -55 & -32 & 53 & s & 1.25 & 2500 \\
    -12.95 & 10 & -38 & 15 & -55 & -16 & 48 & s & 1.3 & 5000 \\
     -9.73 & 27 & -33 & 25 & -30 & -3 & 60 & s & 1.3 & 6000 \\
     -3.61 & 46 & -15 & 44 & -14 & 11 & 61 & s & 1.3 & 5000 \\
     -2.67 & 45 & -11 & 48 & -11 & 25 & 56 & s & 1.3 & 6000 \\
     -2.61 & 49 & -10 & 48 & -11 & 11 & 59 & s & 1.25 & 5000 \\
     -0.64 & 59 & -8 & 50 & -7 & 15 & 67 & s & 1.3 & 5000 \\
     -0.61 & 65 & -4 & 50 & -7 & 35 & 69 & s & 1.3 & 3000 \\
      8.30 & 64 &  2 & 59 & -2 & 37 & 62 & s & 1.15 & 7500 \\
      8.34 & 64 &  0 & 59 & -2 & 53 & 64 & s & 1.25 & 7500 \\
     11.36 & 60 &  2 & 61 & -2 & 44 & 58 & s & 1.25 & 6000 \\
     12.22 & 65 &  3 & 61 & -2 & 36 & 62 & s & 1.3 & 3000 \\
     17.22 & 60 &  0 & 63 & -3 & 43 & 60 & s & 1.2 & 6000 \\
     18.22 & 64 &  2 & 63 & -3 & 39 & 62 & s & 1.2 & 5000 \\
     23.26 & 66 & -3 & 65 & -6 & 35 & 69 & s & 1.3 & 5000 \\
     27.24 & 56 & -5 & 65 & -7 & 31 & 61 & s & 1.25 & 4000 \\
     27.28 & 60 & -7 & 65 & -7 & 31 & 67 & s & 1.25 & 6000 \\
     28.33 & 71 &  4 & 67 & -7 & 33 & 67 & s & 1.25 & 6000 \\
     31.10 & 62 & -4 & 67 & -7 & 35 & 66 & s  & 1.25 & 6000 \\
     34.20 & 60 & -7 & 70 & -9 & 14 & 67 & s & 1.25 & 6500 \\
     36.14 & 61 & -8 & 69 & -12 & 39 & 69 & s & 1.3 & 5000 \\
     36.23 & 68 & -7 & 69 & -12 & 23 & 75 & s & 1.3 & 6000 \\
     39.27 & 60 & -12 & 69 & -11 & 19 & 72 & s & 1.3 & 8000 \\
     40.17 & 56 & -15 & 68 & -11 & 16 & 71 & s & 1.3 & 6000 \\
     43.20 & 60 & -12 & 69 & -12 & 23 & 72 & s & 1.3 & 7000 \\
     47.13 & 64 & -10 & 70 & -13 & 26 & 74 & s & 1.2 & 6500 \\
     51.21 & 71 & -4 & 70 & -13 & 30 & 75 & s & 1.2 & 5000 \\
     52.22 & 75 & -3 & 70 & -13 & 19 & 78 & s & nm & 3000 \\
     53.26 & 62 & -12 & 70 & -14 & 14 & 74 & s & nm & 3000 \\
     54.21 & 73 & -12 & 71 & -14 & 18 & 85 & s & 1.25 & 3000 \\
     57.14 & 70 & -11 & 71 & -14 & 15 & 81 & s & 1.2 & 3500 \\
     58.27 & 69 & -12 & 72 & -15 & 27 & 81 & s & 1.2 & 3000 \\
     62.21 & 68 & -14 & 73 & -15 & 20 & 82 & s & 1.25 & 4000 \\
     63.27 & 66 & -22 & 73 & -15 &  5 & 88 & s & 1.2 & 3000 \\
     72.29 & 61 & -25 & 74 & -15 &  5 & 86 & s & 1.1 & 3000 \\
     76.16 & 71 & -23 & 77 & -15 & nm & 94 & s & 1.1 & 3500 \\
     82.27 & 72 & -38 & 80 & -15 & nm & 110 & s & 1.1 & 2000 \\
    \hline
  \end{tabular}
  \caption{Morphologic information on {\it fan A} in the coma of comet 41P.  The
    information is obtained from SDSS r filter images of the comet. The table
    lists for the various observing epochs (in days to/from perihelion): The
    position angle (PA, counted east over north) with respect to the nucleus of
    the eastern (`\texttt{Aeast}') and western (`\texttt{Awest}') edge of the
    fan cone, measured at 2\,000 km projected distance in the coma for an
    intensity ratio 5\% above the mean coma brightness at this distance;
    the position angles of `\texttt{Aeast}' and `\texttt{Awest}' obtained from
    the dust coma simulations applying the best-matching rotation axis and fan
    location information as mentioned in the text; the position angle of the
    brightest part (Abright) in the fan close to the nucleus (typically at a few
    100 km projected distance); the opening angle of the fan cone; the curvature
    of the fan (s = straight without curvature); the maximum intensity ratio
    with respect to the mean local coma level versus distance from the nucleus;
    the maximum extension at which the fan was still detected in the analyzed
    images. Results from observational data are indicated under column label
    `obs', those from simulations under column label `sim'. String `nm' indicates
    properties that could not be properly  measured in the images of the
    comet. Fan PAs for 2017 March 11 (-32.73 days), 2017 June 21 (70.19 days), 
    2017 July 13 (92.16 days), 2017 July 16 (95.20 days), 2017 July 17 (96.14 days) could not
    be measured. The fan PAs for the WWFI observations on 2017 May 14 (31.26 days)
    are compatible with the ones listed under epoch 31.10 days in the table above.
    The uncertainty of the measured PAs is 3 deg, that of the opening angle is  4 deg, 
    that of the maximum intensity is 0.03, and that of the maximum extension is 
    smaller than 500 km.}
  \label{tab8}
\end{table*}

\begin{table*}
  \centering
  \tiny
  \begin{tabular}{r|rr|rrrrr|l}           
    \hline\hline
    Jet      & Time       & Date       & Jet   & Jet & Jet & Jet & Jet & Comments \\
    Position & to/from    & Start      & Axis & Opening & Max.   & Geometry & Max.  &   \\
             & Perihelion & of Night   & PA & Angle & Intensity &   & Extension &   \\
             & (days)     & (UT)       & (deg) & (deg) & (ratio) &   & (km) &   \\
    \hline
          WJ & -19.64     & 2017/03/23 & 291 & 7 & 1.05 & s & 4000 & strong \\
          WJ & -18.84     & 2017/03/24 & 282 & 16 & 1.00 & s & 5000 & marginal, PAC=295deg ? \\
          WJ & -16.95     & 2017/03/26 & 290 & 7 & 1.00 & s & 3000 & weak \\
          WJ & -15.71     & 2017/03/27 & 305 & 10 & 1.00 & s & 5000 & weak, PAC=290deg ? \\
          WJ & -14.79     & 2017/03/28 & 300 & 15 & 1.02 & s & 3000 & clear \\
          WJ & -13.88     & 2017/03/29 & 303 & 21 & 1.10 & s & 6000 & clear/strong \\
    \hline
          EJ & -13.88     & 2017/03/29 & 31 & 18 & 1.20 & mp & 2000 & strong \\
    \hline
          WJ & -12.95     & 2017/03/30 & 310 & 18 & 1.15 & s & 3500 & strong \\
          WJ & -9.73      & 2017/04/02 & 331 & 13 & 1.00 & s & 7000 & weak \\
    \hline
          EJ & -2.67      & 2017/04/09 & 86 & 11 & 1.05 & mp & 2000 & clear \\
          EJ & -0.64      & 2017/04/11 & 86 & 14 & 1.05 & mp & 3000 & clear \\
          EJ & 8.30       & 2017/04/20 & 73 & 29 & 1.15 & mp & 3000 & clear, overlap with fan A\\
    \hline
          WJ & 8.34       & 2017/04/20 & 5 & 18 & 1.20 & s & 6000 & clear, overlap with fan A\\
          WJ & 11.36      & 2017/04/23 & 353 & 7 & 1.05 & s & 2500 & clear \\
          WJ & 17.22      & 2017/04/29 & 2 & 21 & 1.20 & mp & 4000 & strong \\
          WJ & 27.24      & 2017/05/09 & 357 & 16 & 1.20 & s & 3000 & weak, overlap with fan A \\
          WJ & 31.10      & 2017/05/13 & 11 & 23 & 1.25 & s & 4000 & strong, overlap with fan A \\
    \hline
          EJ & 34.20      & 2017/05/16 & 68* & 14 & 0.90 & s & 1500 & clear \\
          EJ & 39.27      & 2017/05/21 & 126* & 13 & 0.90 & s & 1500 & weak   \\
          EJ & 40.17      & 2017/05/22 & 127* & 15 & 0.90 & s & 1000 & weak   \\
          EJ & 43.20      & 2017/05/25 & 110* & nm & nm & nm & nm & weak, difficult to measure \\
          EJ & 47.13      & 2017/05/29 & 136 & 24 & 1.00 & s & 2000 & clear/weak \\
          EJ & 51.21      & 2017/06/02 & 77 & 7 & 1.05 & s & 3000 & clear \\
    \hline
  \end{tabular}
  \caption{Jet features in the coma of 41P. The table lists information of jets
    identified in the coma of the comet. The jets are seen outside of the region
    of coma {\it fan A}, either east (`EJ' in column `Jet Position') 
    or west of the fan structure (`WJ' in column `Jet Position'). Further 
    information given in the table is: Time to/from
    perihelion (days), start date of the observing night, position angle (PA,
    counted east over north) of the jet axis at 2\,000 km projected distance
    from the nucleus, jet opening angle, maximum intensity ratio seen in the jet
    compared to the mean flux of the coma at the respective radial distance from
    the coma center, jet geometry (s = straight, mp = small positive curvature,
    i.e. increasing PA with radial distance from the nucleus), maximum extension
    found in the processed images. The comment column provides additional
    information on the detectability of the jet structures (strong – very
    prominent, clear – well detectable, weak – weak detection, marginal – barely
    detectable) and on overlaps with {\it fan A}. PAC means PA of the jet at the
    coma center (note: in the cases mentioned the jet is not traceable to the
    coma center since it overlaps with {\it fan A}). Here, `nm' stands for not
    measurable. The uncertainty of the jet axis PA is 3 deg, that of the jet
    opening angle is 4 deg, that of the maximum intensity is 0.03, and that of
    the maximum extension is smaller than 500 km.}
  \label{tab9}
\end{table*}

\end{document}